\documentclass[conference,compsoc]{IEEEtran}
\ifCLASSOPTIONcompsoc
  \usepackage[nocompress]{cite}
\else
  \usepackage{cite}
\fi
\ifCLASSINFOpdf
\else
\fi
\usepackage{wrapfig}
\usepackage{etoolbox}
\makeatletter
\patchcmd{\@makecaption}
  {\scshape}
  {}
  {}
  {}
\makeatother

\newcommand{\comment}[1]{}

\usepackage{balance} 
\usepackage{algorithm}
\usepackage{algorithmic}
\usepackage{cite}
\usepackage{amsmath,amsthm,amssymb,amsfonts}
\usepackage{algorithmic}
\usepackage{subcaption}
\usepackage{graphicx}
\usepackage{textcomp}
\usepackage{multirow}
\usepackage{multicol}
\usepackage{xcolor,colortbl}
\def\BibTeX{{\rm B\kern-.05em{\sc i\kern-.025em b}\kern-.08em
    T\kern-.1667em\lower.7ex\hbox{E}\kern-.125emX}}
    
\usepackage{enumitem}

\theoremstyle{definition}
\newtheorem{property}{Property}

\newtheorem{definition}{Definition}
\newtheorem{lemma}{Lemma}
\newtheorem{theorem}{Theorem}
\newtheorem{corollary}{Corollary}

\begin{document}
%

\title{A Quantitative Theory of Bottleneck Structures for Data Networks}

\author{\IEEEauthorblockN{\newline \newline Technical Report, August 2021, New York, USA \\ \\ Jordi Ros-Giralt$^1$, Noah Amsel$^2$, Sruthi Yellamraju$^3$, James Ezick$^3$, Richard Lethin$^3$ 
\\ Yuang Jiang$^4$, Aosong Feng$^4$, Leandros Tassiulas$^4$ 
}
\IEEEauthorblockA{
{contact: jros@qti.qualcomm.com}, \\ $^1$Qualcomm Europe, Inc \\
$^2$Formerly of Qualcomm Technologies, Inc \\
$^3$Qualcomm Technologies, Inc \\
$^4$Yale, University}}



\maketitle


\begin{abstract}
The conventional view of the congestion control problem in data networks is based on the principle that a flow's performance is uniquely determined by the state of its bottleneck link, regardless of the topological properties of the network. However, recent work has shown that the behavior of congestion-controlled networks is better explained by models that account for the interactions between bottleneck links. These interactions are captured by a latent \textit{bottleneck structure}, a model describing the complex ripple effects that changes in one part of the network exert on the other parts. In this paper, we present a \textit{quantitative} theory of bottleneck structures (QTBS), a mathematical and engineering framework comprising a family of polynomial-time algorithms that can be used to reason about a wide variety of network optimization problems, including routing, capacity planning and flow control. QTBS can contribute to traffic engineering by making clear predictions about the relative performance of alternative flow routes, and by providing numerical recommendations for the optimal rate settings of traffic shapers. A particularly novel result in the domain of capacity planning indicates that previously established rules for the design of folded-Clos networks are suboptimal when flows are congestion controlled. We show that QTBS can be used to derive the optimal rules for this important class of topologies, and empirically demonstrate the correctness and efficacy of these results using the BBR and Cubic congestion-control algorithms.
\end{abstract}


%
\IEEEpeerreviewmaketitle

\section{Introduction}

Most research on the problem of congestion control for data networks is based on the principle that the performance of a flow is solely determined by the state of its bottleneck link. This view was presented in the original congestion control algorithm by Jacobson \cite{Jacobson:1988:CAC:52325.52356}, which helped the Internet recover from congestion collapse in 1988, and it persisted throughout the more than 30 years of research and development that followed, including Google's new BBR algorithm \cite{Cardwell:2016:BCC:3012426.3022184}. While it is certainly true that a flow's performance is limited by the state of its bottleneck link, recent work \cite{Ros-Giralt-SIGMETRICS-10.1145/3366707} reveals a deeper view of network behavior, describing how bottlenecks interact with each other through a latent structure---called the \textit{bottleneck structure}---that depends on the topological, routing and flow control properties of the network. This latent structure explains how the performance of one bottleneck can affect other bottlenecks, and provides a framework to understand how perturbations in the capacity of a link or the rate of a flow propagate through a network, affecting other links and flows. 


While \cite{Ros-Giralt-SIGMETRICS-10.1145/3366707} introduced the concept of bottleneck structure, the analysis provided was qualitative. In this paper we present a \textit{quantitative theory of bottleneck structures} (QTBS), a mathematical framework that yields a set of polynomial time algorithms for quantifying the ripple effects of perturbations in a network. Perturbations can either be unintentional (such as the effect of a link failure or the sudden arrival of a large flow in a network) or intentional (such as the upgrade of a network link to a higher capacity or the modification of a route with the goal of optimizing performance). With QTBS, a network operator can quantify the effect of such perturbations and use this information to optimize network performance.

\noindent The theoretical contributions of this paper are as follows:

\begin{itemize}[leftmargin=*]
    \item A new generalized bottleneck structure called \textit{gradient graph} is studied in detail. A key difference with the bottleneck structure introduced in \cite{Ros-Giralt-SIGMETRICS-10.1145/3366707} is that the gradient graph allows us to not only qualify the influences that flows and bottlenecks exert on each other, but also to quantify them. This leads to the development of a quantitative theory of bottleneck structures (QTBS), introduced in this paper. (Section \ref{ssec:gradgraph})
    \item A novel, fast algorithm to compute the gradient graph is developed. This algorithm constitutes an asymptotic speed-up compared to those presented in \cite{Ros-Giralt-SIGMETRICS-10.1145/3366707}, allowing us to scale our methodology to large production networks (Sections \ref{ssec:gradgraph}.)
    \item The concepts of \textit{link} and \textit{flow gradient} are introduced. These mathematical operators quantify the effects of infinitesimally small perturbations in a network, the core building blocks of QTBS. A new, fast method to efficiently compute the gradients by leveraging the bottleneck structure is presented. (Section \ref{ssec:linkflowgrad}.)
\end{itemize}
Applications demonstrating the practical implications of QTBS are provided in the areas of routing, capacity planning and flow control. In each of these applications, we show how QTBS can potentially alter some of the established conventional best practices. Our practical contributions are as follows:

\begin{itemize}[leftmargin=*]
    \item In the routing application, we introduce an algorithm to find maximal-throughput routes by anticipating the effects of the congestion control algorithm. While in traditional traffic engineering approaches (e.g., \cite{BWE-10.1145/2829988.2787478}) the problems of routing and flow control are considered independently, we show how QTBS can help resolve them jointly, allowing operators to design routes that are efficient from a congestion control standpoint. (Section \ref{ssec:routing}.)
    \item In the capacity planning application, we use QTBS to optimize the bandwidth allocation between the spine and leaf links of a fat-tree (also known as folded-Clos \cite{fat-tree-amin-10.1145/1402958.1402967}). We demonstrate that, due to the effects of congestion control, the optimal design differs from the full fat-tree configuration proposed by Leiserson \cite{leiserson1985fat}. (Section \ref{ssec:clos}.)
    \item In the flow control application, we show that QTBS can be used to precisely compute the rate reduction that a set of traffic shapers must impose on the network's low priority flows in order to achieve a quantifiable positive impact on the high-priority flows. (Section \ref{ssec:timebound}.)
    \item To demonstrate that networks behave according to QTBS, we carry out experiments for each application we consider using production TCP/IP code and the widely adopted BBR \cite{Cardwell:2016:BCC:3012426.3022184} and Cubic \cite{Cubic:Ha:2008:CNT:1400097.1400105} congestion control algorithms. (Section \ref{sec:applications}.)
\end{itemize}

\section{Quantitative Theory of Bottleneck Structures (QTBS)}

\subsection{Network Model}

In their simplest form, networks can be modeled using two kinds of elements: links, which are communication resources with a limited capacity; and flows, which make use of these communication resources. We formalize the definition of a network as follows:

\begin{definition} \textit{Network.} \label{def:network} 
We say that a tuple $ \mathcal{N} = \langle \mathcal{L}, \mathcal{F}, \{c_l,\forall l \in \mathcal{L} \} \rangle$ is a network if:
\begin{itemize}[leftmargin=*] 
    \item $\mathcal{L}$ is a set of links of the form $\{l_1, l_2, ..., l_{|\mathcal{L}|}\}$,
    \item $\mathcal{F}$ is a set of flows of the form $\{f_1, f_2, ..., f_{|\mathcal{F}|}\}$, and
    \item $c_l$ is the capacity of link $l$, for all $l \in \mathcal{L}$.
\end{itemize}
Each flow $f$ traverses a subset of links $\mathcal{L}_{f} \subset \mathcal{L}$ and, similarly, each link $l$ is traversed by a subset of flows $\mathcal{F}_{l} \subset \mathcal{F}$. Finally, each flow $f$ transmits data at a rate $r_f$ and the capacity constraint $\sum_{\forall f \in \mathcal{F}_l} r_f \leq c_l$ must hold for all $l \in \mathcal{L}$. 
\end{definition}



A core concept upon which our mathematical framework rests is the notion of a \textit{bottleneck link}. Intuitively, a flow is bottlenecked at a link if bypassing the link would allow its transmission rate to increase. A link whose capacity is fully utilized is always a bottleneck of at least one flow, though not necessarily of all the flows traversing it. In this work, we adopt the following formal definition:

\begin{definition} \textit{Bottleneck link.} \label{def:bottleneck} 
Let $ \mathcal{N} = \langle \mathcal{L}, \mathcal{F}, \{c_l,\forall l \in \mathcal{L} \} \rangle$ be a network where each flow $f \in \mathcal{F}$ transmits data at a rate $r_f$ as determined by a congestion control algorithm (e.g., TCP's algorithm \cite{Jacobson:1988:CAC:52325.52356}). We will say that flow $f$ is bottlenecked at link $l$---equivalently, that link $l$ is a bottleneck of flow $f$---if and only if: 
\begin{itemize}[leftmargin=*] 
    \item Flow $f$ traverses link $l$.
    \item $\partial r_f / \partial c_l^{-} \neq 0$. That is, the transmission rate of flow $f$ changes upon small reductions in link $l$'s capacity.\footnote{We use the notation $\partial y / \partial x^{-}$ to denote the left derivative. This subtlety is necessary because a flow can have multiple bottleneck links. In this case, decreasing the capacity of only one bottleneck would affect the rate of the flow, while increasing its capacity would not; thus, the (two-sided) derivative would not exist.}
\end{itemize}
\end{definition}

This characterization of bottlenecks is a generalization of some of the classic definitions found in the literature. Unlike previous work, however, it is based on the notion of a \textit{perturbation}, mathematically expressed as a derivative of a flow rate with respect to the capacity of a link ($\partial r_f / \partial c_l$). As an example to illustrate that our definition of bottleneck is relatively loose, in Appendix \ref{app:proof:maxmin-bottleneck} we show that it generalizes the classic max-min definition of Bertsekas and Gallager \cite{Bertsekas:1992:DN:121104}. The generality of the definition of bottlenecks used in this paper suggests that our framework can be applied to a wide variety of rate allocation schemes---not only to max-min fairness \cite{Bertsekas:1992:DN:121104}, proportional fairness \cite{Kelly1998} and specific algorithms (e.g., BBR \cite{Cardwell:2016:BCC:3012426.3022184}, Cubic \cite{Cubic:Ha:2008:CNT:1400097.1400105}, Reno \cite{Reno:Fall:1996:SCT:235160.235162}, etc.), but to other classes of congestion control solutions that meet the conditions of Definition \ref{def:bottleneck}. We leave this promising direction for future work, and focus on the classic max-min setting considered in \cite{Ros-Giralt-SIGMETRICS-10.1145/3366707}.



We complete the description of our network model by defining the concept of a link's \textit{fair share}:

\begin{definition} \textit{Fair share of a link.} \label{def:fairshare} 
Let $ \mathcal{N} = \langle \mathcal{L}, \mathcal{F}, \{c_l,\forall l \in \mathcal{L} \} \rangle$ be a network. The \textit{fair share} $s_l$ of a link $l \in \mathcal{L}$ is the rate of the flows that are bottlenecked at link $l$.
\end{definition}


As we will see throughout this work, the concept of link fair share is dual to the concept of flow rate, in that many of the mathematical properties that are applicable to the rate of a flow are also applicable to the fair share of a link.

\subsection{The Gradient Graph} \label{ssec:gradgraph}

Our objective is to derive a mathematical framework capable not just of detecting but also of quantifying the influences that links and flows exert on each other. In \cite{Ros-Giralt-SIGMETRICS-10.1145/3366707}, the authors introduced two bottleneck structures, the bottleneck precedence graph (BPG) and the gradient graph, and demonstrated that data networks qualitatively operate according to the BPG structure. The authors briefly described the concept of the gradient graph, but their work focused mostly on the mathematical properties of the bottleneck precedence graph. In our paper, we instead focus on a modified version of the gradient graph structure. Our work stems from the insight that, as we will show, this structure enables not just qualitative analysis, as in \cite{Ros-Giralt-SIGMETRICS-10.1145/3366707}, but also quantitative analysis, providing a framework to better understand and optimize network performance.

We start with the definition of the gradient graph:

\begin{definition} \textit{Gradient graph.} \label{def:gradgraph} 
Let $ \mathcal{N} = \langle \mathcal{L}, \mathcal{F}, \{c_l,\forall l \in \mathcal{L} \} \rangle$ be a network. The \textit{gradient graph} is a directed graph such that:
\begin{enumerate} 
    \item There exists a vertex for each bottleneck link and each flow in the network. \label{def:gradgraph:c1}
    \item For every flow $f \in \mathcal{F}$: \begin{enumerate} 
        \item If $f$ is bottlenecked at link $l \in \mathcal{L}$, then there exists a directed edge from $l$ to $f$; \label{def:gradgraph:linkedges}
        \item If $f$ traverses link $l \in \mathcal{L}$, then there exists a directed edge from $f$ to $l$; \label{def:gradgraph:backedges} \label{def:gradgraph:c2b}
    \end{enumerate} 
\end{enumerate}
For ease of exposition, in this paper we will use the terms gradient graph and bottleneck structure interchangeably. This definition is borrowed from \cite{Ros-Giralt-SIGMETRICS-10.1145/3366707}, except for a subtle but relevant modification of \ref{def:gradgraph:backedges}. (The rest of the theoretical developments presented in this work are new contributions.) Previously, edges were only included from flows to links that they traverse, but that do not bottleneck them. In this work, we also include edges from flows to their bottleneck links. We call these ``backward edges'', and we introduce them because they are required by several of our theorems and algorithms. 
\end{definition}

The utility of our definition of gradient graph as a data structure for understanding network performance is captured in the following theorem:

\begin{theorem}{\textit{Propagation of network perturbations.}}\label{lem:propagation}
Let $x, y \in \mathcal L \cup \mathcal F$ be a pair of links or flows in the network. Then a perturbation in the capacity $c_x$ (for $x \in \mathcal L$) or transmission rate $r_x$ (for $x \in \mathcal F$) of $x$ will affect the fair share $s_y$ (for $y \in \mathcal L$) or transmission rate $r_y$ (for $y \in \mathcal F$) of $y$ if only if there exists a directed path from $x$ to $y$ in the gradient graph.
\end{theorem}

\begin{proof} See Appendix \ref{app:proof:propagation}.
\end{proof}

Intuitively, the gradient graph of a network describes how perturbations in link capacities and flow transmission rates propagate through the network. Imagine that flow $f$ is bottlenecked at link $l$. From Definition \ref{def:bottleneck}, this necessarily implies that a perturbation in the capacity of link $l$ will cause a change on the transmission rate of flow $f$, $\partial r_f / \partial c_l \neq 0$. This is reflected in the gradient graph by the presence of a directed edge from a link $l$ to a flow $f$ (Condition \ref{def:gradgraph:linkedges} in Definition \ref{def:gradgraph}). A change in the value of $r_f$, in turn, affects all the other links traversed by flow $f$. This is reflected by the directed edges from $f$ to the links it traverses (Condition \ref{def:gradgraph:backedges}). This basic process of (1) inducing a perturbation in a vertex (either in a link or a flow vertex) followed by (2) propagating the effects of the perturbation along the departing edges of the vertex creates a ripple effect in the bottleneck structure as described in Theorem \ref{lem:propagation}. Leveraging this result, we can formally introduce the concept of \textit{region of influence}:

\begin{definition} \textit{Regions of influence in a network.} \label{def:region_influence} The region of influence of a link or flow $x \in \mathcal L \cup \mathcal F$, denoted $\mathcal{R}(x)$, is the set of links and flows $y$ that are reachable from $x$ in the gradient graph.
\end{definition}


The region of influence is an important concept in network performance analysis and optimization because it describes what parts of a network are affected by perturbations in the performance of a link or a flow. In Section \ref{ssec:linkflowgrad}, we will also see how such influences can be quantified.

We now introduce the \textit{GradientGraph()} algorithm (Algorithm \ref{al:GradientGraph}), a procedure that constructs the gradient graph of a network. The algorithm begins with crude estimates of the fair share rates of the links, and iteratively refines them until all the capacity in the network has been allocated and the rate of each flow reaches its final value. In the process, the gradient graph is constructed level by level. The algorithm starts by initializing the available capacity of each link (line 3), estimating its fair share  (line 4) and adding all links to a min-heap by taking their fair share value as the key (line 5). At each iteration, the algorithm picks the unresolved link with the lowest fair share value from the min-heap (line 8). Once this link is selected, all unresolved flows remaining in the network that traverse it are resolved. That is, their rates are set to the fair share of the link (line 12) and they are added to the set of vertices of the gradient graph $\mathcal{V}$ (line  13). In addition, directed edges are added in the gradient graph between the link and all the flows bottlenecked at it (line 10) and from each of these flows to the other links that they traverse (line 15). Lines 16-17-18 update the available capacity of the link, its fair share, and the position of the link in the min-heap according to the new fair share. Finally, the link itself is also added as a vertex in the gradient graph (line 22).  This iterative process is repeated until all flows have been added as vertices in the gradient graph (line 7). The algorithm returns the gradient graph $\mathcal{G}$, the fair share of each link $\{ s_l, \forall l \in \mathcal{L} \}$ and the rate of each flow $\{ r_f, \forall f \in \mathcal{F} \}$.

We conclude this section stating the time complexity of the \textit{GradientGraph()} algorithm:

\begin{lemma}{\textit{Time complexity of GradientGraph().}}\label{lem:complexity}
The time complexity of running \textit{GradientGraph()} is $O(|\mathcal{L}| \log |\mathcal{L}| \cdot H)$, where $H$ is the maximum number of flows that traverse a single link.
\end{lemma}
\begin{proof} See Appendix \ref{app:proof:complexity}.
\end{proof}

\begin{algorithm}[ht]
{\fontsize{8.0}{8.0}\selectfont
\caption{GradientGraph($ \mathcal{N} = \langle \mathcal{L}, \mathcal{F}, \{c_l,\forall l \in \mathcal{L} \} \rangle$)}
\label{al:GradientGraph}

{

\begin{algorithmic}[1]

\STATE $\mathcal{V} = \emptyset$; $ E = \emptyset$; $r_f = \infty, \forall f \in \mathcal{F}$;

\FOR{$l \in \mathcal{L}$} {
    \STATE $a_l = c_l$; \quad \quad \# available capacity \label{al:GradientGraph:init_a}
    \STATE $s_l = a_l / |\mathcal{F}_l|$; \quad \# fair share
    \STATE $\mathrm{MinHeapAdd}(\text{key = }s_l, \text{value= }l)$; \label{al:GradientGraph:init_push}
}
\ENDFOR

\WHILE{$\mathcal{F} \not\subseteq \mathcal{V}$}
   \STATE $l = \mathrm{MinHeapPop}()$; \label{al:GradientGraph:pop}
    
    \FOR {$f \in \mathcal{F}_l$ \text{ such that } $r_f \geq s_l$} {
        \STATE $E = E \cup \{(l, f), (f,l)\}$; \label{al:GradientGraph:add_bneck_edges}
        \IF {$f \not \in \mathcal{V}$} {
        \STATE $r_f = s_l$;
        \STATE $\mathcal{V} = \mathcal{V} \cup \{f\}$; 
        \FOR {$l' \in \mathcal{L}_f$ such that $r_f < s_{l'}$} {
            \STATE $E = E \cup \{(f, l')\}$ \label{al:GradientGraph:add_downstream_edges}
            \STATE $a_{l'} = a_{l'} - s_l$ \label{al:GradientGraph:set_a}
            \STATE $s_{l'} = a_l / |\mathcal{F}_l \setminus \mathcal{V}|$; \label{al:GradientGraph:set_s}
            \STATE $\mathrm{MinHeapUpdateKey}(\text{value = }l', \text{newKey = } s_{l'})$; \label{al:GradientGraph:update}
        }
        \ENDFOR
        }
        \ENDIF
    }
    \ENDFOR

    \STATE $\mathcal{V} = \mathcal{V} \cup \{l\}$;

\ENDWHILE

\RETURN $ \langle \mathcal{G} = \langle V, E \rangle, \{ s_l, \forall l \in \mathcal{L} \}, \{ r_f, \forall f \in \mathcal{F} \} \rangle $;

\end{algorithmic}
}
}
\end{algorithm}







\subsection{Link and Flow Gradients} \label{ssec:linkflowgrad}

In this section, we focus on the problem of quantifying the ripple effects created by perturbations in a network. Because networks are composed of links and flows, there are two kinds of perturbations: (1) those originating from changes to the capacity of a link and (2) those originating from changes to the rate of a flow. When such changes occur, the congestion control algorithm adjusts its allocation of bandwidth to the flows so as to maintain two objectives: (1) maximizing network utilization while (2) ensuring fairness among competing flows.
The congestion control algorithm acts like a function mapping network conditions (including its topology, link capacities, and flow paths) to rate allocations.
Large changes in any of these inputs can have complicated ripple effects on the flow rates, but for sufficiently small changes, the bandwidth allocation function is linear.\footnote{Technically, it is piecewise linear, like the absolute value function, so picking a linear function that locally approximates it requires knowing the direction of the change.} 
This local linearity property naturally motivates the concept of link and flow gradients:

\begin{definition} \textit{Link and flow gradients.} \label{def:gradients} 
Let $ \mathcal{N} = \langle \mathcal{L}, \mathcal{F}, \{c_l,\forall l \in \mathcal{L} \} \rangle$ be a network. We define:
\begin{itemize}[leftmargin=*] 
    \item The \textit{gradient} of a link $l \in \mathcal{L}$ with respect to another link $l^* \in \mathcal{L}$ as $\nabla_{l^*}(l) = \partial s_l / \partial c_{l^*}$;
    \item The \textit{gradient} of a flow $f \in \mathcal{F}$ with respect to some link $l^* \in \mathcal{L}$ as $\nabla_{l^*}(f) = \partial r_f / \partial c_{l^*}$;
    \item The \textit{gradient} of a link $l \in \mathcal{L}$ with respect to a flow $f^* \in \mathcal{F}$ as $\nabla_{f^*}(l) = \partial s_l / \partial r_{f^*}$;
    \item The \textit{gradient} of a flow $f \in \mathcal{F}$ with respect to another flow $f^* \in \mathcal{F}$ as $\nabla_{f^*}(f) = \partial r_f / \partial r_{f^*}$.
\end{itemize}
\end{definition}

Intuitively, the gradient with respect to a link measures the impact that a small perturbation in its capacity has on another link or flow. In real networks, this corresponds to the scenario of physically upgrading a link or, in programmable networks (e.g., \cite{SDN6994333}), logically modifying the capacity of a virtual link. Thus, link gradients can generally be used to resolve network design and capacity planning problems. Similarly, the gradient with respect to a flow measures the impact that a perturbation in its rate has on a link or another flow. This scenario corresponds, for instance, to the case of traffic shaping a flow to alter its transmission rate or changing the route of a flow---which can be seen as dropping the rate of that flow down to zero and adding a new flow with a different path. Thus, flow gradients can generally be used to resolve traffic engineering problems. In Section \ref{sec:applications} we will see applications in real networks that illustrate each of these scenarios.

We now present an algorithm called \textit{ForwardGrad()} (Algorithm \ref{al:ForwardGrad}) for calculating link and flow gradients. The algorithm takes a set of links and flows, the gradient graph of the corresponding network, a link or flow $x$ with respect to which to compute the gradients, and a direction $\Delta x$ of the perturbation. It outputs the gradients of all links and flows in the network with respect to $x$. \textit{ForwardGrad()} takes inspiration from forward mode automatic differentiation (``Forward Prop'') \cite{10.1137/080743627}, an algorithm that uses directed acyclic graphs to represent complicated mathematical functions as compositions of simpler functions, whose derivatives can be composed by repeatedly applying the chain rule. In the case of congestion control, we do not have a closed-form mathematical formula that relates network conditions (the inputs) to the flow rates and fair share values (the outputs), but we can use the gradient graph to break down and optimize this function. 

The thrust of the algorithm is as follows. For all $l \in \mathcal L$, let $\Delta_l$ be the change in the fair share rate of link $l$. For all $f \in \mathcal F$, let $\Delta_f$ be the change in the rate of flow $f$. We call these variables the ``drifts'' caused by a perturbation. Before the perturbation, $\Delta_l = \Delta_f = 0$ for all links and flows. To begin the algorithm, we make an infinitesimally small perturbation in the independent variable (the one in the ``denominator'' of the derivative) that can be positive or negative. If the independent variable $x$ is a flow $f$, we set $\Delta_f = \delta$ (line \ref{al:ForwardGrad:init_r}). If it is a link $l$, and $S_l$ is the set of direct successors of node $l$ in the gradient graph, we set $\Delta_l = \delta / |S_l|$ (line \ref{al:ForwardGrad:init_s}). This is done since, by definition of the gradient graph, $|S_l|$ is the number of flows bottlenecked at $l$ and the change in $l$'s capacity will be distributed evenly among these flows. To determine how this perturbation propagates to the rest of the network, we follow all directed paths from that vertex and update the drifts according to the following two invariants:
\begin{itemize}[leftmargin=*]
    \item \textit{Invariant 1: Flow Equation}. A flow's drift $\Delta_f$ equals the minimum drift of its bottleneck links. That is, $\Delta_f = \min_{l \in P_f} \Delta_l$, where $P_f$ is the set of links visited directly before flow vertex $f$ on a path from the starting vertex $x$ (the predecessors in the graph).
    \item \textit{Invariant 2: Link Equation}. A link's drift $\Delta_l$ is the negative of the flow drifts entering its vertex, divided by the number of flow drifts leaving it. That is, $\Delta_l = - \sum_{f \in P_l} \Delta_f / |S_l|$, where $P_l$ is the set of flow vertices visited directly before link vertex $l$ and $S_l$ is the set of flow vertices visited directly after link vertex $l$ on a path from the starting vertex $x$.
\end{itemize}
Finally, the derivative of a given variable with respect to the independent variable that we perturbed can be calculated by dividing its drift by $\delta$. In particular, assume the capacity of link $l$ is the independent variable that we perturbed and let the rate of flow $f$ be the dependent variable in which we want to measure the effect of this perturbation.  Then, $ \partial r_{f} / \partial c_{l} = \Delta_{f} / \delta$.

Since the flow and link equations lie at the heart of the algorithm, we provide some further explanation. Invariant 1 ensures that the capacity limits are respected and the network's resources are not wasted. Each flow must use exactly the amount of bandwidth allocated by its bottleneck link, so if the bottleneck's fair share changes, the flow's rate must change too. It also ensures fairness, since each flow bottlenecked at a certain link will experience the same drift. Invariant 2 ensures that capacity is neither created nor destroyed through the process of propagating a perturbation, except at the link whose capacity was initially perturbed. If a link's predecessors are using less bandwidth than before, then the savings must be redistributed evenly among the other flows that traverse the link.

\begin{figure}[t]
\centering
\includegraphics[width=0.8\columnwidth]{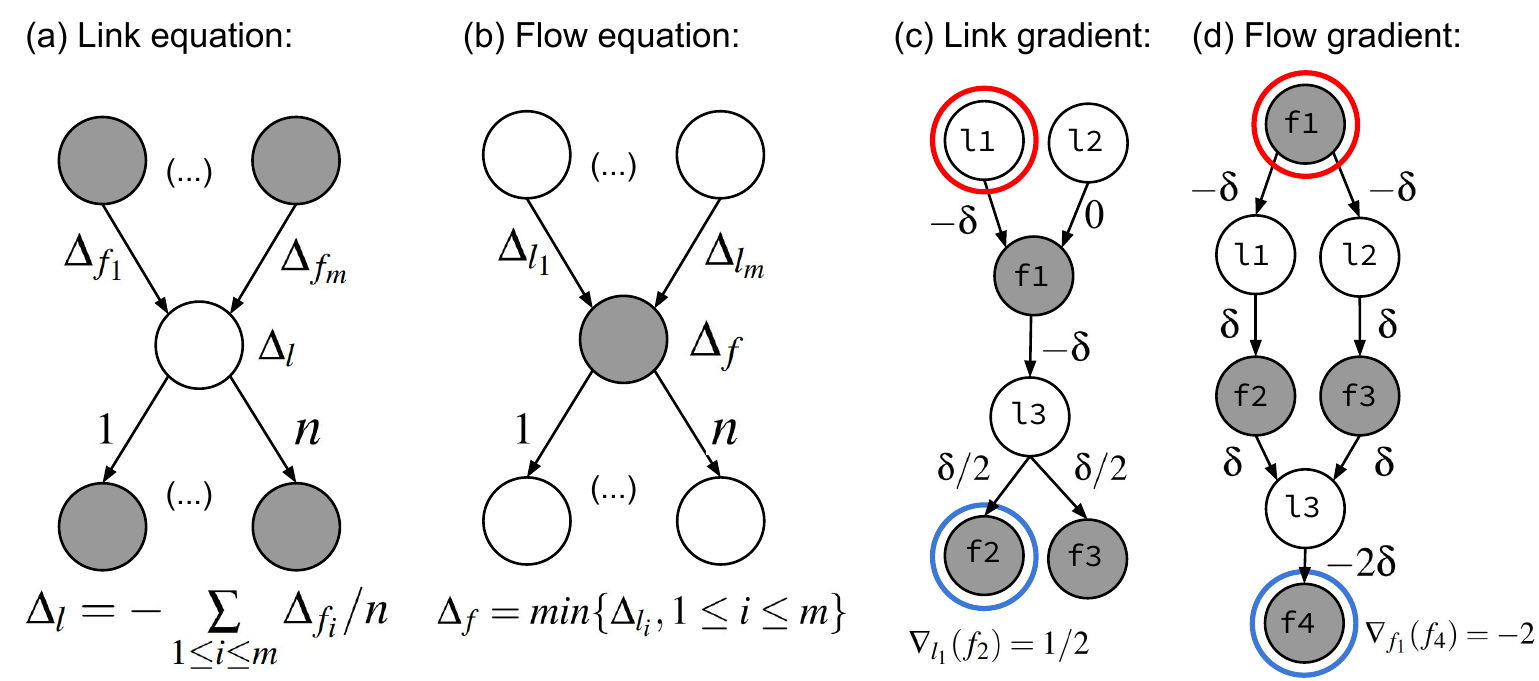}

\caption{(a) Link equation, (b) flow equation, and examples of (c) link gradient and (d) flow gradient. }
\label{fig:fgg_equations}
\end{figure}



Fig. \ref{fig:fgg_equations}(a) and (b) show graphical representations of the link and flow equations.
Fig. \ref{fig:fgg_equations}(c) and (d) present two simple examples of gradient graphs that we use to illustrate how to compute link and flow gradients. Note that throughout the paper, we use white vertices to denote bottleneck links and gray vertices to denote flows. We also omit backward edges for visual simplicity. Fig. \ref{fig:fgg_equations}(c) presents the case of computing the link gradient $\nabla_{l_1}(f_2)$. A perturbation is applied to link $l_1$ that decreases its capacity $c_{l_1}$ by an infinitesimally small amount $\delta$. Since only one flow is bottlenecked at $l_1$, we have $\Delta_{l_1} = - \delta$. This perturbation propagates to flow $f_1$ according to the flow equation: $\Delta_{f_1} = \min \{\Delta_{l_1}, \Delta_{l_2}\} = \min\{-\delta, 0\} = -\delta$. The perturbation is propagated down to link $l_3$ according to the link equation: $\Delta_{l_3} = - \Delta {f_1} / 2 = \delta / 2$. Finally, applying the flow equation for $f_2$, we obtain the flow drift $\Delta_{f_2} = \delta / 2$. Thus, the gradient of flow $f_2$ with respect to link $l_1$ is $\nabla_{l_1}(f_2) = \Delta_{f_2} / \delta = 1/2$. Fig. \ref{fig:fgg_equations}d illustrates a simple example of flow gradient computation. We leave it to the reader to verify that, for this bottleneck structure, the gradient of flow $f_4$ with respect to flow $f_1$ is $\nabla_{f_1}(f_4) = -2$.

To make this process into a precise algorithm, we still must specify the order in which to process the vertices of the graph. At each step, the vertex we process must be a neighbor of one of the vertices we have already visited. Even though backward edges create loops in the gradient graph, we never visit a vertex twice. If multiple vertices meet these criteria, we pick the one with the minimal rate or fair share value. If there are multiple vertices with the minimal rate or fair share value, we pick the one that would receive the minimum drift if it were processed next (see line \ref{al:ForwardGrad:pop} where keys in the heap are ordered pairs of rate/fair share and drift). This reflects the order in which the bottleneck structures are constructed in Algorithm \ref{al:GradientGraph}, which itself reflects the order in which the rates and fair shares converge in congestion controlled networks \cite{Ros-Giralt-SIGMETRICS-10.1145/3366707}. That is, we first visit the vertex that would receive the smallest rate or fair share if the perturbation were applied and bandwidth were reallocated from scratch. This completes the description of the \textit{ForwardGrad()} algorithm.

\begin{algorithm}[ht]
{\fontsize{8.0}{8.0}\selectfont
\caption{ForwardGrad($\mathcal{L}, \mathcal{F}, \mathcal{G}, x \in \mathcal L \cup \mathcal F, \Delta x \in \{\pm1\}$)}
\label{al:ForwardGrad}
{
\begin{algorithmic}[1]
\STATE $\Delta c_l = 0 \qquad \forall l \in \mathcal L$
\STATE $\Delta r_f = 0 \qquad \forall f \in \mathcal F$ \quad \# Drift of flow $f$ \label{al:ForwardGrad:init_r}
\STATE $\Delta s_l = 0 \qquad \forall l \in \mathcal L$ \quad \# Drift of link $l$ \label{al:ForwardGrad:init_s}
\IF{$x \in \mathcal F$}{
    \STATE $\Delta u_x = \Delta x$
    \STATE $\mathrm{MinHeapAdd}(\mathrm{key=} \langle u_x, \Delta u_x \rangle, \mathrm{value=} x)$
}
\ELSIF{$x \in \mathcal L$}{
    \STATE $\Delta c_x = \Delta x$
    \STATE $\Delta s_x = \Delta c_x/ |\mathrm{successors}(x, \mathcal G)|$
    \STATE $\mathrm{MinHeapAdd}(\mathrm{key=} \langle s_x, \Delta s_x \rangle, \mathrm{value=} x)$
}
\ENDIF

\STATE $V = \emptyset$ \qquad \# the set of previously visited nodes

\REPEAT {
    \REPEAT {
        \STATE $y = \mathrm{MinHeapPop}()$;    \# Get the next unvisited node \label{al:ForwardGrad:pop}
    }
    \UNTIL{$y \not \in \emptyset$} \label{al:ForwardGrad:no_repeats}
    \STATE $V = V \cup \{y\}$
    \IF{($y \in \mathcal L$ and $\Delta s_y = 0$) or ($y \in \mathcal F$ and $\Delta u_y = 0$)} {
        \STATE Continue
    }
    \ENDIF
    \FOR{$y' \in \mathrm{successors}(y,\mathcal G)  \setminus V$} {
        \IF{$y' \in \mathcal F$}{
            \STATE $\Delta r_{y'} = \min_{l \in \pi(y', \mathcal G)} \Delta s_l$ \# Flow equation invariant \label{al:ForwardGrad:set_r}
            \STATE $\mathrm{MinHeapAdd}(\mathrm{key=} \langle r_{y'}, \Delta r_{y'} \rangle, \mathrm{value=} y')$ \label{al:ForwardGrad:push_flow}
        }
        \ELSIF{$y' \in \mathcal L$}{
            \STATE $\Delta c_{y'} = \Delta c_{y'} - \Delta u_y$ \label{al:ForwardGrad:set_c}
            \STATE $\Delta s_{y'} = \Delta c_{y'} / |\mathrm{successors}(y', \mathcal G) \setminus V|$ \# Link equation invariant \label{al:ForwardGrad:set_s}
            \STATE $\mathrm{MinHeapAdd}(\mathrm{key=} \langle s_{y'}, \Delta s_{y'} \rangle, \mathrm{value=} y')$ \label{al:ForwardGrad:push_link}
        }
        \ENDIF
    }
    \ENDFOR
}
\UNTIL{$\mathrm{MinHeapEmpty}()$}

\RETURN $\langle \Delta s_l \quad \forall l \in \mathcal L, \quad \Delta r_f \quad \forall f \in \mathcal F \rangle$

\end{algorithmic}
}
}
\end{algorithm}

The next two theorems show that Algorithm \ref{al:ForwardGrad} is both correct and efficient.

\begin{theorem} \textit{Correctness of ForwardGrad().} \label{lem:forward_grad_correct}
Let $\mathcal{N} = \langle \mathcal{L}, \mathcal{F}, \{c_l,\forall l \in \mathcal{L} \} \rangle$ be a network and let $\mathcal G$ be the corresponding gradient graph. Let $x \in \mathcal{L} \cup \mathcal F$. After running Algorithm \ref{al:ForwardGrad}, $\Delta s_l = \nabla_x(l)$ for all $l \in \mathcal L$, and $\Delta r_f = \nabla_x(f)$ for all $f \in \mathcal F$.
\end{theorem}

\begin{proof}
See Appendix $\ref{app:proof:forward_grad}$.
\end{proof}

\begin{theorem} \textit{Time complexity of ForwardGrad().} \label{lem:forward_grad_runtime}
    Let $x \in \mathcal L \cup \mathcal F$. Then Algorithm \ref{al:ForwardGrad} finds the gradients of all links and flows in the network with respect to $x$ in time $O(|\mathcal R(x)| \cdot \log |\mathcal R(x)|)$. 
    \begin{proof}
        See Appendix $\ref{app:proof:forward_grad_runtime}$.
    \end{proof}
\end{theorem}




To conclude and complement this section, we state an upper bound on the value of the gradients:

\begin{property} \textit{Gradient bound.} \label{prop:bound} 
Let $ \mathcal{N} = \langle \mathcal{L}, \mathcal{F}, \{c_l,\forall l \in \mathcal{L} \} \rangle$ be a network and let $ \mathcal{G} $ be its gradient graph. Let $\delta$ be an infinitesimally small perturbation performed on a flow or link $x \in \mathcal{L} \cup \mathcal{F}$, producing a drift $\Delta_y$, for all $y \in \mathcal{L} \cup \mathcal{F}$. Then, $|\nabla_{x}(y)| = |\Delta_y| / \delta \leq d^{D(\mathcal{G})/4}$, where $D(X)$ is the diameter of a graph $X$ and $d$ is the maximum indegree and outdegree of any vertex in the graph.
\end{property}

\begin{proof} See Appendix \ref{app:proof:bound}.
\end{proof}

\section{Applications to Data Networks and Experimental Results} \label{sec:applications}

Because bottleneck structures are a fundamental property intrinsic to any congestion-controlled data network, its applications span a variety of networking problems. In this section, our goal is to present examples and experiments illustrating how QTBS can be used to resolve some of these problems. We will see that in each of them, the framework is able to provide new insights into one or more operational aspects of a network. The examples presented in this section are not exhaustive, but only illustrative. To help organize the applications, we divide them in two main classes: traffic engineering and capacity planning. For each of these classes, we provide specific examples of problems that relate to applications commonly found in modern production networks. 

To experimentally demonstrate that data networks behave qualitatively and quantitatively according to QTBS, we use \textit{Mininet-G2} \cite{g2MininetRepo}, a network emulation framework  developed by our team that consists of a set of software modules and extensions to Mininet \cite{mininetWebsite}. Leveraging software define networking (SDN), \textit{Mininet-G2} enables the creation and analysis of arbitrary network architectures using real production TCP/IP code, including production-grade implementations of congestion control algorithms such as BBR, Cubic or Reno. (See also Appendix \ref{app:g2mininetinfo} for more information.) We are open sourcing \textit{Mininet-G2} and all the experiments presented in this paper, hoping this will also enable the research community to verify our findings and further experiment with the theory of bottleneck structures.

All the experimental results presented in this section are based on Google's BBR congestion control algorithm \cite{Cardwell:2016:BCC:3012426.3022184}. Results for similar experiments using Cubic \cite{Cubic:Ha:2008:CNT:1400097.1400105} can be found in Appendix \ref{app:exp:cubic}. For each experiment, we used Jain's fairness index \cite{jain-fairness-index-journals/corr/cs-NI-9809099} as an estimator to measure how closely the predictions of the theory of bottleneck structure model match the experimental results. For all BBR experiments presented in the next sections, this index was above 0.99 accuracy on a scale from 0 to 1 (See Appendix \ref{app:jainfidx}), reflecting the strength of QTBS in modeling network behavior. 


\subsection{Traffic Engineering: Computation of the Highest-Throughput Route} \label{ssec:routing} 

In traditional IP networks, the problems of flow routing and congestion control are separately resolved by following a two-step process: first, a routing protocol (e.g., BGP \cite{BGP-RFC}, OSPF, etc.) is used to determine the path between any two nodes in a network; then, flows are routed according to such paths and their transmission rates are regulated using a congestion control algorithm (e.g., BBR \cite{Cardwell:2016:BCC:3012426.3022184}). This layered and disjoint approach is known to be scalable but suboptimal because the routing algorithm identifies paths without taking into account the flow transmission rates assigned by the congestion control algorithm \cite{BWE-10.1145/2829988.2787478, JoinRouting, NewtonRouting, TassiulasRouting}.

In this section, we use QTBS to resolve the following joint routing and congestion control problem in a scalable manner:

\begin{definition} \textit{Flow-rate maximal routing.} \label{def:ratemaximal} Let $ \mathcal{N} = \langle \mathcal{L}, \mathcal{F}, \{c_l,\forall l \in \mathcal{L} \} \rangle$ be a network and suppose that a new flow $f$ arrives. We will say that a routing algorithm is \textit{flow-rate maximal} if it routes flow $f$ through a path that maximizes its transmission rate $r_f$.
\end{definition}

In traditional IP routing, all packets transmitted from a source to a destination node follow the same \textit{lowest-cost} route \cite{BGP-RFC}. This rigidity leads to the well-known fish problem \cite{Bertsekas:1992:DN:121104}, whereby certain paths in a network become congested while other paths are underutilized. A flow-rate maximal algorithm, instead, is able to bypass points of congestion by assigning new flows to the highest-throughput path available given the current usage of the network. 

One might mistakenly think that the least congested path can be identified by looking for links with small fair shares (Definition \ref{def:fairshare}). However, the placement of a new flow onto a given path will itself alter the state of the network, changing those fair shares and potentially rendering the chosen path sub-optimal. In this section, we show that QTBS can be used to identify the maximal-rate path for a flow while taking into account the perturbations created by the placement of the flow itself, thus solving the flow-rate maximal routing problem.


\textit{MaxRatePath()} (Algorithm \ref{al:MaxRatePath}) is an algorithm that uses QTBS to compute flow-rate maximal paths. It takes the following inputs: a network $\mathcal{N} = \langle \mathcal{L}, \mathcal{F}, \{c_l,\forall l \in \mathcal{L} \} \rangle$, the set of routers $\mathcal{U}$, and the source and the destination routers of the flow we intend to route, $u_s$ and $u_d$. By convention, a link $l \in \mathcal{L}$ is identified with the tuple $l = (u_x, u_y)$, where $u_x, u_y \in \mathcal{U}$ are the two routers connected by link $l$. The algorithm returns the new flow $f$, expressed as the set of links it traverses, guaranteeing they form a path from $u_s$ to $u_d$ that yields the maximal rate $r_f$ for $f$.

As the pseudocode shows, \textit{MaxRatePath()} is based on Dijkstra's shortest path algorithm, with routers as vertices and links as edges in the network topology graph. The difference resides in the way the ``distance'' to a neighboring router $u'$ is calculated (lines 12-14). In \textit{MaxRatePath()}, this value represents not the number of hops on the shortest path from $u_s$ to $u'$, but the inverse of the largest possible rate that a flow would experience if it were added on some path from $u_s$ to $u'$. That is, the distance to $u'$ is the smallest possible time needed to send 1 bit of information from $u_s$ to $u'$. Unlike in the standard Dijkstra's algorithm, this value cannot be computed by adding an edge length to $d_u$, the distance to a neighbor of $u'$. Instead, we create a new flow $f$ by extending the optimal path from $u_s$ to $u$. So at each iteration of the algorithm, $f$ takes the path $u_s \to \cdots \to u \to u'$ (line 12). We then construct the gradient graph that would correspond to this network if the new flow $f$ were added (line 13). Finally, we use the inverse of the the rate assigned to the new flow $r_f$ as the distance value (line 14). In the pseudocode, we invoke the \textit{GradientGraph()} algorithm in line 13, reconstructing the gradient graph from scratch to include the new flow. However, we can get this result more efficiently by updating the initial gradient graph (the one corresponding to the network before adding the new flow), since the new flow will only affect a subset of the existing links and flows. We leave the precise algorithm for performing this update to future work.

\begin{algorithm}
{\fontsize{8.0}{8.0}\selectfont
\caption{$\mathrm{MaxRatePath} (\mathcal{N} = \langle \mathcal{L}, \mathcal{F}, \{c_l,\forall l \in \mathcal{L} \} \rangle, \mathcal{U} , u_s\in \mathcal{U}, u_d \in \mathcal{U})$}
\label{al:MaxRatePath}
{
\begin{algorithmic}[1]

\STATE F = MinHeap();  \quad \# Frontier set
\STATE C = Set();  \quad \quad \quad \quad \# Converged set

\STATE $d_u = \infty, \forall u \in \mathcal{U}$;  \quad \# Initialize distance metric

\STATE $d_{u_s} = 0$;
\STATE $\mathrm{F.insert}(\text{key = }0, \text{value= }u_s)$;

\WHILE{$u = \mathrm{F.extractMin}()$}

    \STATE $\mathrm{C.insert}(u)$; \quad \# Has converged
    \IF{ $u == u_d$ } 
        \STATE $\mathrm{break}$;
    \ENDIF
    \FORALL{$(u, u') \in \mathcal{L}$ \AND $u' \notin C $}
        \STATE $f = \{(x_1, x_2), (x_2, x_3), ..., (x_{i-1}, x_i) \mid x_j \in C, (x_j, x_{j+1}) \in \mathcal{L}, x_1 = u_s, x_{i-1}=u, x_i = u' \}$;
        \STATE $ \langle \mathcal{G}, \{s_l\}, \{r_f\} \rangle  = \mathrm{GradientGraph}(\mathcal{N} = \langle \mathcal{L}, \mathcal{F} \cup \{f\}, \{c_l,\forall l \in \mathcal{L} \} \rangle)$;
        \STATE $distance = 1 / r_f$;  \quad \# Flow completion time to send 1 bit
        \IF{$d_{u'} \leq distance$}
            \STATE $\mathrm{continue}$;
        \ENDIF
        \STATE $d_{u'} = distance$;
        \IF{$u' \notin F$}
            \STATE $\mathrm{F.insert}(\text{key = }d_{u'}, \text{value= }u')$;
        \ELSE
            \STATE $\mathrm{F.update}(\text{key = }d_{u'}, \text{value= }u')$;
        \ENDIF
    \ENDFOR

\ENDWHILE

\STATE $f = \{(x_1, x_2), (x_2, x_3), ..., (x_{i-1}, x_i) \mid x_j \in C, (x_j, x_{j+1}) \in \mathcal{L}, x_1 = u_s, x_i = u_d \}$;

\RETURN $f$;

\end{algorithmic}
}
}
\end{algorithm}

\begin{lemma}{\textit{Correctness of the MaxRatePath algorithm.}}\label{lem:maxratepath}
Let $\mathcal{N} = \langle \mathcal{L}, \mathcal{F}, \{c_l,\forall l \in \mathcal{L} \}\rangle$ be a network and $\mathcal{U}$ the set of its routers. Suppose that $f$ and $f'$ are two flows not in $\mathcal{F}$ that originate at router $u_s$ and end at router $u_d$. Then $f = \mathrm{MaxRatePath}(\mathcal{N}, \mathcal{U} , u_s, u_d)$ implies $r_f \geq r_{f'}$.
\end{lemma}
\begin{proof} The proof of this lemma is constructive and describes a procedure to efficiently compute the maximal-rate path of a flow using the bottleneck structure. See Appendix \ref{app:proof:maxratepath} for details.
\end{proof}

\begin{figure}[ht]
\centering

\begin{subfigure}[b]{0.4\columnwidth}
\centering
\includegraphics[width=\columnwidth]{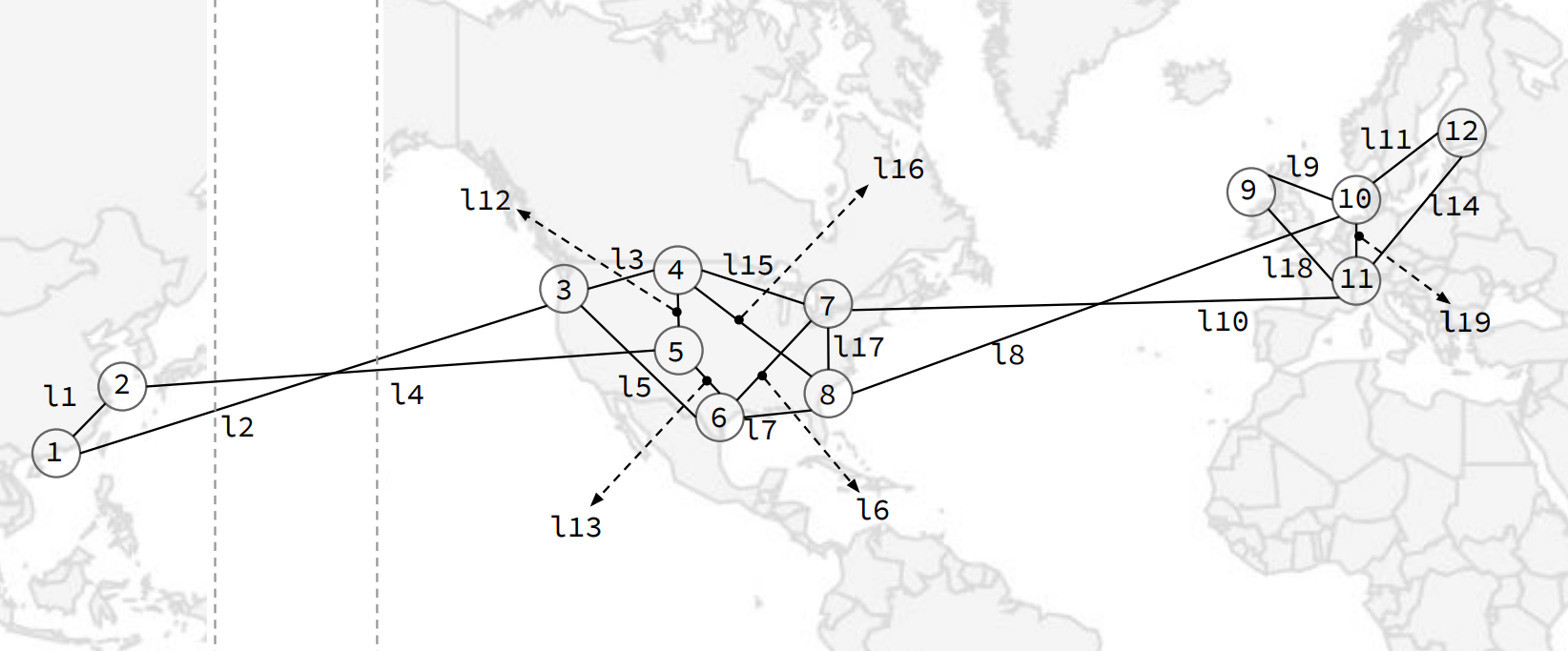}
\caption{Network topology. }
\label{fig:net_b4_routing}
\end{subfigure}
\begin{subfigure}[b]{0.49\columnwidth}
\includegraphics[width=\columnwidth]{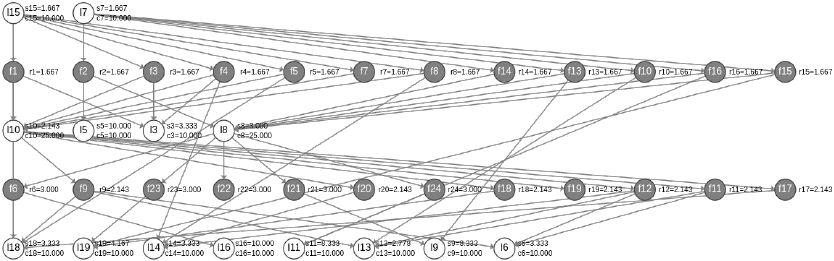}
\caption{Base bottleneck structure.}
\label{fig:ex_routing:1}
\end{subfigure}
\begin{subfigure}[b]{0.49\columnwidth}
\includegraphics[width=\columnwidth]{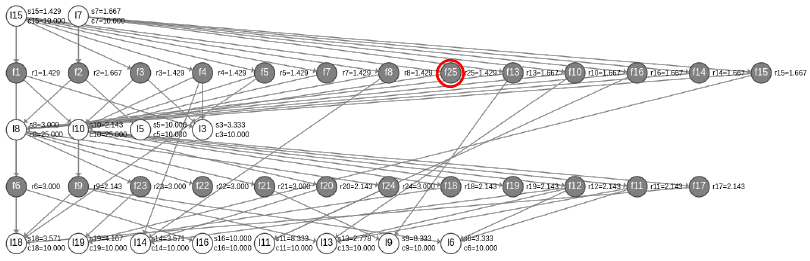}
\caption{Bottleneck structure using the shortest path.}
\label{fig:ex_routing:2}
\end{subfigure}
\begin{subfigure}[b]{0.49\columnwidth}
\includegraphics[width=\columnwidth]{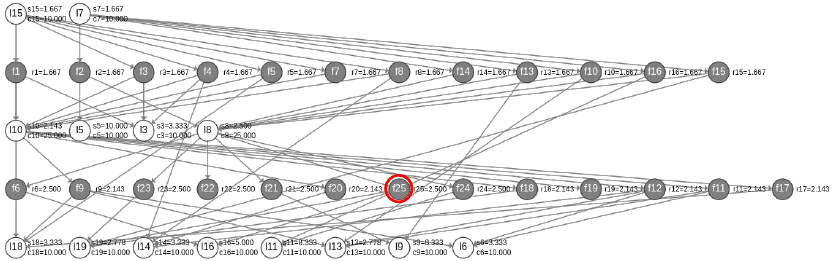}
\caption{Bottleneck structure using the maximal-rate path.}
\label{fig:ex_routing:3}
\end{subfigure}
\caption{Computation of maximal-throughput path for flow $f_{25}$.}
\label{fig:ex_routing}
\end{figure}

To illustrate how we can use QTBS and the MaxRatePath algorithm to compute the highest-throughput path for a given flow, consider the network shown in Fig. \ref{fig:net_b4_routing}. This topology corresponds to Google's B4 network as described in \cite{B4-Jain:2013:BEG:2534169.2486019}, the SDN-WAN network that connects Google's data centers globally. For the sake of illustration, we will assume there are two flows (one for each direction) connecting every data center in the US with every data center in Europe, with all flows routed along a shortest path from source to destination. Since there are six data centers in the US and four in Europe, this configuration has a total of $48$ flows ($|\mathcal{F}| = 6 \times 4 \times 2 = 48$). (See Table \ref{tab:flowpaths_routing} in Appendix~\ref{app:flowpaths} for a description of the exact path followed by each flow.) All links are assumed to have a capacity of $10$ Gbps except for the transatlantic links, which are configured at $25$ Gbps (i.e., $c_{l} = 10$, for all $l \notin \{l_8, l_{10}\}$, $c_{l_8} = c_{l_{10}} = 25$). While obviously production networks operate with a much higher number of flows, in our example we use a reduced number to simplify the descriptions of the bottleneck structures and the steps followed to resolve the given problem. This simplification is without loss of generality, and the same approach is applicable to large scale operational networks. (See Appendix \ref{app:g2prodnws} for notes on integration with production networks.)

Fig. \ref{fig:ex_routing:1} shows the corresponding bottleneck structure obtained from running Algorithm \ref{al:GradientGraph} on the proposed network configuration. This structure shows that flows are organized in two levels: the top-level includes flows $\{f_1, f_2, f_3, f_4, f_5, f_7,$ $f_8, f_{10}, f_{13}, f_{14}, f_{15}, f_{16} \}$ and the low-level includes flows $\{ f_6, f_9, f_{11}, f_{12}, f_{17}, f_{18}, f_{19}, f_{20},f_{21}, $ $f_{22}, f_{23}, f_{24}\}$. Note that because each pair of data centers is connected via two flows (one for each direction), without loss of generality, in Fig. \ref{fig:ex_routing:1} we only include the first 24 flows (flows transferring data from US to Europe), since the results are symmetric for rest of the flows---i.e., flow $f_i$ has the same theoretical transmission rate and is positioned at the same level in the bottleneck structure as flow $f_{i+24}$ for all $1 \leq i \leq 24$. Note also that all the top-level flows operate at a lower transmission rate (with all rates at 1.667) than the bottom-level flows (with rates between 2.143 and 3). As was proven in \cite{Ros-Giralt-SIGMETRICS-10.1145/3366707}, this is in fact a property of all bottleneck structures: flows operating at lower levels of the bottleneck structure have greater transmission rates than those operating at higher levels. 

Under this configuration, suppose that we need to initiate a new flow $f_{25}$ to transfer a large data set between data centers $4$ and $11$. For instance, this flow could correspond to the transmission of a terabyte data set from a data center in the US to another in Europe. Our objective in this exercise is to identify a high-throughput route to minimize the time required to transfer the data. 


In Fig. \ref{fig:ex_routing:2} we show the bottleneck structure obtained for the case that $f_{25}$ uses the shortest path $l_{15} \rightarrow l_{10}$. For instance, this corresponds to the solution obtained from running BGP \cite{BGP-RFC} with a link cost metric equal to 1. Using this path, the new flow would be placed at the upper bottleneck level---i.e., the lower-throughput level---in the bottleneck structure, receiving a theoretical rate of $r_{25}= 1.429$. Note that the presence of this new flow slightly modifies the performance of some of the flows on the first level (flows $\{f_1, f_3, f_4, f_5, f_7, f_8\}$ experience a rate reduction from 1.667 to 1.429), but it does not modify the performance of the flows operating at the bottom level. This is because, for the given configuration, the new flow only creates a shift in the distribution of bandwidth on the top level, but the total amount of bandwidth used in this level stays constant. (In Fig. \ref{fig:ex_routing:1}, the sum of all the flow rates on the top bottleneck level is $1.667 \times 12 = 20$, and in Fig. \ref{fig:ex_routing:2} this value is the same: $1.429 \times 7 + 1.667 \times 6 = 20$.) As a result, the ripple effects produced from adding flow $f_{25}$ into the network cancel each other out without propagating to the bottom level. 

While $l_{15} \rightarrow l_{10}$ is the shortest path, it is not the path with the highest throughput. To find such a path, we run the MaxRatePath procedure (Algorithm \ref{al:MaxRatePath}) and obtain the solution $l_{16} \rightarrow l_{8} \rightarrow l_{19}$. The resulting bottleneck structure is shown in Fig. \ref{fig:ex_routing:3}. Using this path, flow $f_{25}$ would now be placed at the bottom level---the higher-throughput level---in the bottleneck structure, thus resulting in a rate value $r_{25} = 2.5$, an increase of 74.95\% with respect to the shortest path solution. Another positive outcome of this solution is that none of the flows operating at the upper level (the flows that receive less bandwidth) see their rate reduced. This is a direct consequence of Theorem \ref{lem:propagation}, since a perturbation on lower levels can have no ripple effects on upper levels. It constitutes also a natural fairness property of the MaxRatePath algorithm: as the procedure assigns maximal-throughput paths to new incoming flows, such flows tend to be placed at the bottom of the bottleneck structure (where the high-throughput links are located), thus tending to create no negative impact on the lower-throughput flows located at the top of the structure.


In the remainder of this section, we set out to empirically confirm these results. We start by creating the B4 network configuration shown in Fig. \ref{fig:net_b4_routing} using \textit{Mininet-G2}. Following our example, we deploy a total of 48 shortest-path flows connecting every pair of nodes (in both directions) between the US and Europe. (Table \ref{tab:flowpaths_routing} in Appendix \ref{app:flowpaths} presents the exact path followed by each flow.) We then add two extra flows labeled $f_{25}$ and $f_{50}$ (one for each direction) to connect data centers 4 and 11 and perform two separate experiments: one placing the flows on the shortest path $l_{15} \leftrightarrow l_{10}$ and another one placing them on the longer path $l_{16} \leftrightarrow l_{8} \leftrightarrow l_{19}$.

\begin{figure}[ht]
\centering
\begin{subfigure}[b]{0.4\linewidth}
\includegraphics[width=\columnwidth]{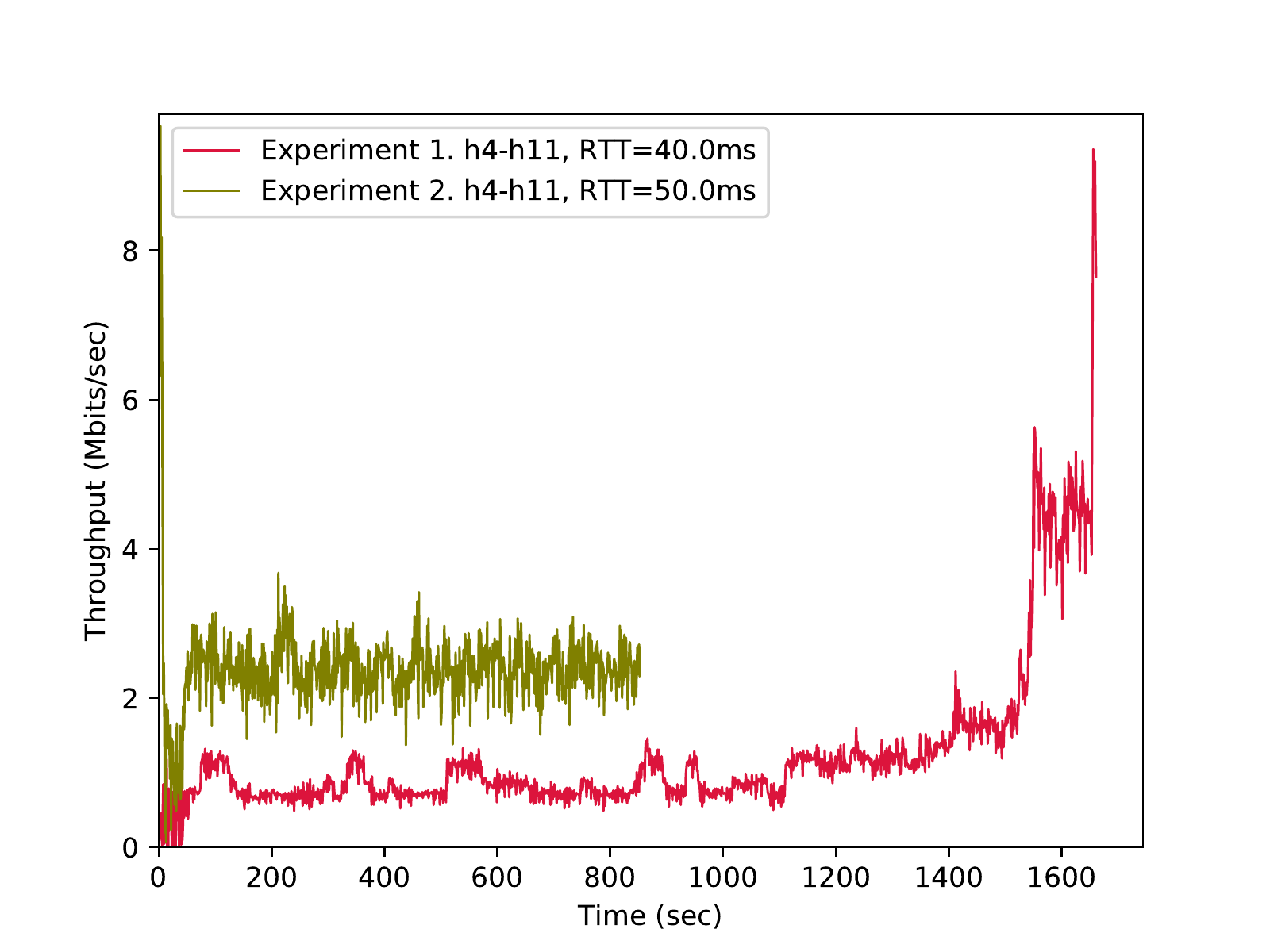}

\caption{Acceleration of flow $f_{25}$.}
\label{fig:ex_routing_flow25}
\end{subfigure}
\begin{subfigure}[b]{0.49\columnwidth}
\includegraphics[width=\columnwidth]{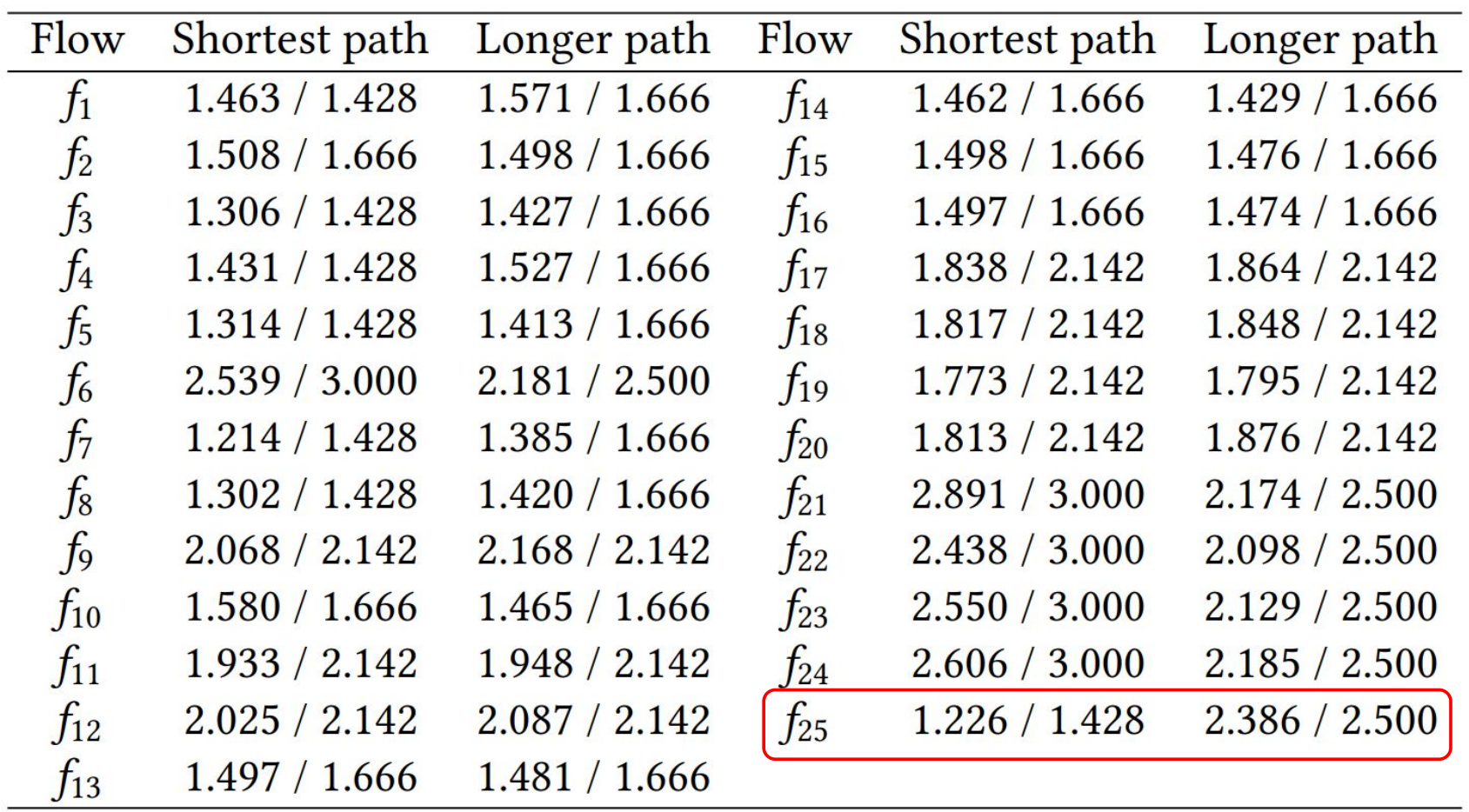}

\caption{Experimental vs theoretical flow rates (Mbps).}
\label{tab:route}
\end{subfigure}

\caption{Identification of a high-bandwidth route to accelerate the performance of flow $f_{25}$.}

\label{fig:gg_clos}
\end{figure}

Fig. \ref{fig:ex_routing_flow25} shows the rate of flow $f_{25}$ for the two experiments (very similar results are obtained for the reverse-path flow $f_{50}$, see Appendix \ref{app:routing}). In the legend of this plot, experiment~1 and~2 correspond to the shortest and the (longer) maximal-throughput path configurations, respectively. As predicted by the bottleneck structure, the longer path achieves a higher throughput and, thus, a lower flow completion time. Fig. \ref{tab:route} presents the average throughput obtained for all twenty-five flows from the US to Europe and for each of the two experiments, alongside the theoretical values according to the bottleneck structure. (The results obtained from the other twenty-five flows on the reverse path are similar and can be found in Appendix \ref{app:routing_flows_2650}.) As shown, flow $f_{25}$ achieves a performance of 1.226 and 2.386 Mbps for the shortest and longer paths, respectively---with the theoretical rates being 1.428 and 2.5 Mbps, respectively. Thus, the longer path yields a 94\% improvement on flow throughput compared to the shortest path. For all the experiments run in this section, Jain's fairness index was above 0.99 (see Appendix \ref{app:jainfidx}), indicating the accuracy of QTBS in predicting flow performance.

This experiment illustrates that using QTBS, it is possible to identify routes that are highly efficient from a congestion control standpoint. Note that this contrasts with traditional approaches that perform traffic engineering by separating the routing and congestion control problems, so that the routing algorithm is unaware of the choices made by the congestion control algorithm and vice versa. See for instance Section 5.3.1 in \cite{BWE-10.1145/2829988.2787478}, which discusses the potential advantages of performing joint routing and congestion control in Google's WAN, but leaves this direction as future work. We reason that QTBS provides a mathematical framework to connect both problems, identifying routes that are globally efficient from both a topological and a congestion control standpoints.

\subsection{Capacity Planning: Design of Optimal Fat-Tree Networks in Data Centers} \label{ssec:clos} 

As Leiserson demonstrated in his seminal paper \cite{leiserson1985fat}, fat-trees are universally efficient networks in the following sense: for a given network size $s$, a fat-tree can emulate any other network that can be laid out in that size $s$ with a performance slowdown at most logarithmic in $s$. This property makes fat-tree topologies highly competitive and is one of the reasons they are so widely used in large-scale data centers \cite{fat-tree-amin-10.1145/1402958.1402967} and high-performance computing (HPC) networks \cite{band-steer-10.1145/3295500.3356145}\footnote{In the context of data centers, fat-tree networks are also known as folded-clos or spine-and-leaf networks \cite{fat-tree-amin-10.1145/1402958.1402967}.}. In this experiment, we use QTBS to demonstrate that, due to the effects of the congestion control algorithm, there exists an optimal trade-off in the allocation of capacity at the top levels of the fat-tree. Further, we show that the optimal bandwidth allocation on the top level deviates from commonly accepted best practices in the design of full fat-tree networks that tend to equate the amount of bandwidth going up and down the tree at each switch \cite{full-fat-tree-denzel2010framework}.

\begin{figure}[ht]
\centering
\begin{subfigure}[b]{0.4\linewidth}
\includegraphics[width=0.6\columnwidth]{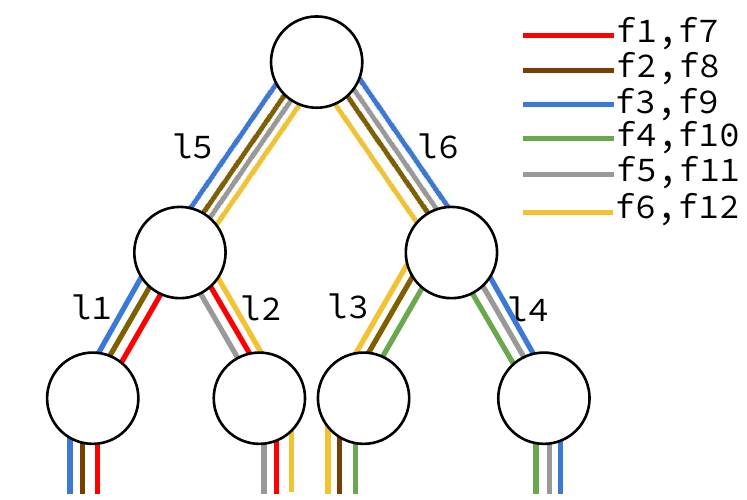}
\caption{Fat-tree network. }
\label{fig:net_clos}
\end{subfigure}
\begin{subfigure}[b]{0.49\columnwidth}
\includegraphics[width=\columnwidth]{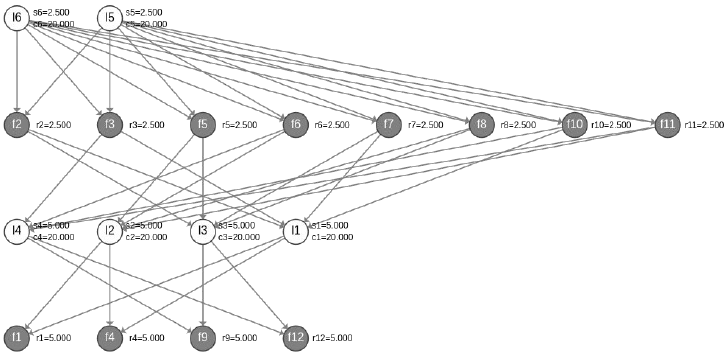}
\caption{Tapering parameter $\tau = 1$.}
\label{fig:gg_clos:1}
\end{subfigure}

\begin{subfigure}[b]{0.49\columnwidth}
\includegraphics[width=\columnwidth]{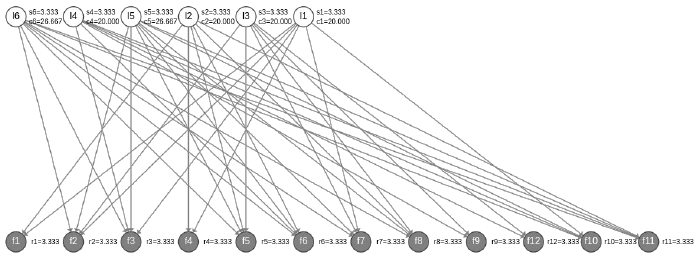}
\caption{Tapering parameter $\tau = 4/3$.}
\label{fig:gg_clos:2}
\end{subfigure}
\begin{subfigure}[b]{0.49\columnwidth}
\includegraphics[width=\columnwidth]{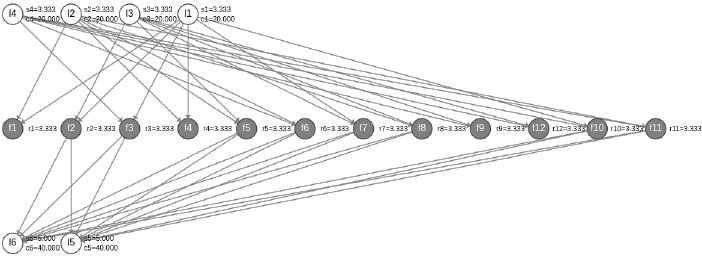}
\caption{Tapering parameter $\tau = 2$.}
\label{fig:gg_clos:3}
\end{subfigure}
\caption{Design of optimal 3-level binary fat-trees.}
\label{fig:gg_clos}
\end{figure}

Consider the network topology in Fig. \ref{fig:net_clos}, which corresponds to a binary fat-tree with three levels and six links ($\mathcal{L} = \{l_1, l_2, ..., l_6\}$). Assume also that there are two flows (one for each direction) connecting every pair of leaves in the fat-tree network, providing bidirectional full-mesh connectivity among the leaves. Since there are four leaves, that's a total of $4 \times 3 \times 2 = 24$ flows. All of the flows are routed following the shortest path. (See Table \ref{tab:flowpaths_clos} in Appendix \ref{app:flowpaths} for a description of the exact path followed by each flow.) For the sake of convention, we will adopt the terminology from data center architectures and use the names \textit{spine} and \textit{leaf} links to refer to the upper and lower links of the fat-tree network, respectively \cite{fat-tree-amin-10.1145/1402958.1402967}.

We fix the capacity of the leaf links to a value $\lambda$ (i.e., $c_{l_1} = c_{l_2} = c_{l_3} = c_{l_4} = \lambda$) and the capacity of the spine links to $\lambda \times \tau$ (i.e., $c_{l_5} = c_{l_6} = \lambda \times \tau$), where $\tau$ is used as a design parameter enabling a variety of network configurations. For instance, in our binary fat-tree example, the case $\tau = 2$ corresponds to a full fat-tree network \cite{full-fat-tree-denzel2010framework}, because the total aggregate bandwidth at each level of the tree is constant, $c_{l_1} + c_{l_2} + c_{l_3} + c_{l_4} = c_{l_5} + c_{l_6} = 4 \lambda$. Similarly, the case $\tau = 1$ corresponds to a thin-tree network, since it results with all the links having the same capacity, $c_{l_i} = \lambda$, for all $1 \leq i \leq 6$. The technique of optimizing the performance-cost trade-off of a fat-tree network by adjusting the capacity of the spine links is sometimes known as \textit{bandwidth tapering} \cite{band-steer-10.1145/3295500.3356145, leon2016characterizing}. The focus of our experiment is to use the bottleneck structure analysis to identify optimized choices for the tapering parameter $\tau$.

In Fig. \ref{fig:gg_clos} we present a sequence of bottleneck structures (obtained from running Algorithm \ref{al:GradientGraph}) corresponding to our fat-tree network with three different values of the tapering parameter $\tau$ and fixing $\lambda = 20$. (Note that the fixing of $\lambda$ to this value is without loss of generality, as the following analysis applies to any arbitrary value $\lambda > 0$.) The first bottleneck structure (Fig. \ref{fig:gg_clos:1}) corresponds to the case $\tau=1$ (i.e., all links have the same capacity, $c_{l_i} = 20$, for all $1 \leq i \leq 6$), which has all flows confined in one of two possible levels: a top level, where flows perform at a lower rate, $r_{f_2} = r_{f_3} = r_{f_5} = r_{f_6} = r_{f_7} = r_{f_8} = r_{f_{10}} = r_{f_{11}} = 2.5$; and a bottom level, where flows perform at twice the rate of the top-level flows, $r_{f_1} = r_{f_4} = r_{f_9} = r_{f_{12}} = 5$. This configuration is thus unfair to those flows operating at the top bottleneck, which receive half the bandwidth of the flows at the bottom level. Furthermore, this configuration is also inefficient at supporting applications with symmetric workload patterns---where all nodes approximately send the same amount of bytes to each other---because the completion time of the slowest flows is significantly higher (twice as high since they get half the rate) than the faster flows. Let us next consider how we can use QTBS to identify a value of $\tau$ that minimizes the maximum completion time of any of the flows under the assumption of symmetric workloads. 

By looking at the bottleneck structure in Fig. \ref{fig:gg_clos:1}, we see that the slowest flows are confined in the top bottleneck level. In order to increase the rates of these flows, we need to increase the tapering parameter $\tau$ that controls the capacity of the spine links $l_5$ and $l_6$. This change transforms the bottleneck structure by bringing the two levels closer together, until, for a large enough value of $\tau$, they fold. We can find this collision point using the link gradients as follows. Using \textit{ForwardGrad()} (Algorithm \ref{al:ForwardGrad}), we obtain a link gradient value of $\nabla_{l}(f) = 0.125$ for all spine links $l \in \{l_5, l_6\}$ and top-level flows $f \in \{f_2, f_3, f_5, f_6, f_7, f_8, f_{10}, f_{11}\}$. On the other hand, the link gradient of any of the \emph{low}-level flows with respect to any of the spine links is $\nabla_{l}(f) = -0.25$, for all $l \in \{l_5, l_6\}$ and $f \in \{f_1, f_4, f_9, f_{12}\}$. That is, an increase by one unit on the capacity of the spine links increases the rate of the top-level flows by 0.125 and decreases the rate of the low-level flows by 0.25. Since the rates of the top and low-level flows are 2.5 and 5, respectively, this means that the two levels will fold at a point where the tapering parameter satisfies the equation $2.5 + 0.125 \cdot \tau \cdot  \lambda = 5 - 0.25 \cdot \tau \cdot \lambda$, yielding $\tau = 4/3$ and $c_{l_5} = c_{l_6} = 26.667$. The resulting bottleneck structure for this configuration is shown in Fig. \ref{fig:gg_clos:2}, confirming the folding of the two levels. This fat-tree configuration is optimal in that the flow completion time of the slowest flow is minimal. Because the bottleneck structure is folded into a single level, this configuration also ensures that all flows perform at the same rate, $r_{f_i} = 3.333$, for all $1 \leq i \leq 6$.

What is the effect of increasing the tapering parameter above $4/3$? This result is shown in Fig. \ref{fig:gg_clos:3} for the case of a full fat-tree $\tau = 2$, i.e., $c_{l_5} = c_{l_6} = 40$. In this case, the two spine links are no longer bottlenecks to any of the flows (since these links are leaves in the bottleneck structure), but all flows continue to perform at the same rate, $r_{f_i} = 3.333$, for all $1 \leq i \leq 6$. Thus, increasing the capacity of the upper-level links above $26.667$ does not yield any benefit, but increases the cost of the network. This result indicates that the fat-tree network shown in Fig. \ref{fig:net_clos} should not be designed with an allocation of capacity on the spine links higher than $\tau = 4/3$ times the capacity of the leaf links. In summary, we have that:

\begin{itemize}[leftmargin=*] 
    \item A tapering parameter $\tau \geq 4/3$ should not be used, since the resulting network is just as efficient as a design with $\tau=4/3$, but more costly.
    \item A tapering parameter $\tau=4/3$ is optimal in that it minimizes the flow completion time of the slowest flow. This should be the preferred design in symmetric workloads that transfer about the same amount of data between all pairs of nodes. 
    \item A tapering parameter $\tau < 4/3$ can be used if workloads are asymmetric, identifying the right value of $\tau$ that produces the right amount of bandwidth at each level of the bottleneck structure according to the workload.
\end{itemize}

\begin{figure*}[ht]
\centering

\begin{subfigure}[b]{.3\linewidth}
\includegraphics[width=\linewidth]{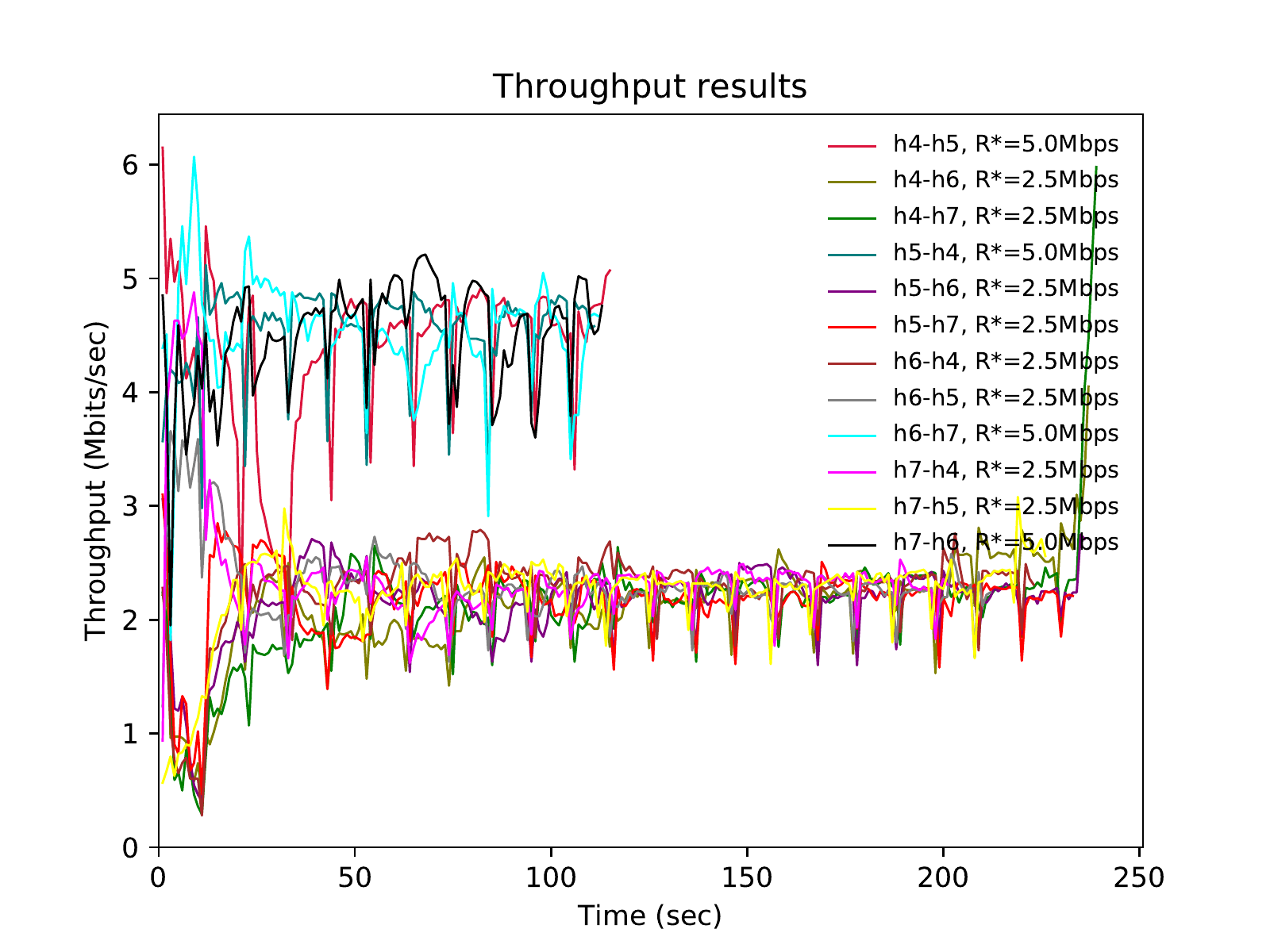}
\caption{Tapering parameter $\tau=1$}
\label{fig:ex_clos:1}
\end{subfigure}%
\begin{subfigure}[b]{.3\linewidth}
\includegraphics[width=\linewidth]{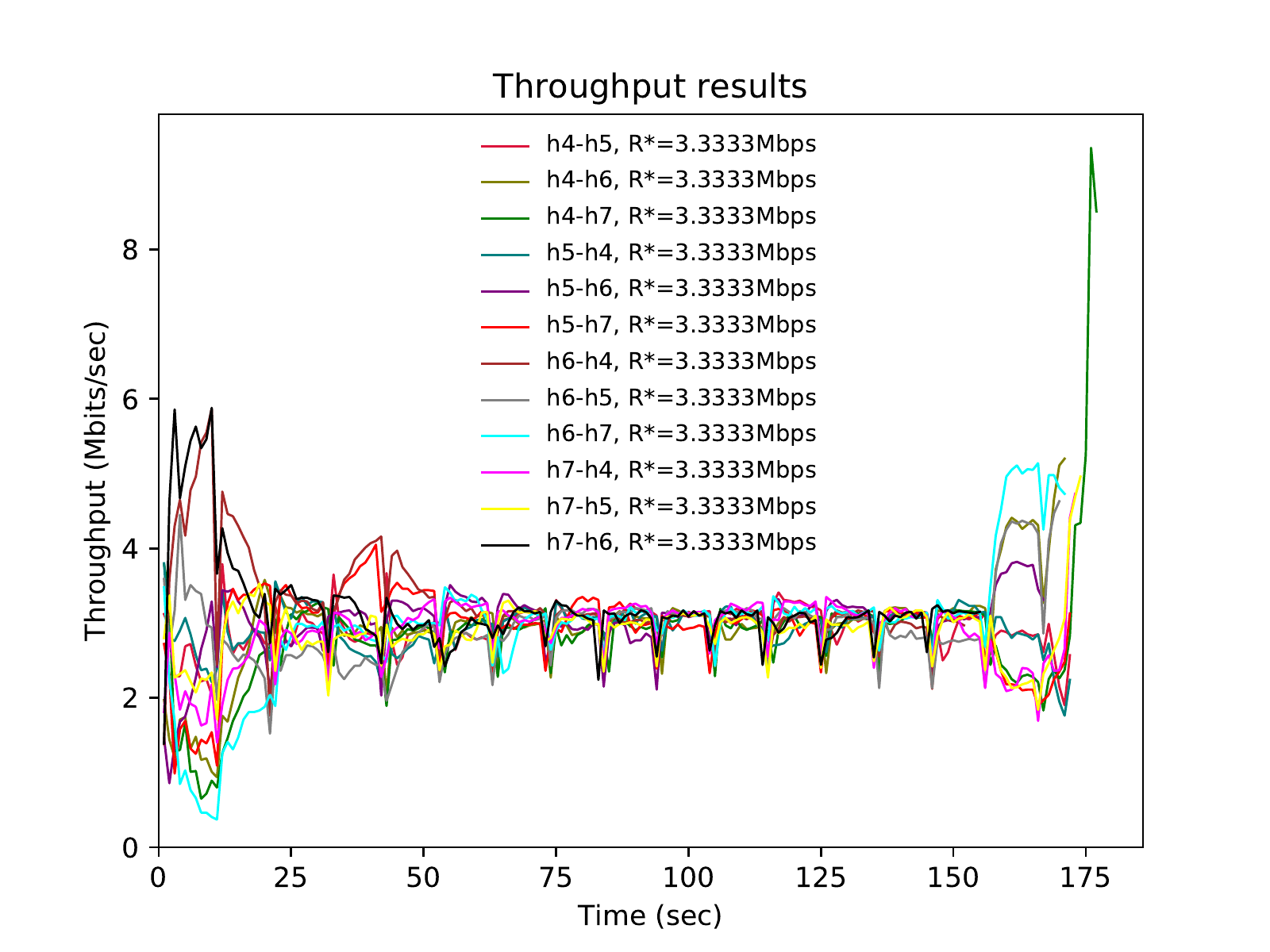}
\caption{Tapering parameter $\tau=4/3$}
\label{fig:ex_clos:2}
\end{subfigure}
\begin{subfigure}[b]{.3\linewidth}
\includegraphics[width=\linewidth]{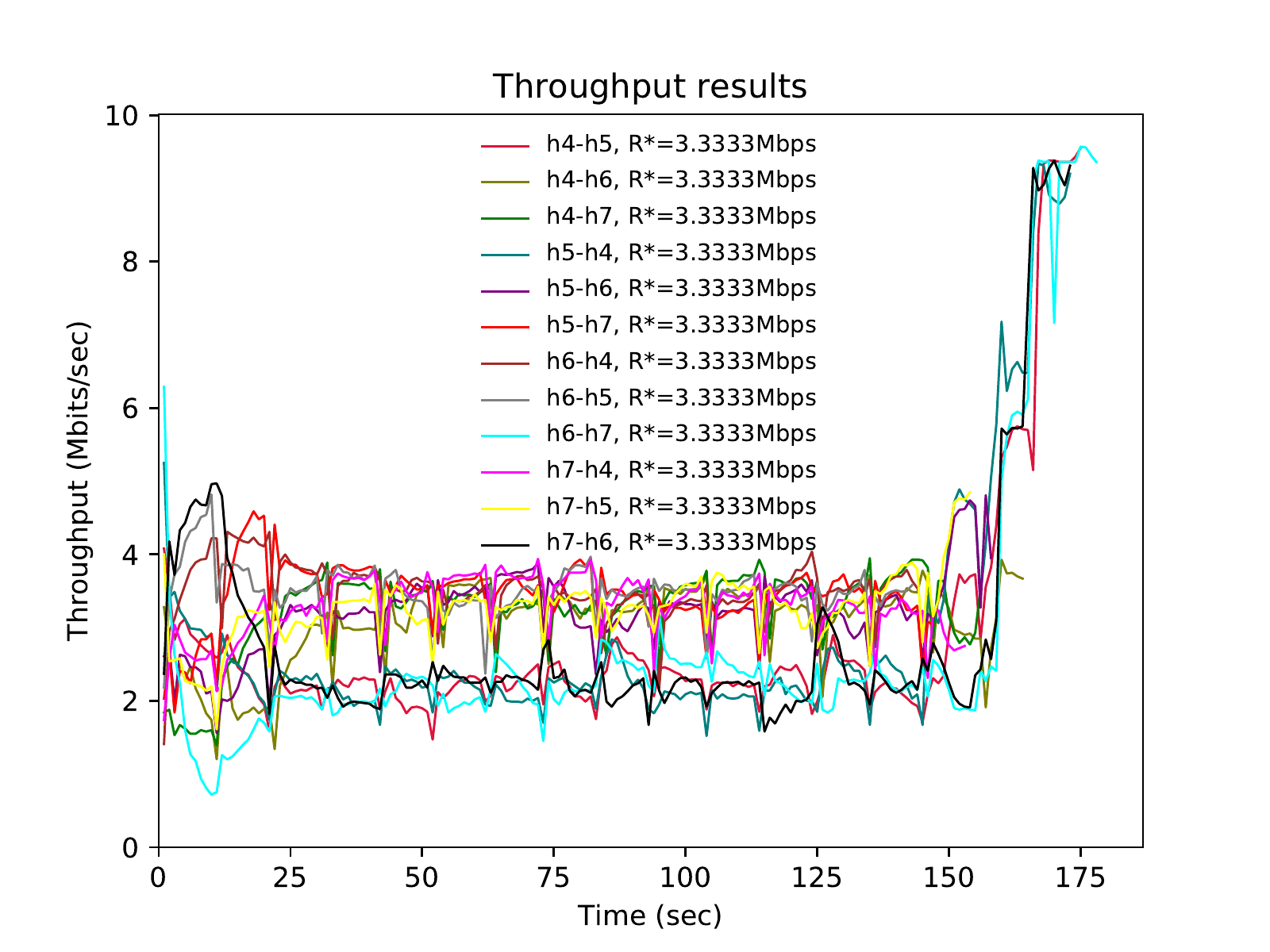}
\caption{Tapering parameter $\tau=2$}
\label{fig:ex_clos:3}
\end{subfigure}
\caption{Optimizing bandwidth tapering on a 3-level binary fat-tree.}
\label{fig:ex_clos}
\end{figure*}

In the rest of this section, we empirically demonstrate the existence of an optimal fat-tree design at $\tau = 4/3$ using Mininet-G2 \cite{g2MininetRepo} configured with the congestion control algorithm BBR. Fig. \ref{fig:ex_clos} presents the results of the experiments for the three values of the tapering parameter, $\tau \in \{1, 4/3, 2\}$. Each plot shows the transmission rate of all twelve flows as part of the network configuration, with each flow transmitting a total of 64 MB of data. Following the example in Section \ref{ssec:clos}, the link capacities are set as follows: $c_{l_1} = c_{l_2} = c_{l_3} = c_{l_4} = \lambda$ = 20 Mbps and $c_{l_5} = c_{l_6} = \lambda \times \tau = 20 \times \tau$ Mbps. 

\begin{table}[bh]
\center

\caption{Flow completion times (seconds) of the fat-tree experiments.}
\label{tab:clos}
\scalebox{0.8}{
\begin{tabular} {  c  c  c  c  c  c  c  c}
\hline
Flow & $\tau = 1$ & $\tau = 4/3$ & $\tau = 2$ & Flow & $\tau = 1$ & $\tau = 4/3$ & $\tau = 2$ 

\\ \hline
$f_1$ &  115  & 172  & 175 & $f_7$ &  223  & 152  & 144 \\
$f_2$ &  237  & 171  & 164 & $f_8$ &  212  & 170  & 143 \\
$f_3$ &  239  & 177  & 156 & $f_9$ &  112  & 171  & 178 \\
$f_4$ &  111  & 172  & 173 & $f_{10}$ &  201  & 173  & 153 \\
$f_5$ &  236  & 167  & 158 & $f_{11}$ &  226  & 174  & 154 \\
$f_6$ &  233  & 172  & 147 & $f_{12}$ &  113  & 155  & 173 \\
\hline
max() & & & & & 239 & 177 & 178 \\
\hline
\end{tabular}
}
\end{table}

As predicated by QTBS, the case $\tau = 1$ has flows operating at one of two bottleneck levels, close to the rates predicted by the bottleneck structure (2.5 Mbps for the upper-level flows and 5 Mbps for the lower-level flows, see Fig. \ref{fig:gg_clos:1}). This fat-tree design is inefficient for symmetric workloads since the flow completion time of the slowest flow is not minimal. Under this configuration, flow $f_{3}$ is the slowest flow and its completion time is $239$ seconds. (See Table \ref{tab:clos} for all flow completion time values.) If we want to maximize the rate of the slowest flow, QTBS tells us that the right tapering parameter value is $4/3$. This case is presented in Fig. \ref{fig:ex_clos:2}, which indeed shows how all flows perform at a very similar rate close to the theoretical value of $3.333$ Mbps (see Fig. \ref{fig:gg_clos:2}). This configuration is optimal in that it minimizes the maximum completion time of any of the flows. In this experiment, the completion time of the slowest flow is $177$ seconds, an improvement of $25.9\%$ with respect to the case of $\tau = 1$. Fig. \ref{fig:ex_clos:3} shows the results for the case of a full fat-tree network, $\tau = 2$. Once again, as predicted by QTBS, this solution achieves about the same completion time as the case $\tau = 4/3$ (the slowest flow completes in $178$ seconds), since in this configuration the leaf links become the bottlenecks and the extra bandwidth added in the spine links does not produce any net benefit, as shown by the bottleneck structure in Fig. \ref{fig:gg_clos:3}. In summary, as predicted by QTBS, the case $\tau=4/3$ constitutes an optimal design in that it is the least costly network that minimizes the maximum completion time of any of the flows.

Note that the existence of an optimal design with a tapering parameter $\tau=4/3$ argues against some of the established conventional best practices in fat-tree networks. For instance, while a full fat-tree ($\tau = 2$) is considered to be universally efficient \cite{leiserson1985fat}, the analysis of its bottleneck structure demonstrates that such design is in general inefficient when flows are regulated by a congestion-control protocol. This is because the fairness and throughput maximization objectives targeted by the congestion control algorithm effectively \textit{bends} the solution space and, as a result, the optimal fat-tree design deviates from the general full fat-tree configuration. This result has implications in the design of data centers that use fat-tree topologies (also known as folded-Clos \cite{fat-tree-amin-10.1145/1402958.1402967}). While in this section we have illustrated how QTBS can be used to optimize a simple fat-tree topology for the case of a symmetric workload pattern, the authors are currently working on deriving the general equations for the optimal design of fat-trees with arbitrary number of spine and leaf links and for generalized (non-symmetric) workload patterns. We will be presenting these results in a forthcoming paper.

\subsection{Traffic Engineering: Accelerating Time-Bound Constrained Flows} \label{ssec:timebound} 


Suppose now that our goal is to accelerate a certain flow $f_t \in \mathcal{F} $ in a network $\mathcal{N}$ so it completes before a target time. A common application for the optimization of time-bound flows can be found in research and education networks, where users need to share data obtained from their experiments, often involving terabytes (or even petabytes) of information, with collaborators around the globe---e.g., when scientists at the European Organization for Nuclear Research (CERN) need to share data with other researchers using the LHCONE network \cite{martelli2015lhcopn} across the globe. Another common use case can be found in large scale data centers, where massive data backups need to be transferred between sites to ensure redundancy \cite{B4-Jain:2013:BEG:2534169.2486019}. In this context, assume operators are only allowed to sacrifice the performance of a subset of flows $\mathcal{F}' \subset \mathcal{F} \setminus \{f_t\}$, considered to be of lower priority than $f_t$. What flows in $\mathcal{F}'$ constitute an optimal choice to traffic-shape so as to accelerate $f_t$? By what amount should the rate of such flows be reduced? And by what amount will flow $f_t$ be accelerated?

\begin{figure}[t]
\centering
\includegraphics[width=0.5\columnwidth]{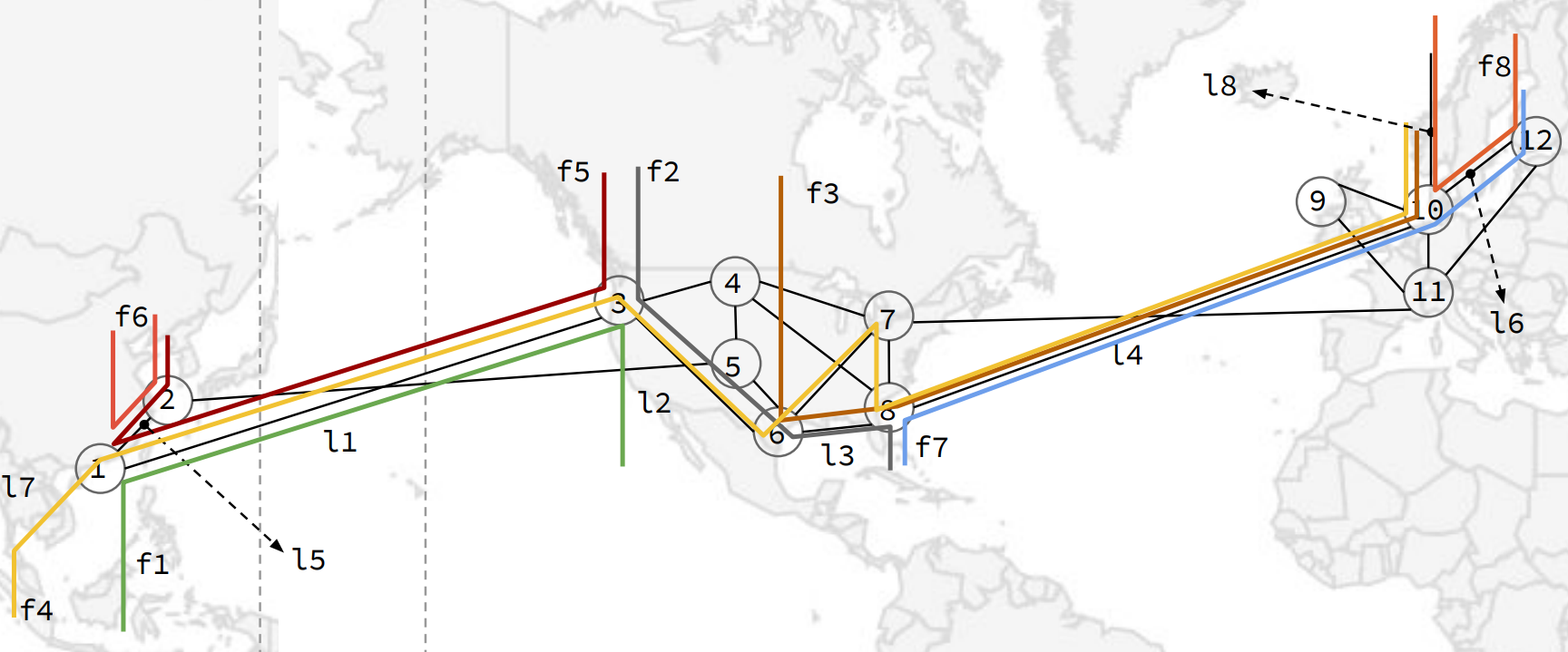}
\caption{Network configuration used in Section \ref{ssec:timebound}. }
\label{fig:net_b4_timebound}
\end{figure}

To illustrate that we can use QTBS to resolve the above problem, we will use the topology of Google's B4 network (Fig. \ref{fig:net_b4_routing}) introduced in Section \ref{ssec:routing}. Assume the network is transporting eight flows, $F = \{f_1, f_2, ..., f_8\}$, routed as shown in the Fig. \ref{fig:net_b4_timebound}. This is without loss of generality as we can apply the same procedure to optimize networks with arbitrary number of flows and topology. We will use the network's bottleneck structure to identify an optimal strategy for accelerating an arbitrary flow in a network. Assume that our objective is to accelerate flow $f_7$ (i.e., $f_t = f_7$) in Fig. \ref{fig:net_b4_timebound}---i.e., the transatlantic flow that connects data centers 8 and 12---to meet a certain flow completion time constraint. Assume also that in order to maximize the performance of $f_7$ we are allowed to traffic shape any of the flows in the set $\mathcal{F}' = \{f_1, f_3, f_4, f_8 \}$. In other words, the set of flows in $\mathcal{F}'$ are considered by the network operator to be of lower priority. 

\begin{figure}[t]
\centering

\begin{subfigure}[b]{.3\columnwidth}
\includegraphics[width=0.8\linewidth]{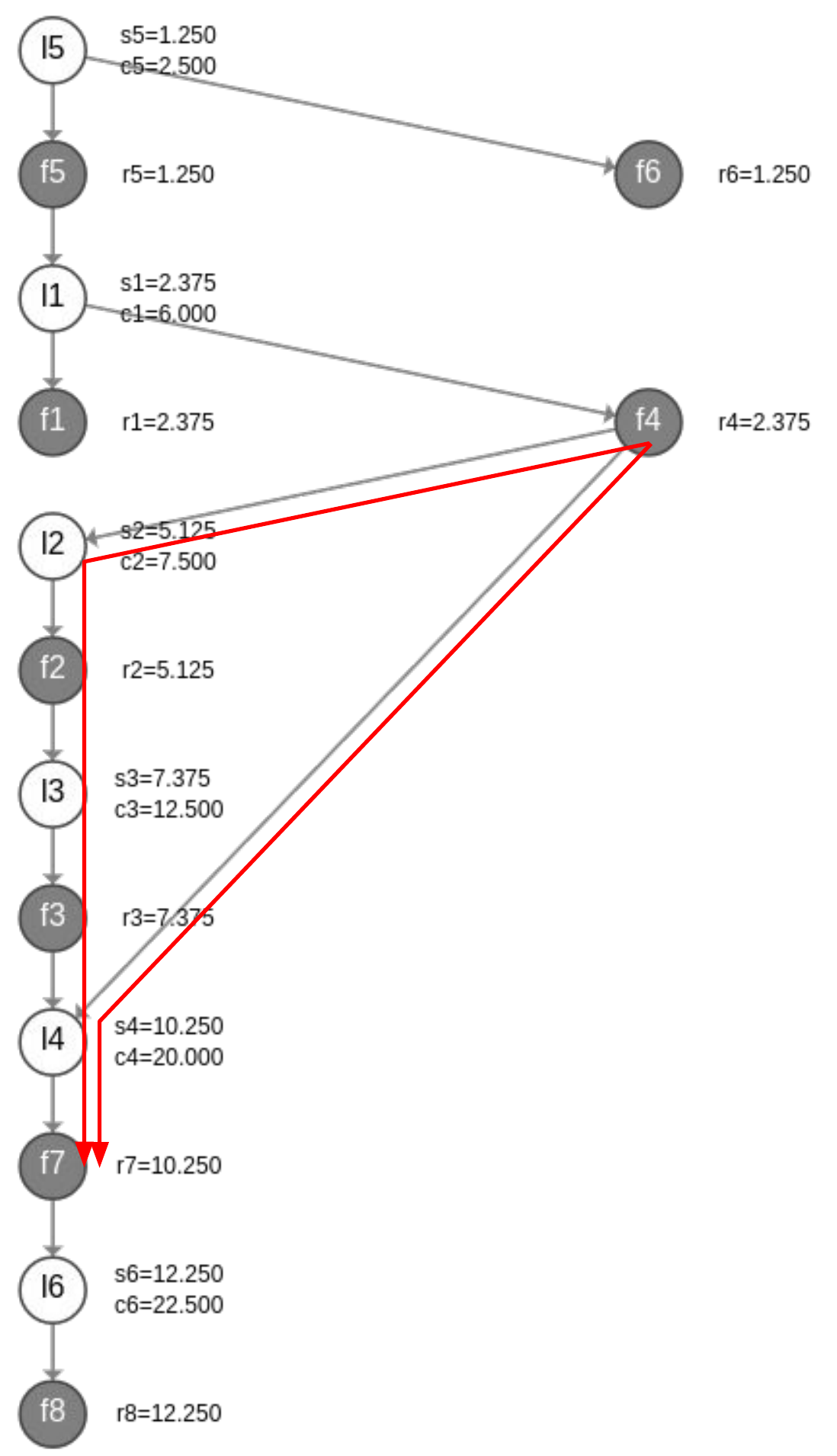}
\caption{Without any traffic shaping.}
\label{fig:gg_timebound_1:1}
\end{subfigure}%
\begin{subfigure}[b]{.3\columnwidth}
\includegraphics[width=0.8\linewidth]{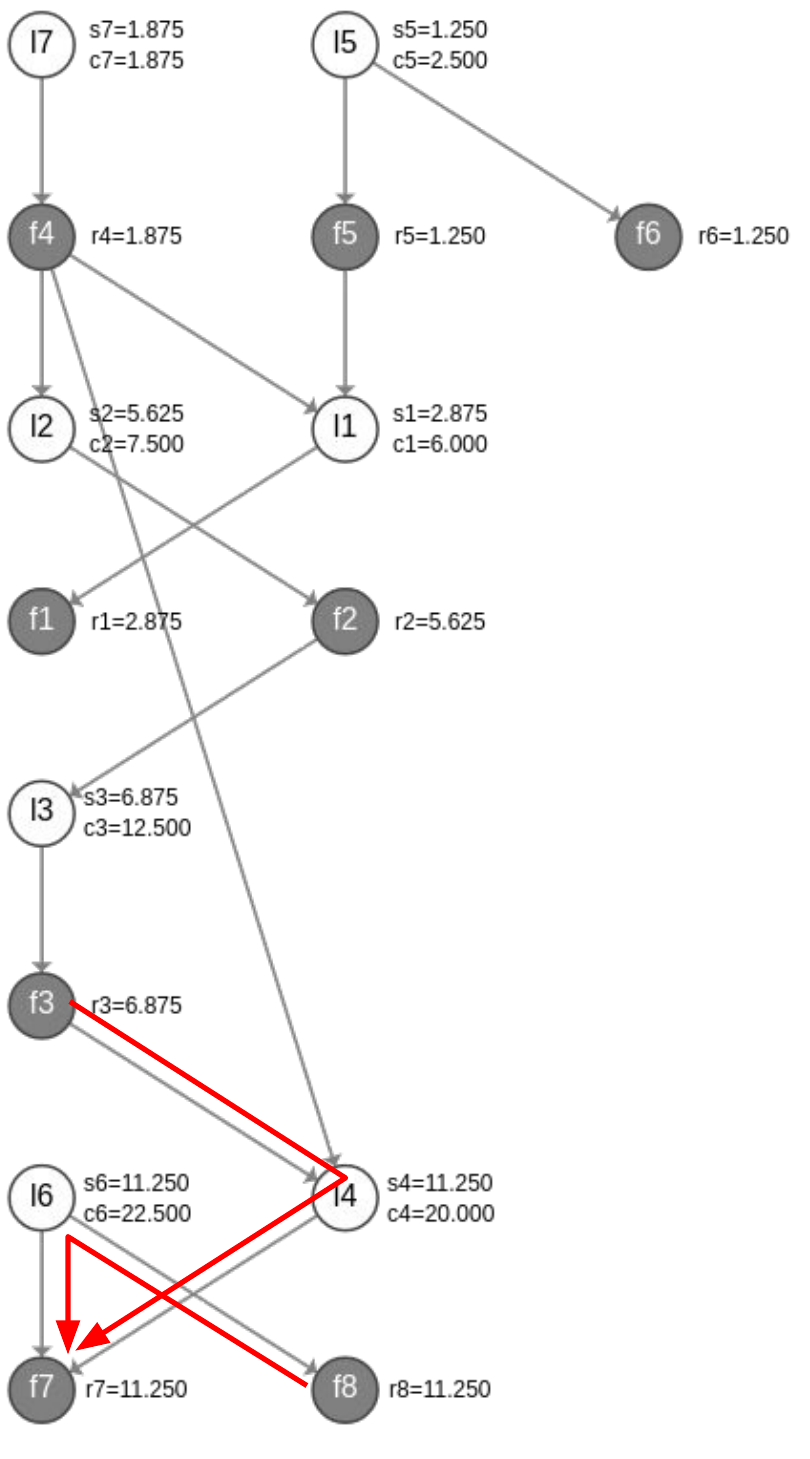}
\caption{Traffic shaping $f_4$.}
\label{fig:gg_timebound_1:2}
\end{subfigure}
\begin{subfigure}[b]{.3\columnwidth}
\includegraphics[width=\linewidth]{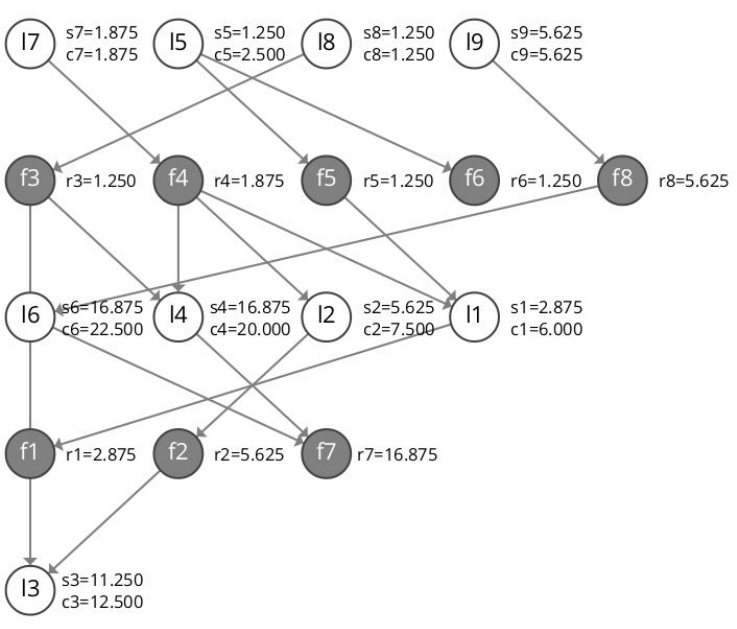}
\caption{Traffic shaping $f_3$, $f_4$ and $f_8$.}
\label{fig:gg_timebound_1:3}
\end{subfigure}

\caption{Bottleneck structures for each of the traffic shaping configurations used to accelerate flow $f_7$.}
\label{fig:gg_timebound_1}
\end{figure}

Fig. \ref{fig:gg_timebound_1} displays the sequence of gradient graphs that lead to the acceleration of flow $f_7$ to meet its time constraint. The graphs include the values of the capacity $c_l$ and fair share $s_l$ next to each link vertex $l$ and the rate $r_f$ next to each flow vertex $f$. Fig. \ref{fig:gg_timebound_1:1} corresponds to the gradient graph of the initial network configuration shown in Fig. \ref{fig:net_b4_timebound} as computed by Algorithm \ref{al:GradientGraph}. From Theorem \ref{lem:propagation}, we know that only the flows that are ancestors to $f_7$ can have an effect on its performance. That means we can discard traffic shaping flow $f_8$ as that will have no impact. We can use the \textit{ForwardGrad()} algorithm (Algorithm \ref{al:ForwardGrad}) to obtain the gradients of flow $f_7$ with respect to the flows in the low priority set $\mathcal{F}'$: $\nabla_{f_1}({f_7}) = 2$, $\nabla_{f_3}({f_7}) = -1$, $\nabla_{f_4}({f_7}) = -2$, and $\nabla_{f_8}({f_7}) = 0$. We are interested in finding the gradient of a flow in $\mathcal{F'}$ that has the highest negative value, so that the traffic shaping of such a flow (i.e., the reduction of its rate) creates a maximal positive increase in the rate of $f_7$. We have that flow $f_4$ has the highest negative gradient with a value of $-2$, yielding an optimal traffic shaping decision. From Fig. \ref{fig:gg_timebound_1:1}, it can be observed that the reduction of flow $f_4$'s rate creates a perturbation that propagates through the bottleneck structure via two different paths: $f_4 \rightarrow l_2 \rightarrow f_2 \rightarrow l_3 \rightarrow f_3 \rightarrow l_4 \rightarrow f_7$ and $f_4 \rightarrow l_4 \rightarrow f_7$. Each of these paths has an equal contribution to the gradient of value $-1$, resulting in $\nabla_{f_4}({f_7}) = -2$. 


We can use the bottleneck structure again to calculate the exact value of the traffic shaper---i.e., the rate reduction applied to flow $f_4$. The core idea is that traffic shaping flow $f_4$ constitutes an optimal decision as long as the bottleneck structure does not change, since a change in the structure would also imply a change in the gradients. As the rate of flow $f_4$ is reduced, some levels in the bottleneck structure will become further away from each other, while the others will become closer to each other. Thus, the latter set will fold if the rate reduction imposed by the traffic shaper is large enough. The speed at which two links in the bottleneck structure get closer to (or further away from) each other is given by their gradients. In particular, if the traffic shaper reduces the rate of flow $f_4$ by an amount of $\rho$ bps, then two links $l$ and $l'$ in the bottleneck structure will collide at a value of $\rho$ that satisfies the equation $s_l - \rho \cdot \nabla_{f_4}({l}) = s_{l'} - \rho \cdot \nabla_{f_4}({l'})$. From the bottleneck structure (Fig. \ref{fig:gg_timebound_1:1}) we can obtain the fair share values $s_l$ and using the \textit{ForwardGrad()} algorithm we can compute the link gradients $\nabla_{f_4}({l})$: $s_{l_2} = 5.125; s_{l_3} = 7.375; s_{l_4} = 10.25; s_{l_6} = 12.25; \nabla_{f_4}({l_2}) = -1; \nabla_{f_4}({l_3}) = 1; \nabla_{f_4}({l_4}) = -2; \nabla_{f_4}({l_6}) = 2$. Using these values, we have that the smallest value of $\rho$ that satisfies the collision equation corresponds to the case $l=l_4$ and $l'=l_6$, yielding a $\rho$ value of $0.5$ (since $10.25 - \rho \cdot (-2) = 12.25 - \rho \cdot 2 \implies \rho = 0.5$). Thus, we conclude that to maximally increase the rate of flow $f_7$, an optimal strategy is to decrease the rate of flow $f_4$ by an amount of $0.5$ units of bandwidth. The resulting bottleneck structure is presented in Fig. \ref{fig:gg_timebound_1:2}, where a new link $l_7$ has been added that corresponds to the new traffic shaper set to reduce the rate of flow $f_4$ by an amount of $0.5$ (from $2.375$ down to $1.875$). Note that as expected, in this new bottleneck structure links $l_4$ and $l_6$ are folded into the same level and have the same fair share: $s_4=s_6=11.25$.

Since $f_7$ has now two bottleneck links ($l_4$ and $l_6$), we cannot accelerate it further unless we increase the fair-shares of both. Using the new bottleneck structure (Fig. \ref{fig:gg_timebound_1:2}), it is easy to see that this can be achieved by decreasing the rate of flows $f_3$ and $f_8$, since the resulting link gradients are each negative $\nabla_{f_3}(l_4) = \nabla_{f_8}(l_6) = -1$. Thus, we add two new traffic shapers $l_8$ and $l_9$ to throttle the rate of flows $f_3$ and $f_8$, respectively, down from their current rates of 6.875 and 11.25. That is: $c_{l_8} = 6.875 - \rho$ and $c_{l_9} = 11.25 - \rho$, for some traffic shaping amount $\rho$. In Fig. \ref{fig:gg_timebound_1:3}, we show the resulting bottleneck structure when choosing a value of $\rho=5.625$ (so $c_{l_8} = 1.25$ and $c_{l_9} = 5.625$), which further accelerates the rate of flow $f_7$ to $r_7 = s_{l_4} - \rho \cdot \nabla_{f_3}(l_4) = s_{l_6} - \rho \cdot \nabla_{f_8}(l_6)  = 11.25 - 5.625 \cdot (-1) = 16.875$. Note that there is some flexibility in choosing the value of this parameter, depending on the amount of acceleration required on flow $f_7$. In this case, we chose a value that maximally accelerates flow $f_7$ while ensuring none of the flows that are traffic shaped receives a rate lower than any other flow. With this configuration, flow $f_3$'s rate is reduced to the lowest transmission rate among all flows in the network, but this value is no lower than the rate of flows $f_5$ and $f_6$ ($r_{f_3} = r_{f_5} = r_{f_6} = 1.25$). Thus, the flow completion time of the slowest flow is preserved throughout the transformations performed in this example.

In summary, a strategy to accelerate the performance of flow $f_7$ consists in traffic shaping the rates of flows $f_3$, $f_4$ and $f_8$ down to $1.25$, $1.875$ and $5.625$, respectively. Such a configuration results in a theoretical increase to the rate of flow $f_7$ from $10.25$ to $16.875$, while ensuring no flow performs at a rate lower than the slowest flow in the initial network configuration. Note that among all the low priority flows in $\mathcal{F}'$, in the above process we opted for not reducing the rate of flow $f_1$. Indeed, the three bottleneck structures computed by this algorithm (Fig. \ref{fig:gg_timebound_1}) tell us that choosing to reduce the rate of flow $f_1$ would in fact have either a negative effect or no effect at all on the rate of flow $f_7$, since the gradients $\nabla_{f_1}({f_7})$ for each structure are $2$, $0$ and $1$, respectively---that is, a reduction on the rate of flow $f_1$ produces a non-positive impact on the rate of flow $f_7$ in all cases. Thus, the quantitative analysis resulting from the bottleneck structure of the network reveals not only the set of flows that should be traffic shaped, but also the flows that should \emph{not} be traffic shaped, as doing so would actually hurt the performance of the flow we intend to accelerate. Note that this result challenges some of the established best practices for traffic engineering flows, which include many proposed algorithms that focus on reducing the rate of the heavy-hitter flows to improve high-priority flows. As shown in this example, without taking into account the bottleneck structure of a network, such algorithms may recommend a traffic shaping configuration that actually has the opposite of the intended effect.

\begin{figure*}[ht]
\centering
\begin{subfigure}[b]{.3\linewidth}
\includegraphics[width=\linewidth]{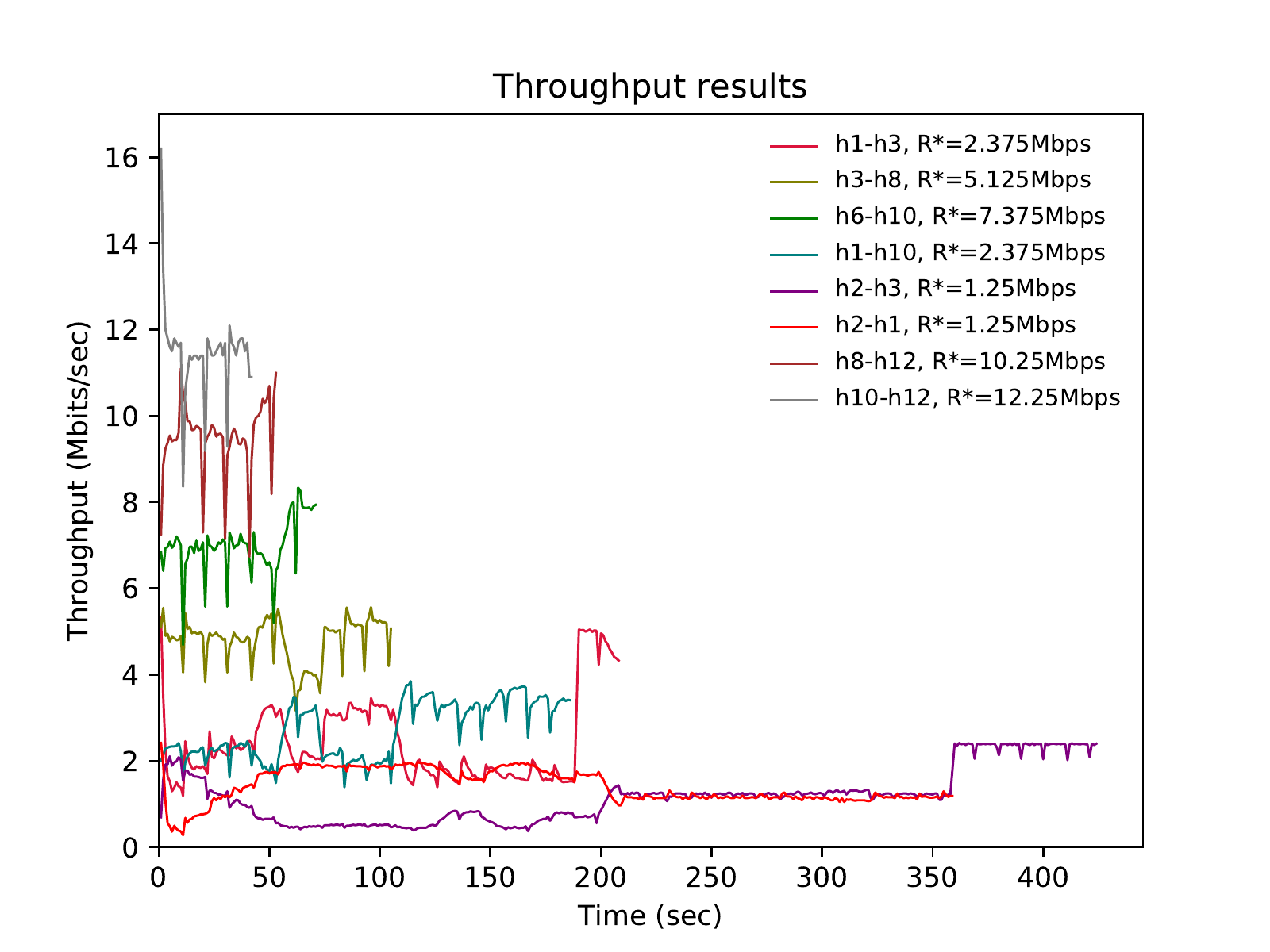}
\caption{Without any traffic shaping.}
\label{fig:ex_timebound_1:1}
\end{subfigure}%
\begin{subfigure}[b]{.3\linewidth}
\includegraphics[width=\linewidth]{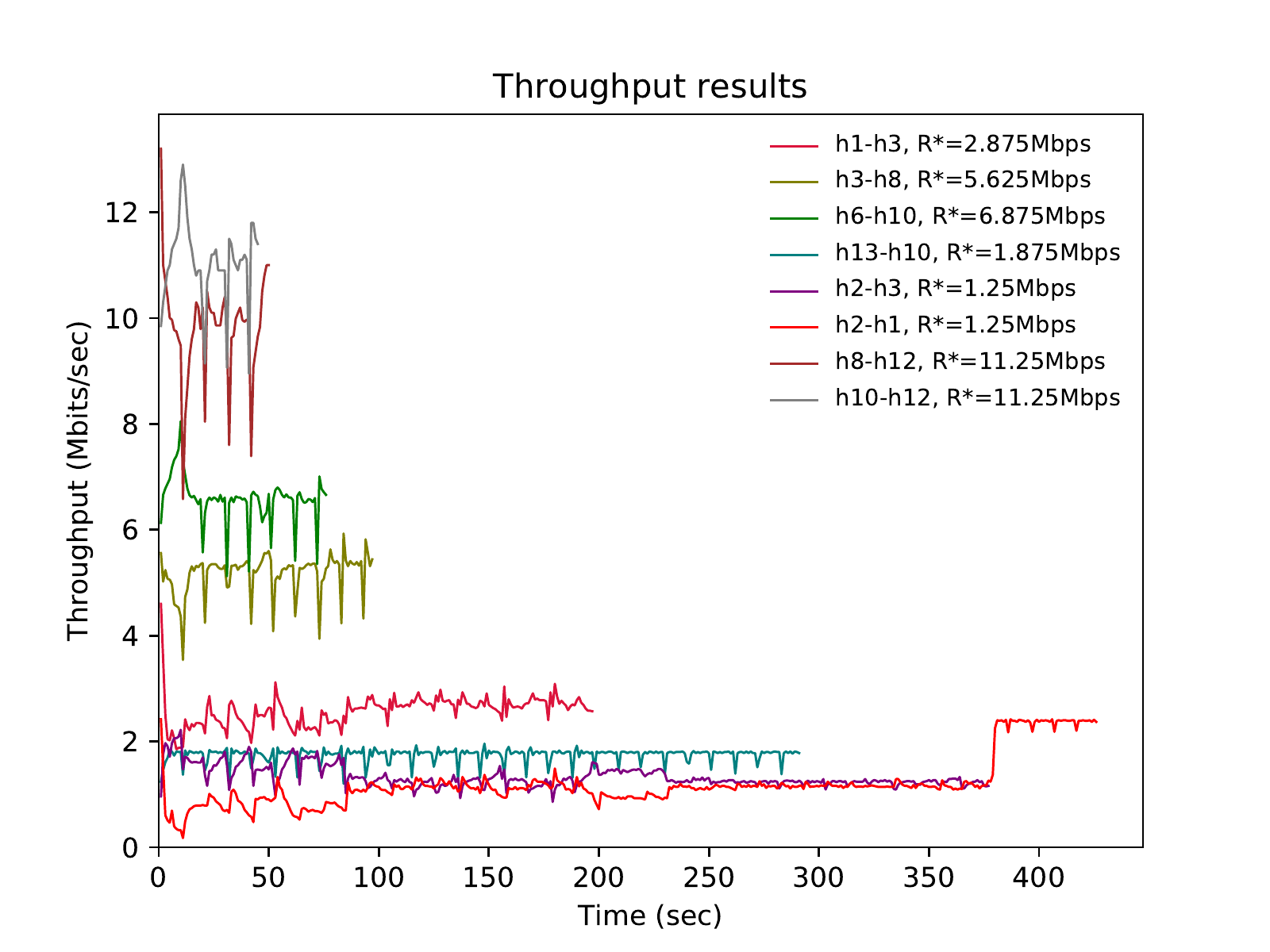}
\caption{Traffic shaping $f_4$.}
\label{fig:ex_timebound_1:2}
\end{subfigure}
\begin{subfigure}[b]{.3\linewidth}
\includegraphics[width=\linewidth]{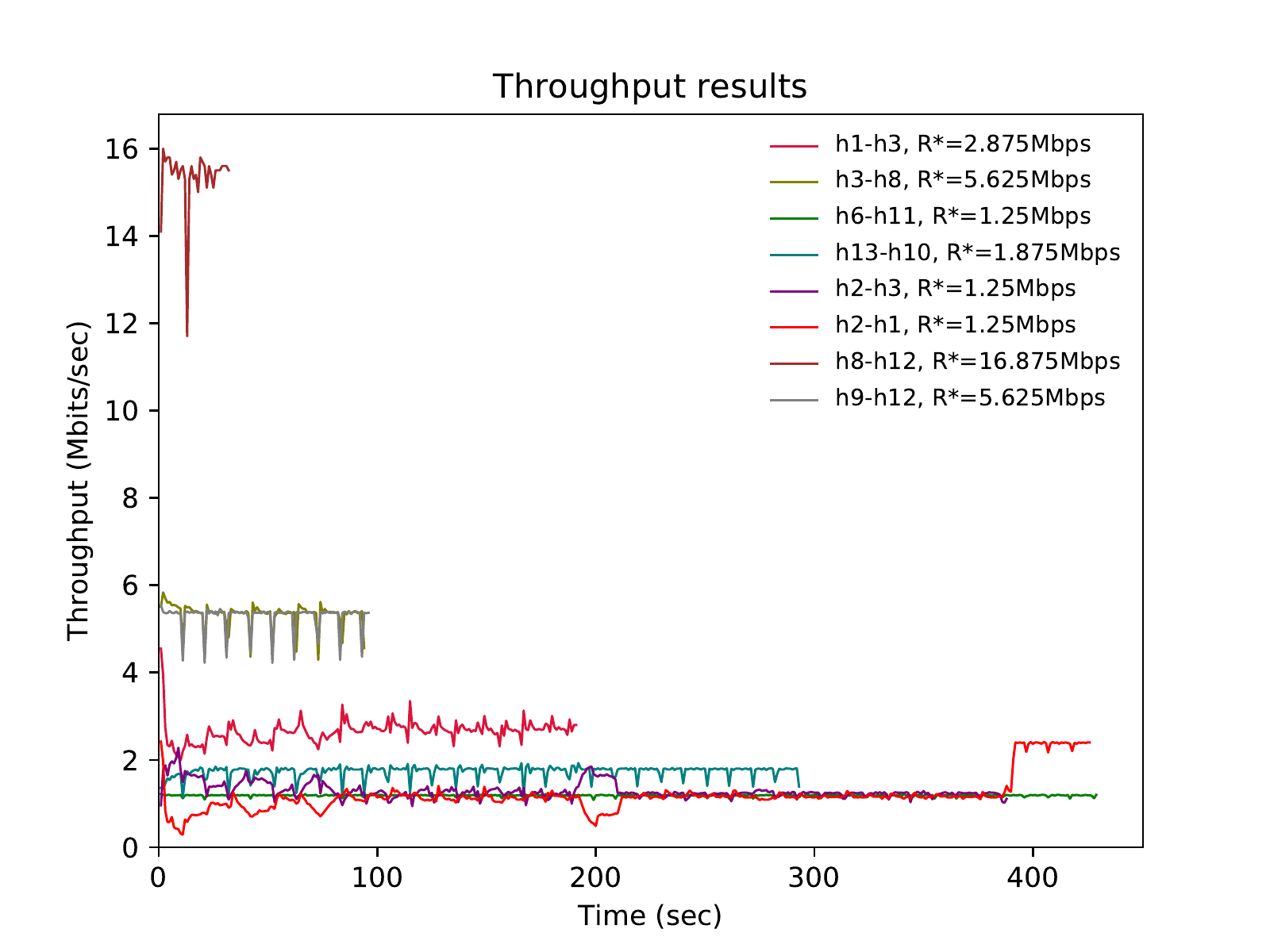}
\caption{Traffic shaping $f_3$, $f_4$ and $f_8$.}
\label{fig:ex_timebound_1:3}
\end{subfigure}

\caption{Flow performance obtained when deploying the traffic shapers to accelerate flow $f_7$ ($h_{8}-h_{12}$).}
\label{fig:ex_timebound_1}
\end{figure*}

To empirically demonstrate the accuracy of QTBS in identifying the set of traffic shapers and their optimal rate, we reproduce the experiments described in this section using  \textit{Mininet-G2}. Fig. \ref{fig:ex_timebound_1} illustrates the performance of the flows for each of the three traffic shaping configurations shown in Fig. \ref{fig:gg_timebound_1} using the BBR congestion control algorithm. The legends in these figures describe the flows, where the notation $h_x-h_y$ means that the flow goes from host $h_x$ to host $h_y$. To map the flows according to Fig. \ref{fig:net_b4_timebound}, we use the convention that host $h_x$ is located in data center $x$. For instance, flow $h_8-h_{12}$ in Fig. \ref{fig:ex_timebound_1} corresponds with flow $f_7$ in Fig. \ref{fig:net_b4_timebound}, which starts at datacenter 8 and ends at datacenter 12. Table \ref{tab:time-bound} shows the average transmission rate obtained for each of the flows and for each of the three experiments. Next to each experimental rate value, this table also includes the theoretical flow transmission rate according to the bottleneck structure. It is easy to see that these values match the transmission rate $r_f$ shown next to each flow vertex (gray vertices) from the corresponding bottleneck structures in Fig. \ref{fig:gg_timebound_1}.

Fig. \ref{fig:ex_timebound_1:1} shows the results of running the initial network without any traffic shapers, corresponding to the bottleneck structure in Fig. \ref{fig:gg_timebound_1:1}. From Table \ref{tab:time-bound}, we see that all experimentally measured flow rates usually track their theoretical value from slightly below. Such an offset between experimental and theoretical rates is a characteristic that holds for all experiments, and is due to imperfections in the distributed nature of the congestion control algorithm (e.g., due to its inability to instantaneously converge to the optimal transmission rate or due to statistical packet drops produced by the asynchronous nature of the network). However, the table clearly demonstrates that the experimental rates behave according to the bottleneck structure of the network. This result is also reinforced by the fact that Jain's fairness index is above 0.99 for all experiments, as shown in Appendix \ref{app:jainfidx}.

Fig. \ref{fig:ex_timebound_1:2} shows the result of adding the first traffic shaper, configured to reduce the rate of flow $f_4$ by an amount of 0.5 Mbps. As predicted by QTBS, this increases the rate of flow $f_7$ (the purple flow $h_8-h_{12}$ in Fig. \ref{fig:ex_timebound_1}), in this case from 9.51 to 9.81 Mbps (Table \ref{tab:time-bound}). Fig. \ref{fig:ex_timebound_1:3} shows the result of adding two additional traffic shapers to reduce the rate of flows $f_3$ and $f_4$ by a an amount of 5.625 Mbps, according to our quantitative analysis of the bottleneck structure. Recall that this configuration was designed to ensure a maximal increase in the rate of flow $f_7$ without decreasing any of the flows' rate below the rate of the slowest flow. We see this behavior in Fig. \ref{fig:ex_timebound_1:3}, where flow $f_7$ (purple flow) has now the highest rate, while the flow completion time of the slowest flow remains at slightly above 400 seconds, throughout the three experiments (Fig. \ref{fig:ex_timebound_1:1}, \ref{fig:ex_timebound_1:2} and \ref{fig:ex_timebound_1:3}). In summary, the combined effect of the three traffic shapers accelerates the observed rate of flow $f_7$ from 9.51 to 15.34 Mbps. As shown in Table \ref{tab:time-bound}, this result closely matches the behavior predicted by the bottleneck structure---that the rate would increase from 10.25 to 16.87 Mbps, while the observed maximum flow completion time of the network remains constant throughout the three experiments.

\begin{table}
\center
\caption{Experimental versus theoretical average flow transmission rate (units in Mbps).}
\label{tab:time-bound}
\scalebox{1}{
\begin{tabular} {  c  c  c  c  c  c  c  c}
\hline
Flow & Experiment 1 & Experiment 2 & Experiment 3
\\ \hline
$f_1$ & 2.44 / 2.37 & 2.57 / 2.87 & 2.65 / 2.87 \\
$f_2$ & 4.78 / 5.12 & 5.16 / 5.62 & 5.33 / 5.62 \\
$f_3$ & 6.99 / 7.37 & 6.57 / 6.87 & 1.18 / 1.25 \\
$f_4$ & 2.72 / 2.37 & 1.74 / 1.87 & 1.73 1.87 \\
$f_5$ & 1.18 / 1.25 & 1.33 / 1.25 & 1.29 / 1.25 \\
$f_6$ & 1.42 / 1.25 & 1.19 / 1.25 & 1.19 / 1.25 \\
$f_7$ & 9.51 / 10.25 & 9.81 / 11.25 & 15.34 / 16.87 \\
$f_8$ & 11.48 / 12.25 & 11.06 / 11.25 & 5.27 / 5.62 \\

\hline
\end{tabular}
}

\end{table}



\section{Related Work} \label{sec:assumptions}

The problem of congestion control is one of the most widely studied areas in data networks. The first congestion control algorithm for the Internet was introduced by Jacobson in \cite{Jacobson:1988:CAC:52325.52356} and implemented as part of the TCP protocol, initiating a more than three-decade long period of intense research. This has resulted in a long list of congestion control algorithms (e.g., \cite{Reno:Fall:1996:SCT:235160.235162}, \cite{Cubic:Ha:2008:CNT:1400097.1400105}), including the BBR algorithm recently proposed by Google \cite{Cardwell:2016:BCC:3012426.3022184}. All of these algorithms are based on the belief that the performance of a flow is solely characterized by the state of its bottleneck. In our work, we show that QTBS reveals a richer story of how such bottlenecks perform and interact with each other from a system-wide performance standpoint. 

A well-known example of the traditional single-bottleneck view is the Mathis equation \cite{Mathis:1997:MBT:263932.264023}, which models the performance of a single TCP flow based on the equation $MSS/(RTT \cdot \sqrt{p})$, where $MSS$ is the maximum segment size, $RTT$ is the round trip time of the flow and $p$ is the packet loss probability. This equation, however, does not take into account the system-wide properties of a network, including its topology, the routing and the interactions between flows. QTBS addresses this gap and provides a methodology to numerically estimate flow throughput. In future research we plan to incorporate the effects of latency and packet loss to QTBS.

The concept of bottleneck structure was recently introduced in \cite{Ros-Giralt-SIGMETRICS-10.1145/3366707}. That work focused on the qualitative properties of the bottleneck precedence graph (BPG), a structure that organizes the relationships among links. Our work focuses on the analysis of a bottleneck structure called the gradient graph. The key difference between the gradient graph and the BPG is that the gradient graph describes the relationships among flows and links, not just links, providing a more comprehensive view of the network. As a result, the gradient graph provides a framework to quantify the interactions among flows and links, resulting in a new class of algorithms to optimize network performance. To the best of our knowledge, this paper presents the first quantitative theory of the analysis of bottleneck structures in data networks.

The problem of traffic engineering (TE) has also been widely studied and continues to be a very active area of research and development. Because QTBS provides a new approach to network optimization, it can be used co-located with existing TE frameworks, augmenting them with detailed information about the interactions among bottleneck links and flows. For instance, in \cite{BWE-10.1145/2829988.2787478}, Google introduces Bandwidth Enforcer (BwE), a centralized bandwidth allocation infrastructure for wide area networking that targets high network utilization. QTBS complement tools like BwE by providing a view of the network's bottleneck structure (for online or offline analysis) and providing traffic shaping and flow routing recommendations such as those presented in Sections \ref{ssec:timebound} and \ref{ssec:routing}.




\section{Conclusions} \label{sec:conclusions}



The analytical strength of a bottleneck structure stems from its ability to capture the solution-space produced by a congestion-control algorithm taking into account the topological and routing constraints of the network. Based on this concept, we develop a quantitative theory of bottleneck structures (QTBS), a new mathematical framework that allows to optimize congestion-controlled networks by providing very efficient algorithms to compute derivatives on the performance parameters of links and flows. To explore the analytical power of QTBS, we use it to reveal insights in traffic engineering and network design problems that are themselves contributions to the literature. In one experiment, we use QTBS to develop a novel routing algorithm that identifies maximal throughput paths, enabling a scalable methodology to jointly solve the problems of routing and congestion control. In another experiment, we use QTBS to reveal the existence of optimal capacity allocations in the spine links of a fat-tree network that outperform (in cost and/or performance) the traditional full fat-tree network designs found in some large-scale data centers and supercomputers. In a third experiment, we demonstrate how to use bottleneck structures to compute the numerical values of optimal rate settings in traffic shapers to help improve the performance of high-priority flows. This paper positions the concept of bottleneck structures as a promising analytical framework to optimize 
network performance.



%

\bibliographystyle{IEEEtran}
\bibliography{qostheory}

\section{Appendices}

\appendix

\renewcommand{\thefigure}{A\arabic{figure}}
\setcounter{figure}{0}

\section{Mathematical Proofs}

\subsection{Generalization to Max-min Fairness} \label{app:proof:maxmin-bottleneck} 

\begin{lemma}
\textit{
If a link is a bottleneck in the max-min sense \cite{Bertsekas:1992:DN:121104}, then it is also a bottleneck according to Definition \ref{def:bottleneck}, but not vice-versa.
}
\end{lemma}
\begin{proof} 
Bertsekas and Gallager \cite{Bertsekas:1992:DN:121104} proved that if a flow $f$ is bottlenecked at link $l$ in the max-min sense, then such a flow must traverse link $l$ and its rate is equal to the link's fair share, $r_f = s_l$. Since a change in the capacity of a link always leads to a change in its fair share, i.e. $\partial s_l / \partial c_l \neq 0$, this necessarily implies $\partial r_f / \partial c_l \neq 0$. Thus, $f$ is also bottlenecked at link $l$ in the sense of Definition \ref{def:bottleneck}.
The reverse, however, does not hold because Definition \ref{def:bottleneck} does not require that $r_f = s_l$ for a flow $f$ bottlenecked at link $l$. (It can be seen that this is also true for other definitions of bottleneck. For instance, a flow that is bottlenecked at a link according to proportional fairness \cite{Kelly1998}, is also bottlenecked according to Definition \ref{def:bottleneck}, but the reserve is also not true.)
\end{proof}

\subsection{Proof of Theorem \ref{lem:propagation}: Propagation of Network Perturbations} \label{app:proof:propagation} 
\textit{
Let $x, y \in \mathcal L \cup \mathcal F$ be a pair of links or flows in the network. Then a perturbation in the capacity $c_x$ (for $x \in \mathcal L$) or transmission rate $r_x$ (for $x \in \mathcal F$) of $x$ will affect the fair share $s_y$ (for $y \in \mathcal L$) or transmission rate $r_y$ (for $y \in \mathcal F$) of $y$ if only if there exists a directed path from $x$ to $y$ in the gradient graph.
}
\begin{proof} 
Consider the case $x = l \in \mathcal{L}$ and assume link $l$ is affected by a perturbation in its capacity. From Definition \ref{def:bottleneck}, we have that $\partial r_{f^*} / \partial c_l \neq 0$, for any flow $f^*$ bottleneck at link $l$. From Definition \ref{def:gradgraph}, these correspond to all flows $f^*$ for which there exists an edge $(l, f^*)$ in $\mathcal{G}$. Let $f_1$ be any of these flows and assume $\Delta_{f_1}$ is its drift. Such drift will induce a perturbation in all the links traversed by $f_1$. From Definition \ref{def:gradgraph}, this corresponds to all the links $l^*$ for which there exists an edge $(f_1, l^*)$ in $\mathcal{G}$. This process of perturbation followed by a propagation repeats itself, affecting all the link and flow vertices that can be reached from link $l$ through a directed path in the gradient graph, and ending at leaf vertices. This demonstrates the sufficient condition of the theorem. The necessary condition is also true because, from the definition of region of influence (Definition \ref{def:region_influence}), none of the links and flows outside $\mathcal{R}(l)$ will be affected by the perturbation. The proof for the case $x = f \in \mathcal{F}$ follows the same argument except that the initial perturbation is applied on the rate of flow $f$. 
\end{proof}

\subsection{Proof of Theorem \ref{lem:forward_grad_correct}: Correctness of \textit{GradientGraph()}} \label{app:proof:forward_grad} 

Let $\mathcal{N} = \langle \mathcal{L}, \mathcal{F}, \{c_l,\forall l \in \mathcal{L} \} \rangle$ be a network and let $\mathcal G$ be the corresponding gradient graph. Let $x \in \mathcal{L} \cup \mathcal F$. After running Algorithm \ref{al:ForwardGrad}, $\Delta s_l = \nabla_x(l)$ for all $l \in \mathcal L$, and $\Delta r_f = \nabla_x(f)$ for all $f \in \mathcal F$.
\begin{proof}
Assume without loss of generality that $x \in \mathcal L$. Our goal is to prove:
\begin{align*}
    \frac{\partial r_f}{\partial c_x} = \Delta r_f \qquad \forall f \in \mathcal F \\
    \frac{\partial s_l}{\partial c_x} = \Delta s_l \qquad \forall l \in \mathcal L
\end{align*}
This is equivalent to showing that, for sufficiently small $\delta$, if we perturbed the capacity of link $x$, $c_x' = c_x + \delta$, and recomputed all the flows' rates and links' fair shares using $\textit{GradientGraph()}$, we would get:
\begin{align*}
    r_f' = r_f + \Delta r_f \cdot \delta \qquad \forall f \in \mathcal F \\
    s_l' = s_l + \Delta s_l \cdot \delta \qquad \forall l \in \mathcal L
\end{align*}
Let $y \in \mathcal L \cup \mathcal F$. There are two cases. First, assume $y \notin \mathcal{R}(x)$, where $\mathcal{R}(x)$ is the region of influence of $x$, then by definition there is no directed path from $x$ to $y$. The algorithm only processes vertices that lie on a directed path from vertex $x$. Thus, if $y \in \mathcal L$, $\Delta s_y = 0$ (line \ref{al:ForwardGrad:init_s}) and if $y \in \mathcal F$, $\Delta r_y = 0$ (line \ref{al:ForwardGrad:init_r}). Moreover, by Theorem \ref{lem:propagation}, $y$ is not affected by the perturbation of $c_x$. That is, $s_l' = s_l$ or $r_f' = r_f$. Thus the equations hold. \\

Now let $y \in \mathcal{R}(x)$. We proceed by induction. As a base case, let $y = x$. The amount of leftover bandwidth at node $x$, which is called $a_x$ in the \textit{GradientGraph()} Algorithm, is the capacity $c_x$ minus the rates of vertices that lie outside the region of influence of $x$, since they are not bottlenecked at $x$ (see \textit{GradientGraph()} lines \ref{al:GradientGraph:init_a} and \ref{al:GradientGraph:set_a}). Thus $a_x' = a_x + (c'_x - c_x) = a_x + \delta$. The fair share rate of $x$ is $s_x = a_x / |\mathcal F_x \setminus \mathcal V|$ (line \ref{al:GradientGraph:set_s} of \textit{GradientGraph()}), where $\mathcal V$ is the set of vertices that were added to the gradient graph before $x$. These necessarily lie outside the region of influence of $x$, so this set is the same after the perturbation. Indeed, $\mathcal F_x \setminus \mathcal V|$ is the set of flows which are bottlenecked at $x$, which is also the set of successors of $x$. Thus,
\[ s_x' = \frac{a_x + \delta}{|\mathcal F_x \setminus \mathcal V|} = s_x + \frac{\delta}{\sigma(x, \mathcal G)} = s_x + \Delta s_x \cdot \delta \]
where $\sigma(x, \mathcal G)$ corresponds to the set of successors of node $x$ as indicated by the gradient graph $\mathcal G$. Thus, the equations hold for $x$. \\

Now assume the equations hold for all vertices which are added to the new gradient graph prior to vertex $y$. First, assume $y \in \mathcal F$. Then $r_y'$ will be the minimum fair share rate of the links it traverses, that is, $\min s_l'$ for all $l \in \mathcal L_y$. However, it suffices to take the minimum over links which were predecessors (bottlenecks) of $y$ in the original gradient graph (these links' fair shares were strictly smaller than those of the other vertices that $y$ traverses, so after an infinitesimally small perturbation they will remain strictly smaller). Thus,
\[ r'_y = \min_{l \in \pi(y, \mathcal G)} s'_l = \min_{l \in \pi(y, \mathcal G)} (s_l + \Delta s_l \cdot \delta) \]
where $\pi(y, \mathcal G)$ corresponds to the set of predecessors of node $y$ as indicated by the gradient graph $\mathcal G$. Now we can substitute $s'_l; = s_l + \Delta s_l \cdot \delta$ because of the induction hypothesis. If $y$ had multiple bottleneck links in the original graph, then they all had the same fairshare. That is, $s_l = r_y$ for all $l \in \pi(y, \mathcal G)$ So
\[ r'_y = r_y + \left[\min_{l \in \pi(y, \mathcal G)} \Delta s_l \right] \cdot \delta \]
Combining this with line \ref{al:ForwardGrad:set_r} of \textit{ForwardGraph()},
\begin{align*} 
    &\Delta r_{y} = \min_{l \in \pi(y, \mathcal G)} \Delta s_l \\
    \implies& r'_y = r_y + \Delta r_{y} \cdot \delta
\end{align*}
which is what we wanted to prove under the assumption that $y \in \mathcal F$. Now, assume $y \in \mathcal L$. The old fair share of $y$ was its available capacity (called $a_y$ in \textit{GradientGraph()}) divided by the number of flows that it bottlenecked, $|\sigma(y, \mathcal G)|$. That is,
\[ s_y = \frac{a_y}{|\sigma(y, \mathcal G)|} = \frac{c_y - \sum_{f \in \mathcal F_y \setminus \sigma(y, \mathcal G)} r_f}{|\sigma(y, \mathcal G)|} \]
(See \textit{GradientGraph()} lines \ref{al:GradientGraph:init_a}, \ref{al:GradientGraph:set_a}, and \ref{al:GradientGraph:set_s}). The new fair share is 
\begin{align*}
    s'_y = \frac{a'_y}{|\sigma(y, \mathcal G')|} = \frac{c'_y - \sum_{f \in \mathcal F_y \setminus \sigma(y, \mathcal G')} r'_f}{|\sigma(y, \mathcal G')|}
\end{align*}
The capacity of $y$ has not changed (unless $y = x$, which we already considered) so $c'_y = c_y$. As a result of the perturbation, some flows which used to be bottlenecked at $y$ may no longer be (if they were perturbed directly, or if they had a second bottleneck that was affected). But no new bottlenecked flows were created, since flows that were not bottlenecked at $y$ before had rates strictly smaller than $s_y$, and the perturbation is infinitesimally small. Let
\[ T_y = \sigma(y, \mathcal G) \setminus \sigma(y, \mathcal G') \]
be the flows that were bottlenecked at $y$, but whose rate was reduced an infinitesimal amount by the perturbation, so that they no longer are.
Then with some algebraic manipulation,
\begin{align*}
    s'_y 
    &= \frac{c_y - \sum_{f \in \mathcal F_y \setminus \sigma(y, \mathcal G')} r'_f}{|\sigma(y, \mathcal G')|} \\
    &= \frac{c_y - \left(\sum_{f \in \mathcal F_y \setminus \sigma(y, \mathcal G')} r_f\right) - \left(\sum_{f \in \mathcal F_y \setminus \sigma(y, \mathcal G')} r'_f - r_f \right)}{|\sigma(y, \mathcal G')|} \\
    &= \frac{c_y - \left(\sum_{f \in \mathcal F_y \setminus \sigma(y, \mathcal G)} r_f\right) - \left(\sum_{f \in T_y} r_f\right)}{|\sigma(y, \mathcal G')|} \\
    &- \frac{\left(\sum_{f \in \mathcal F_y \setminus \sigma(y, \mathcal G')} r'_f - r_f \right)}{|\sigma(y, \mathcal G')|} \\
    &= \frac{a_y - |T_y|s_y}{|\sigma(y, \mathcal G)| - |T_y|} - \frac{\left(\sum_{f \in \mathcal F_y \setminus \sigma(y, \mathcal G')} r'_f - r_f \right)}{|\sigma(y, \mathcal G')|} 
\end{align*}
where in the last line, we have used the fact that $r_f = s_y$ for all $f \in T_y$, since each of these flows used to be bottlenecked at at $y$ before the perturbation. Furthermore, since
\[ s_y = \frac{a_y}{|\sigma(y, \mathcal G)|} \]
the first term in the last line above simplifies as follows:
\begin{align*}
    \frac{a_y - |T_y|s_y}{|\sigma(y, \mathcal G)| - |T_y|}
    = \frac{|\sigma(y, \mathcal G)|s_y - |T_y|s_y}{|\sigma(y, \mathcal G)| - |T_y|} = s_y
\end{align*}
Thus
\[ s_y' = s_y - \frac{\left(\sum_{f \in \mathcal F_y \setminus \sigma(y, \mathcal G')} r'_f - r_f \right)}{|\sigma(y, \mathcal G')|} \]
In $\textit{ForwardGrad()}$, each of the flow vertices in $F_y \setminus \sigma(y, \mathcal G')$ has a smaller key in the heap than $y$ does, since they either had a smaller rate than $s_y$ before, or they have a more negative drift than $y$ does (so their $\Delta r_f < \Delta s_y$). This means that they will be processed before $y$, and that $\Delta r_f = r_f' - r_f$. Thus, by the time $y$ is processed,
\[ \Delta c_y = - \sum_{f \in \mathcal F_y \setminus \sigma(y, \mathcal G')} \Delta r_f = - \sum_{f \in \mathcal F_y \setminus \sigma(y, \mathcal G')} (r'_f - r_f)/\delta \]
(see line \ref{al:ForwardGrad:set_c}) where we have $r_f' = r_f + \Delta r_f \cdot \delta$ by the induction hypothesis. As we just reasoned, by the time $y$ is processed, all the nodes in $T_y$ have been visited already for the same reason. Thus,
\[ |\sigma(y, \mathcal G) \setminus V| = |\sigma(y, \mathcal G) \setminus T_y| = |\sigma(y, \mathcal G') \]
Combining with line \ref{al:ForwardGrad:set_s}, 
\begin{align*}
    \Delta s_{y} = \frac{\Delta c_{y} }{|\sigma(y, \mathcal G) \setminus V|} \\
    \implies s_y = s_y + \frac{\Delta c_{y} \cdot \delta}{|\sigma(y, \mathcal G) \setminus V|}= s_y + \Delta s_y \cdot \delta
\end{align*}
which is what we wanted to prove. By induction, $\Delta s_l = \partial s_l / \partial c_x$ and $\Delta r_f = \partial r_f / \partial c_x$ for all links $l$ and all flows $f$ in the region of influence of $x$.
\end{proof}

\subsection{Proof of Theorem \ref{lem:forward_grad_runtime}: Time Complexity of \textit{ForwardGrad()}} \label{app:proof:forward_grad_runtime}
Let $x \in \mathcal L \cup \mathcal F$. Then Algorithm \ref{al:ForwardGrad} finds the gradients of all links and flows in the network with respect to $x$ in time $O(|\mathcal R(x)| \cdot \log |\mathcal R(x)|)$. 
\begin{proof}
    Algorithm \ref{al:ForwardGrad} only adds vertices to the heap (line \ref{al:ForwardGrad:push_flow} and \ref{al:ForwardGrad:push_link}) if they are neighbors of a previously visited vertex, so it only visits vertices in the region of influence $\mathcal{R}(x)$. Moreover, the algorithm only visits each node once (lines \ref{al:ForwardGrad:pop} - \ref{al:ForwardGrad:no_repeats}). The only operation that is not constant-time is updating the heap (lines \ref{al:ForwardGrad:push_flow} and \ref{al:ForwardGrad:push_link}). Since the heap has at most $|\mathcal{R}(x)|$ elements, each of these operations takes $\log |\mathcal{R}(x)|$, so the total runtime of the algorithm is $O(|\mathcal{R}(x)| \cdot \log |\mathcal{R}(x)|)$.
\end{proof}

\subsection{Proof of Lemma \ref{lem:complexity}: Time Complexity of GradientGraph()} \label{app:proof:complexity} 

\textit{
The time complexity of running \textit{GradientGraph()} is $O(|\mathcal{L}| \log |\mathcal{L}| \cdot H)$, where $H$ is the maximum number of flows that traverse a single link.
}
\begin{proof} 
Note that each statement in the algorithm runs in constant time except for lines \ref{al:GradientGraph:init_push}, \ref{al:GradientGraph:pop}, and \ref{al:GradientGraph:update}. Each is an operation on a heap of size at most $|\mathcal{L}|$, so each will run in $\log|\mathcal{L}|$ time. Lines \ref{al:GradientGraph:init_push} and \ref{al:GradientGraph:pop} will each run $|\mathcal{L}|$ times, since the two outer loops run at most once for each link. Line \ref{al:GradientGraph:update} will run at most once for every pair of a link with a flow that traverses it. Note that this value is less than the number of edges that are added to the gradient graph in lines \ref{al:GradientGraph:add_bneck_edges} and \ref{al:GradientGraph:add_downstream_edges}. Thus, the number of times line \ref{al:GradientGraph:update} is run is bounded by $|\mathcal{L}| \cdot H$, where $H$ is the maximum number of flows that traverse a single link. Thus, in total, the algorithm runs in time $O(H |\mathcal{L}| \log (|\mathcal{L}|))$. 
\end{proof}

\subsection{Proof of Property \ref{prop:bound}: Gradient Bound} \label{app:proof:bound} 
\textit{
Let $ \mathcal{N} = \langle \mathcal{L}, \mathcal{F}, \{c_l,\forall l \in \mathcal{L} \} \rangle$ be a network and let $ \mathcal{G} $ be its gradient graph. Let $\delta$ be an infinitesimally small perturbation performed on a flow or link $x \in \mathcal{L} \cup \mathcal{F}$, producing a drift $\Delta_y$, for all $y \in \mathcal{L} \cup \mathcal{F}$. Then, $\nabla_{x}(y) = \Delta_y / \delta \leq d^{D(\mathcal{G})/4}$, where $D(X)$ is the diameter of a graph $X$ and $d$ is the maximum indegree and outdegree of any vertex in the graph.
}







\begin{proof}

From the invariants of the flow and link equations, we observe that the absolute value of a perturbation can only increase when traversing a link vertex. This is because the flow equation $\Delta_f = \min_{l \in P_f} \Delta_l$ necessarily implies that the size of the perturbation will either stay the same or decrease. The link equation $\Delta_l = - \sum_{f \in P_l} \Delta_f / |S_l|$, however, allows perturbations to grow in absolute value. This will happen whenever the sum of the flow drifts arriving at a link vertex is larger than the outdegree of such vertex: $\sum_{f \in P_l} \Delta r_f > |S_l|$. The size of the perturbation will in fact maximally increase when the link outdegree is 1 and the sum of the flow drifts arriving at it is maximal. This is achieved when the bottleneck structure is configured with flows having an outdegree of $d$ and links having an indegree of $d$, connected by a stage of inter-medium links and flows of indegree and outdegree equal to 1, as shown in Fig. \ref{fig:gg_bound}. Concatenating this bottleneck structure block, we have that at each block the perturbation increases $d$ times. Because the length of this block is 4, there are a maximum of $D(\mathcal{G})/4$ blocks, where $D(\mathcal{G})$ is the diameter of the gradient graph. This leads to the upper bound $\nabla_{x}(y) = \Delta_y / \delta \leq d^{D(\mathcal{G})/4}$.

\begin{figure}[t]
\centering
\includegraphics[width=0.25\columnwidth]{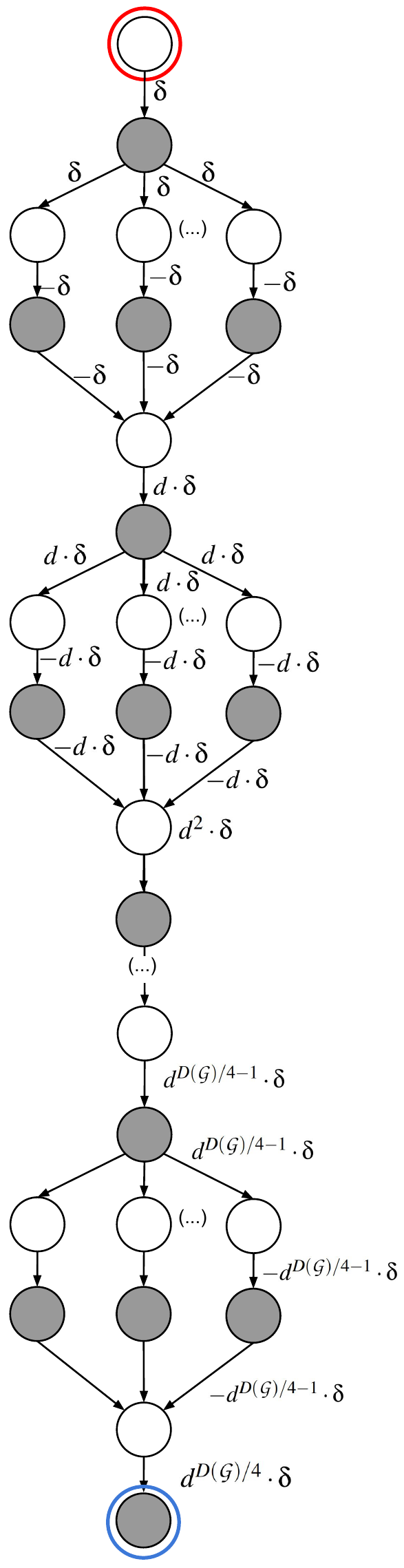}
\caption{Bottleneck structure with maximal drift used to prove the gradient bound lemma.}
\label{fig:gg_bound}
\end{figure}
\end{proof}


\subsection{Lemmas to Demonstrate the Correctness of the MaxRatePath Algorithm} \label{app:proof:maxratepath} 

\begin{lemma}{\textit{Flow rate decay with incremental hop count.}}\label{lem:maxratepath-1} 
Let $\mathcal{N} = \langle \mathcal{L}, \mathcal{F}, \{c_l,\forall l \in \mathcal{L} \}\rangle$ be a network and let $r_f$ be the transmission rate of a flow $f \in \mathcal{F}$. Let $r_f'$ be the new transmission rate of flow $f$ after we extend it to traverse an additional link $l^* \in \mathcal{L} \setminus f$. Then, $r_f' \leq r_f$.
\end{lemma} 
\begin{proof} 
Fig. \ref{fig:routing-proof}-a shows the initial situation of the lemma, with flow $f$ bottlenecked at a link $l$. Since the transmission rate of flow $f$ is $r_f$, we have that $s_{l} = r_f$. Suppose that we extend flow $f$ to traverse an extra link $l^* \in \mathcal{L} \setminus f$. Now consider the next set of transformations applied on the network:

\begin{enumerate}
    \item Create a new flow $f^*$ configured to only traverse link $l^*$.
    \item Add a traffic shaper $l_s$ to flow $f^*$ and set its rate to zero, i.e. $s_{l_s} = c_{l_s} = 0$.
    \item Increase the rate of the traffic shaper until either (a) $l^*$ becomes a bottleneck of $f^*$ or (b) $s_{l_s} = s_{l}$.
    \item Connect flows $f$ and $f^*$ together.
    \item Remove the traffic shaper $l_s$.
\end{enumerate}

It is easy to see that the above process yields a bottleneck structure that is the same as if flow $f$ had been extended to traverse the extra link $l^*$, since at the end of these steps the two flows $f$ and $f^*$ are merged into a single flow (effectively extending flow $f$ to traverse the additional link $l^*$) and the traffic shaper is removed.

Let us now derive the bottleneck structure of the network after applying the above transformations. Steps (1) and (2) are shown in Fig. \ref{fig:routing-proof}-b. In step (3), as we increase the capacity of the traffic shaper $s_{l_s}$, the rate of flow $f^*$ increases at the same pace. Suppose that condition (3-a) holds so that $l^*$ becomes a bottleneck of $f^*$. This situation is shown in Fig. \ref{fig:routing-proof}-c.1. Since flow $f^*$ is bottlenecked at both links $l^*$ and $l_s$, we have that $s_{l^*}=s_{l_s}$. Because condition (3-b) does not hold, it must also be that $s_{l_s} < s_{l}$, which then implies $s_{l^*} < s_{l}$. This necessarily means that, in step (4), the merging of the two flows $f$ and $f^*$ leads to the bottleneck structure shown in Fig. \ref{fig:routing-proof}-c.2, whereby the new flow $f$ is no longer bottlenecked at link $l$ and, instead, it becomes bottlenecked at link $l^*$. Thus, we have that $r_f' = s_{l^*} < s_{l} = r_f$. Note that in this case, the ripple effects of extending flow $f$ to traverse the extra link $l^*$ affected the performance of link $l$, resulting in an increase of its fair share $s_{l}$ value.

Assume instead that condition (3-b) holds so that $s_{l_s} = s_{l}$. In this case, the merging of flows $f$ and $f^*$ in step (4) leads to the bottleneck structure in Fig. \ref{fig:routing-proof}-d, whereby flow $f$ continues to be bottleneked at link $l$ and, thus, $r_f' = s_{l} = r_f$. Note that in this case, link $l$ was unaffected by the ripple effects of extending flow $f$ to traverse the extra link $l^*$.

Finally, in step 5 we can freely remove the traffic shaper $l_s$ from the network without producing any ripple effect, since flow $f$ is also bottlenecked at either link $l^*$ (case 3-a) or link $l$ (case 3-b). 

In conclusion, we have that at the end of this process, $r_f' \leq r_f$.

\begin{figure}[t]
\centering
\includegraphics[width=0.8\columnwidth]{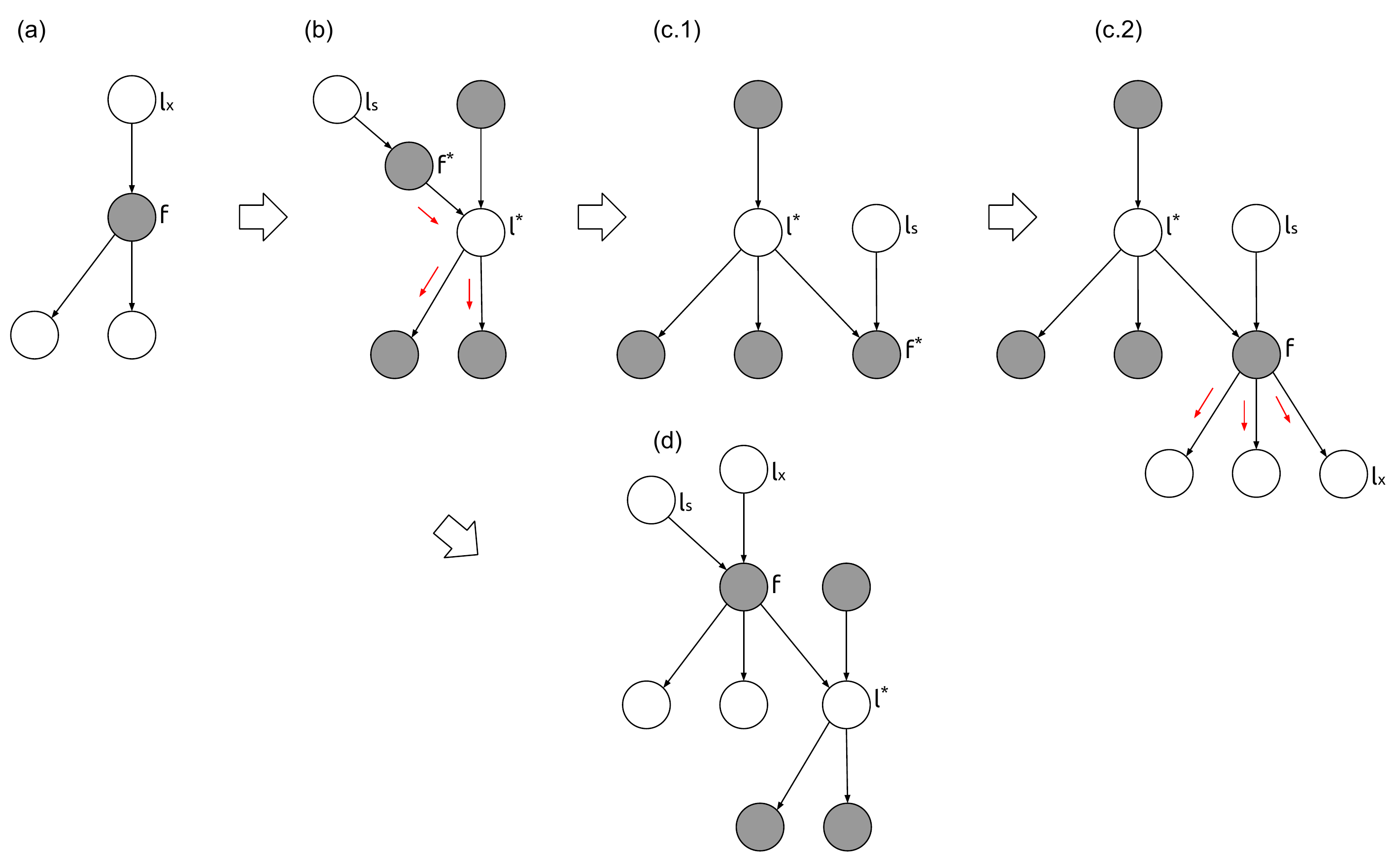}
\caption{Bottleneck structures transformations used to demonstrate that adding a new link to a flow cannot increment its transmission rate.}
\label{fig:routing-proof}
\end{figure}

\end{proof}

\begin{corollary}{\textit{New bottleneck with incremental hop count.}}\label{cor:maxratepath-2}
Let $\mathcal{N} = \langle \mathcal{L}, \mathcal{F}, \{c_l,\forall l \in \mathcal{L} \}\rangle$ be a network and let $r_f$ be the transmission rate of a flow $f \in \mathcal{F}$. Let $r_f'$ be the new transmission rate of flow $f$ after we extend it to traverse an additional link $l^* \in \mathcal{L} \setminus f$. If, $r_f' < r_f$, then the newly extended flow is bottlenecked at link $l^*$. 
\end{corollary}
\begin{proof} 
This result is a direct consequence of Lemma \ref{lem:maxratepath-1}, since the condition $r_f' < r_f$ corresponds to case (3-a), which leads to the bottleneck structure in Fig. \ref{fig:routing-proof}-c.2.
\end{proof}

\textsc{Lemma} \ref{lem:maxratepath}. \textit{
Let $\mathcal{N} = \langle \mathcal{L}, \mathcal{F}, \{c_l,\forall l \in \mathcal{L} \}\rangle$ be a network and $\mathcal{U}$ the set of its routers. Suppose that $f$ and $f'$ are two flows not in $\mathcal{F}$ that originate at router $u_s$ and end at router $u_d$. Then $f = \mathrm{MaxRatePath}(\mathcal{N}, \mathcal{U} , u_s, u_d)$ implies $r_f \geq r_{f'}$.
}
\begin{proof} 

Consider the network configuration in Fig. \ref{fig:routing-proof-2} and assume that routing data from router $u_s$ to $u_x$ using flow $f_1$ leads to a higher transmission rate than using flow $f_2$, $r_{f_1} > r_{f_2}$. Since the MaxRatePath algorithm uses the inverse of the rate as the path cost metric, this implies that $d_{f_1} < d_{f_2}$, where we use the notation $d_f$ to denote the cost of using flow $f$ to route traffic through the network. To demonstrate the correctness of the algorithm, we need to show that $d_{f_1} < d_{f_2}$ implies $d_{f_1 \cup \{l^*\}} \leq d_{f_2\cup \{l^*\}}$, since this condition is enough to demonstrate convergence in the Dijkstra algorithm \cite{Cormen2009}.

We will assume that $d_{f_1} < d_{f_2}$ and $d_{f_1 \cup \{l^*\}} > d_{f_2\cup \{l^*\}}$ are both true and arrive at a contradiction. From Lemma \ref{lem:maxratepath-1} we have that $d_{f_2} \leq d_{f_2\cup \{l^*\}}$. This implies that $d_{f_1 \cup \{l^*\}} > d_{f_2\cup \{l^*\}} \geq d_{f_2} > d_{f_1}$. Using Corollary \ref{cor:maxratepath-2}, it must be that $l^*$ is the bottleneck of the flow $f_1 \cup \{l^*\}$. Now since flow $f_2 \cup \{l^*\}$ also traverses link $l^*$, it must be that its rate cannot be higher than that of flow $f_1 \cup \{l^*\}$. But this implies $d_{f_1 \cup \{l^*\}} \leq d_{f_2 \cup \{l^*\}}$, arriving at a contradiction.

\begin{figure}[t]
\centering
\includegraphics[width=0.4\columnwidth]{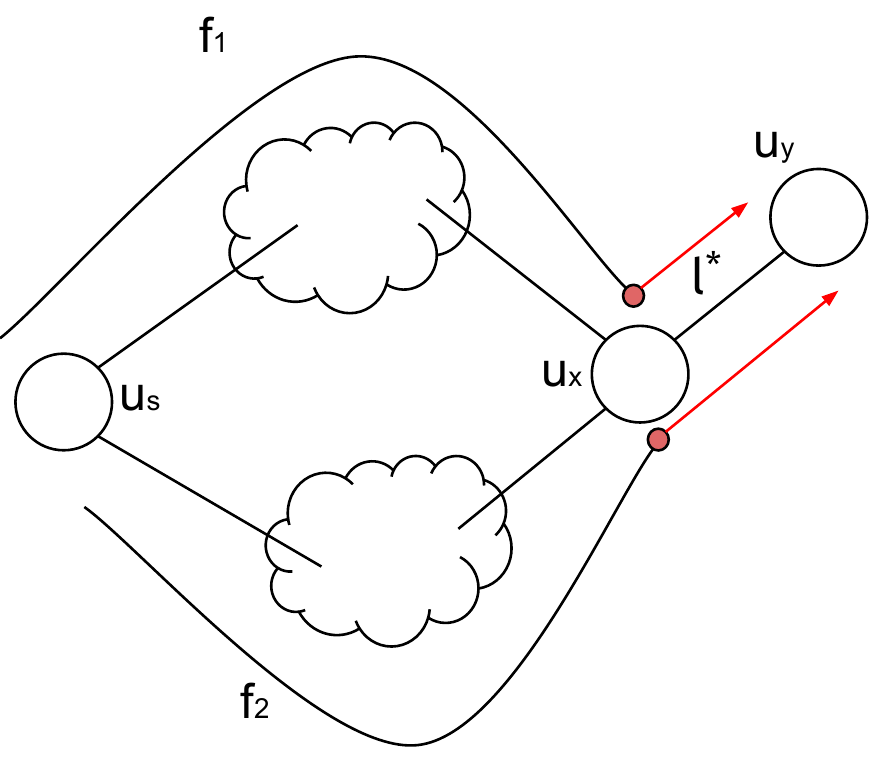}
\caption{In the MaxRatePath algorithm, adding a new link to a flow preserves the correctness of the previous high-throughput path decisions.}
\label{fig:routing-proof-2}
\end{figure}

\end{proof}

\section{Mininet-G2} 
\label{app:g2mininetinfo}
The Mininet-G2 tool \cite{g2MininetRepo} provides a powerful, flexible interface to emulate networks of choice with customizable topology, routing and traffic flow configurations. It uses Mininet \cite{mininetWebsite} and the POX SDN controller \cite{poxWiki} to create such highly customizable networks. It also uses iPerf \cite{iperfWebsite} internally to generate network traffic and offers an interface to configure various flow parameters such as the source and destination hosts, start time, and data size, among others. Mininet-G2 also offers an integration with sFlow-RT \cite{sflow01} agent that enables real-time access to traffic flows from Mininet emulated network. Since Mininet uses real, production grade TCP/IP stack from the Linux kernel, it proves to be an ideal testbed to run experiments using congestion control protocols such as BBR and Cubic to study bottleneck structures and flow performance in a realistic way. Apart from its flexible configuration interface, Mininet-G2 also offers a set of useful utilities to compute and plot various performance metrics such as instantaneous network throughput, flow convergence time, flow completion time, Jain's fairness index among others for a given experiment.

\section{Experiments with TCP Cubic}
\label{app:exp:cubic} 
\begin{figure*}
\centering

\begin{subfigure}[b]{.3\linewidth}
\includegraphics[width=\linewidth]{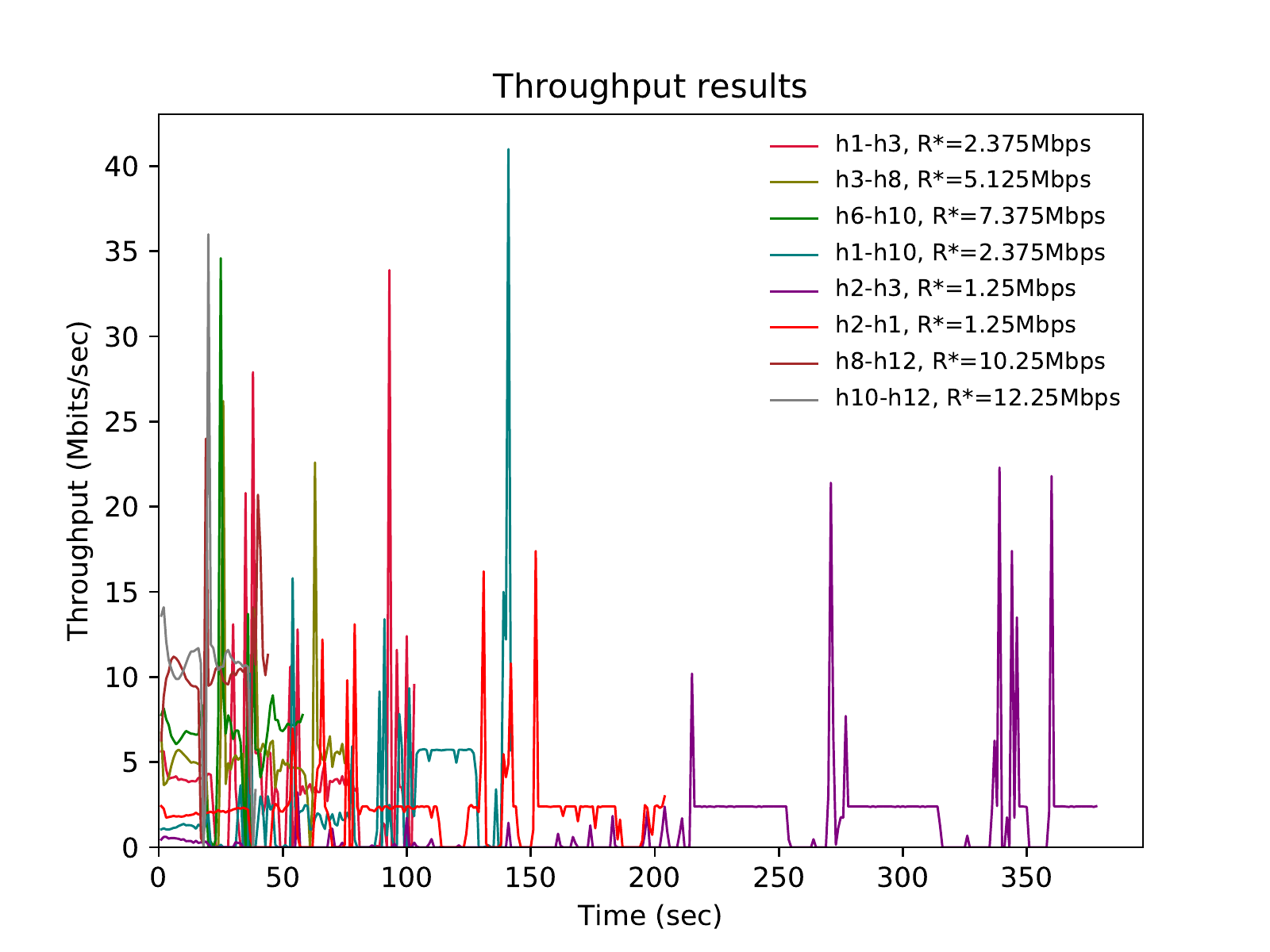}
\caption{Without any traffic shaping.}
\end{subfigure}%
\begin{subfigure}[b]{.3\linewidth}
\includegraphics[width=\linewidth]{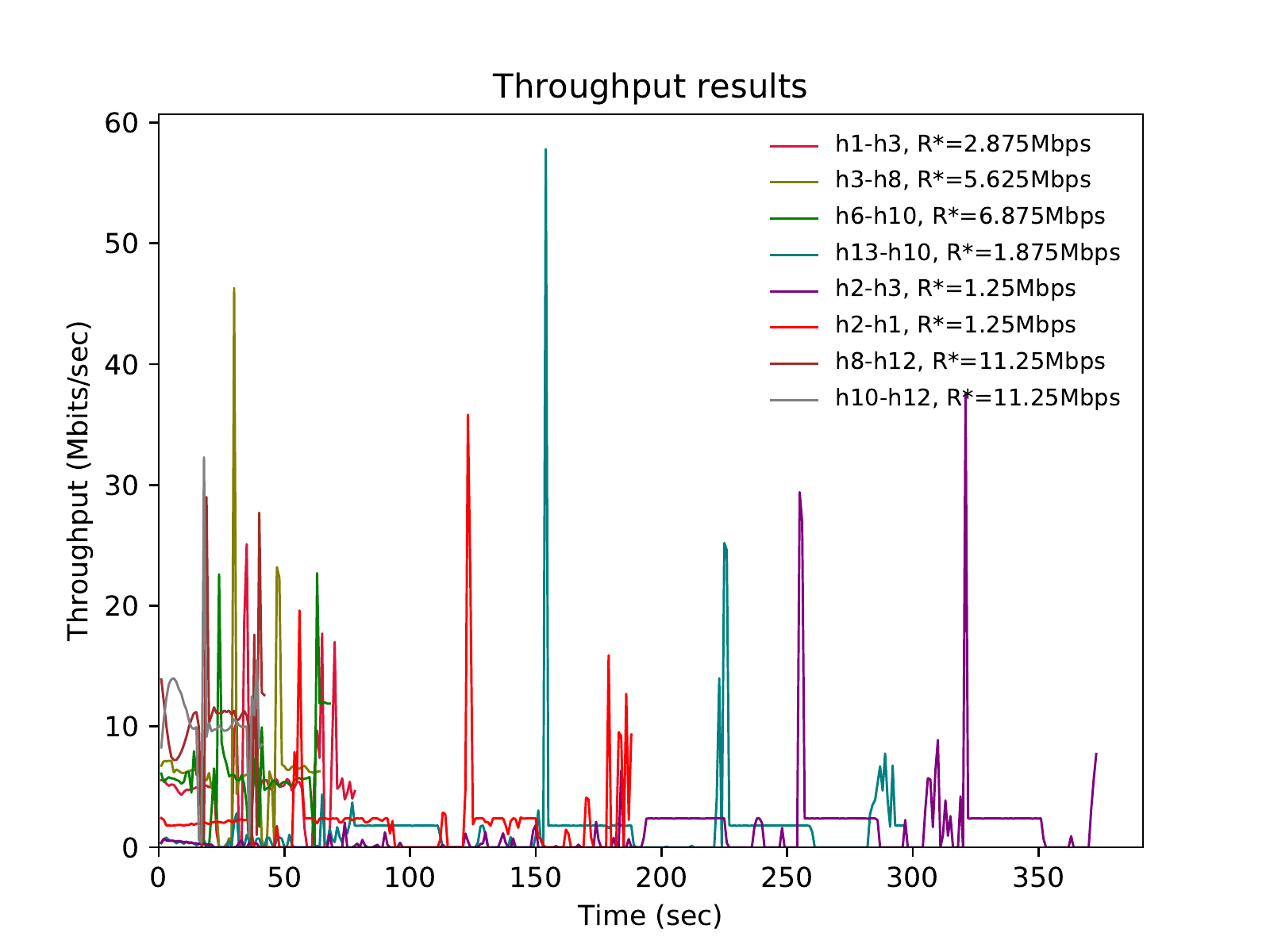}
\caption{Traffic shaping $f_4$.}
\end{subfigure}
\begin{subfigure}[b]{.3\linewidth}
\includegraphics[width=\linewidth]{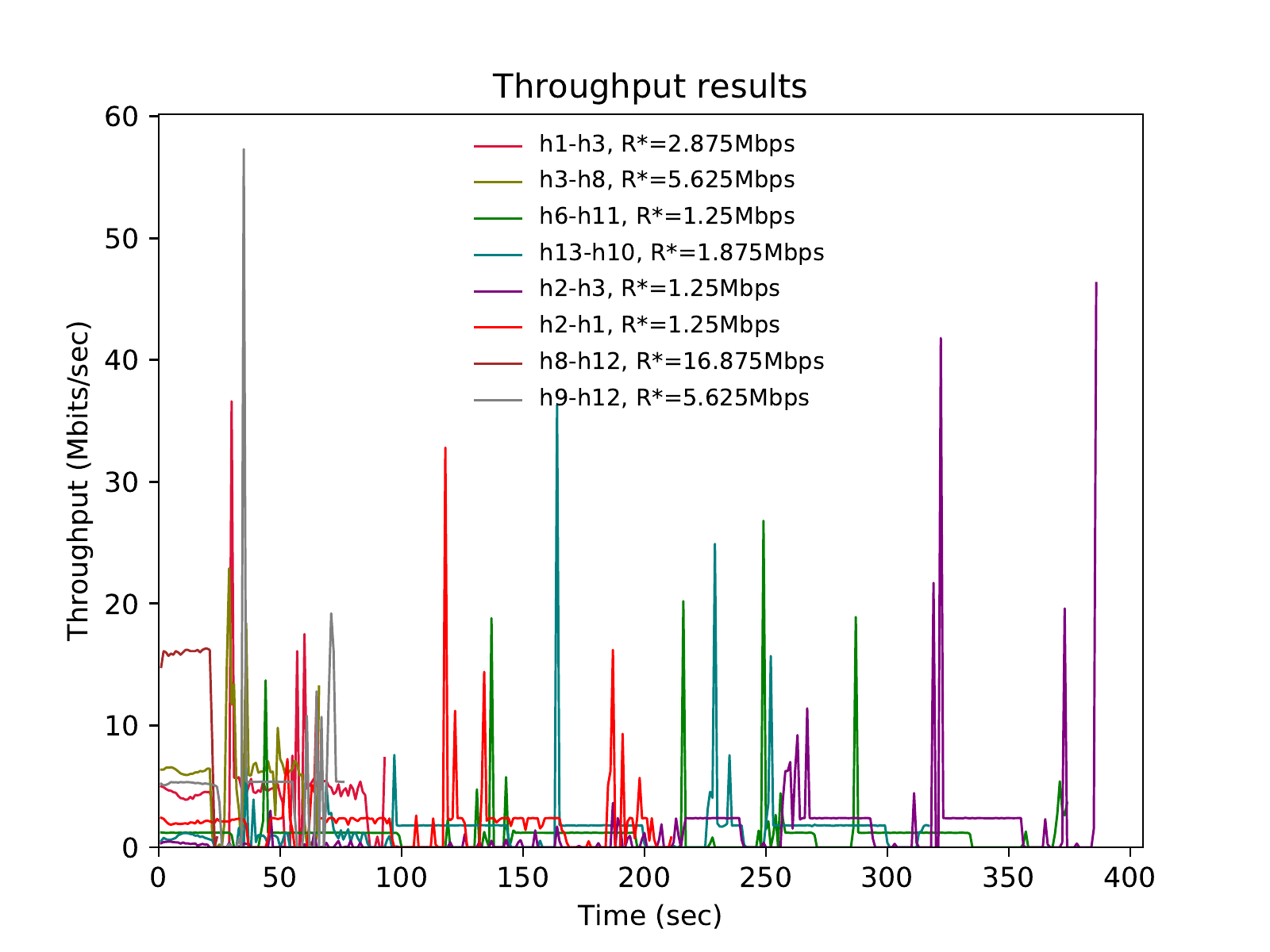}
\caption{Traffic shaping $f_3$,$f_4$, and $f_8$.}
\end{subfigure}

\caption{Traffic shaping schedule to accelerate flow $f_7$ ($h_{8}-h_{12}$) (TCP Cubic).}
\label{fig:ex_timebound_1_cubic}
\end{figure*}

\begin{figure*}
\centering

\begin{subfigure}[b]{.3\linewidth}
\includegraphics[width=\linewidth]{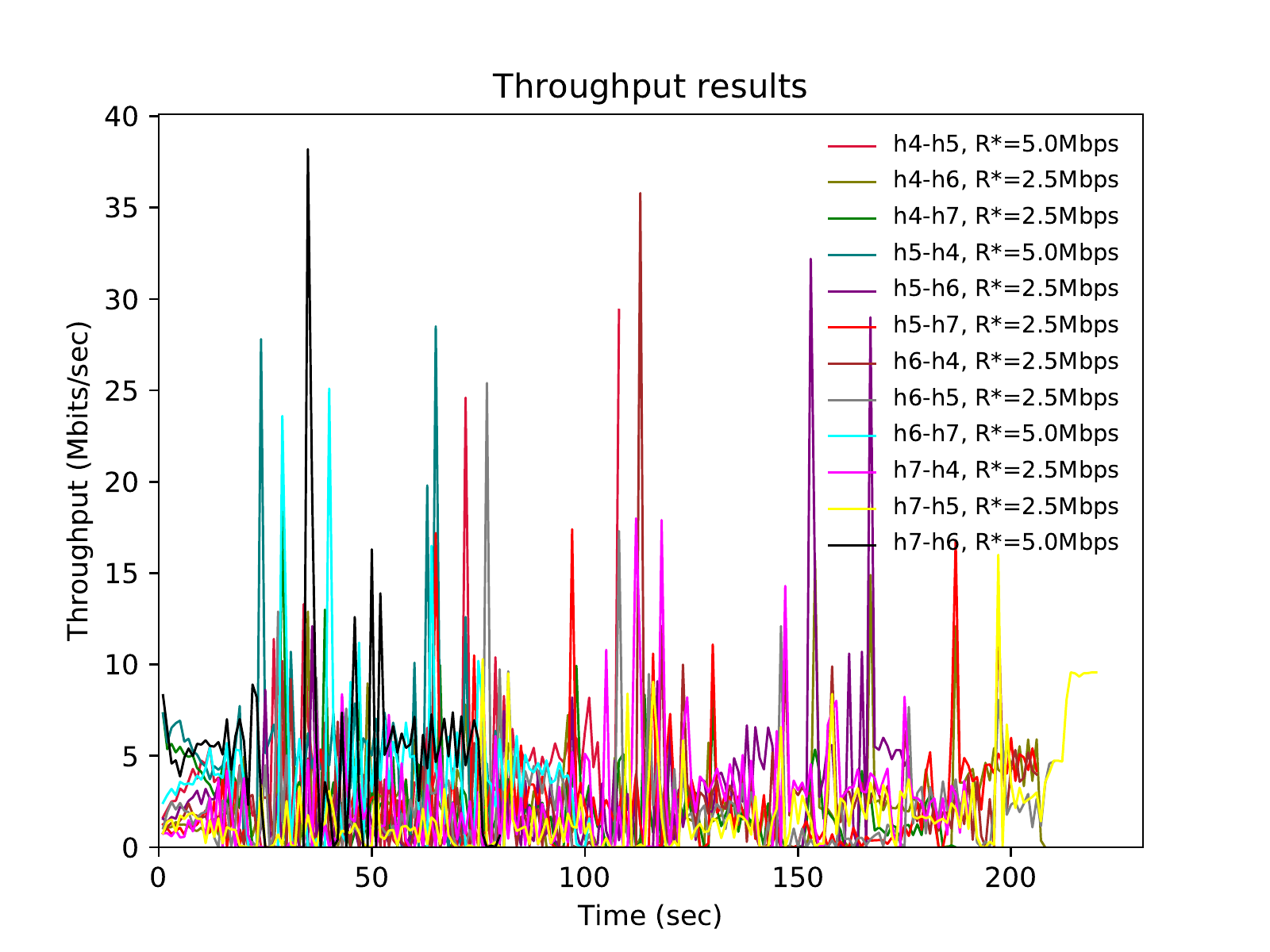}
\caption{Tapering parameter $\tau=1$.}
\end{subfigure}%
\begin{subfigure}[b]{.3\linewidth}
\includegraphics[width=\linewidth]{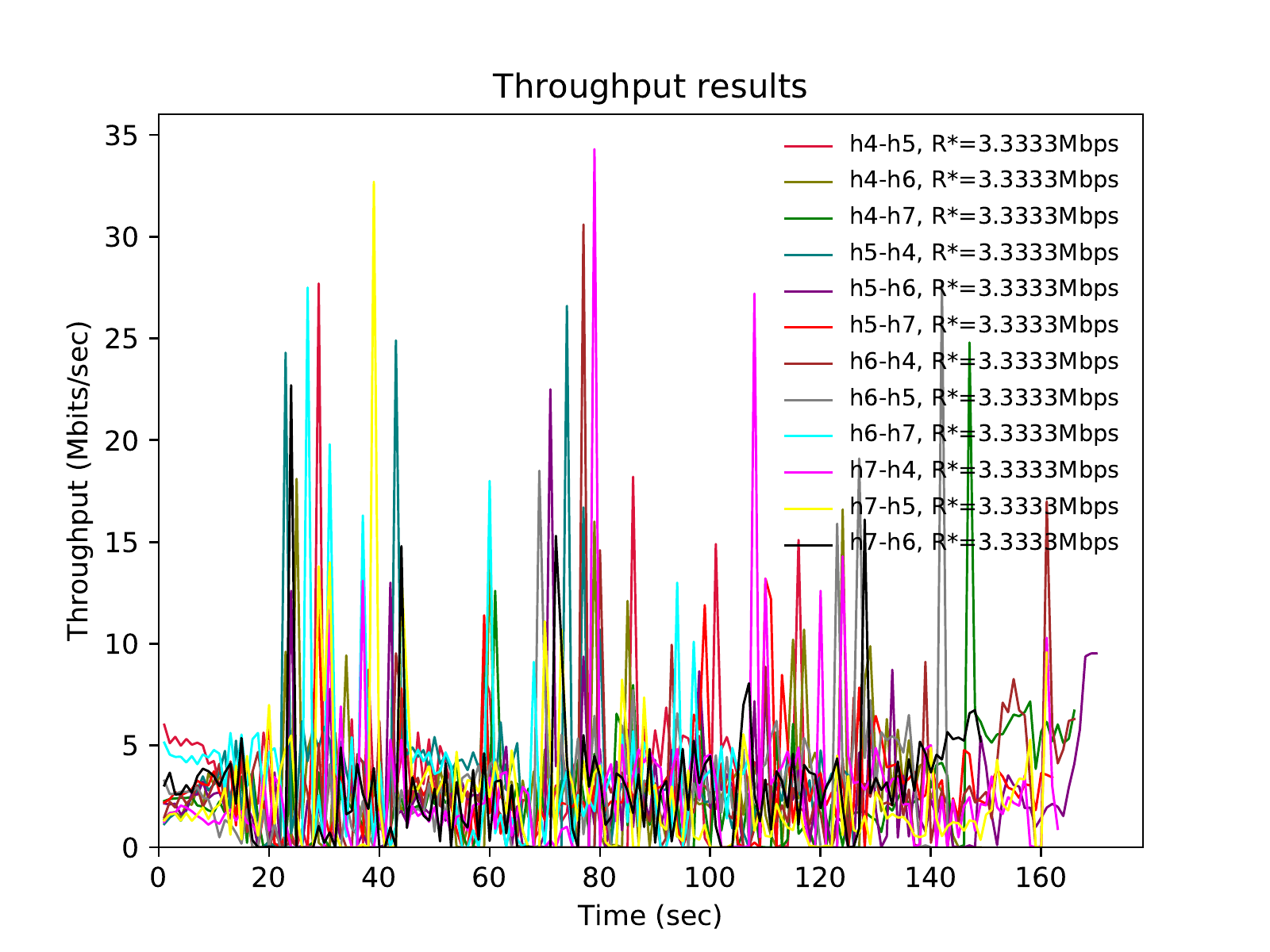}
\caption{Tapering parameter $\tau=4/3$.}
\end{subfigure}
\begin{subfigure}[b]{.3\linewidth}
\includegraphics[width=\linewidth]{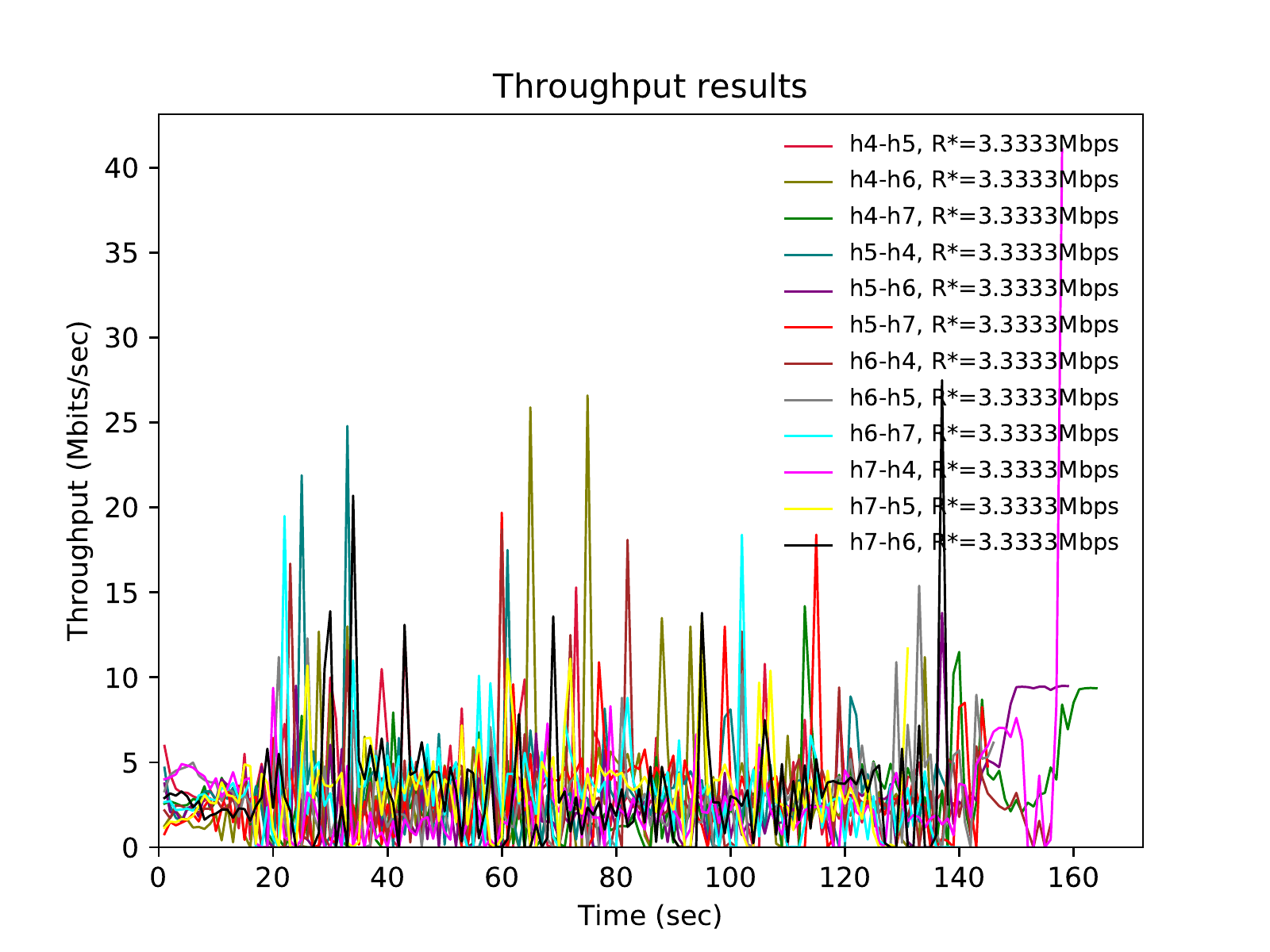}
\caption{Tapering parameter $\tau=2$.}
\end{subfigure}

\caption{Optimizing bandwidth tapering on a 3-level binary fat-tree (TCP Cubic).}
\label{fig:ex_clos_1_cubic}
\end{figure*}

Tables \ref{tab:time-bound-cubic}, \ref{tab:route_cubic} and \ref{tab:clos_cubic} and Fig. \ref{fig:ex_timebound_1_cubic} and \ref{fig:ex_clos_1_cubic} present the results for the experiments described in Sections. \ref{ssec:timebound}, \ref{ssec:routing} and \ref{ssec:clos} when using Cubic as the congestion control protocol. 

\begin{table}
\center
\caption{Experimental versus theoretical average flow transmission rate (units in Mbps) for Section. \ref{ssec:timebound} when using TCP Cubic.}
\label{tab:time-bound-cubic}
\scalebox{1}{
\begin{tabular} {  c  c  c  c  c  c  c  c}
\hline
Flow & Experiment 1 & Experiment 2 & Experiment 3
\\ \hline
$f_1$ & 3.91 / 2.37 & 5.10 / 2.87 & 4.43 / 2.87 \\
$f_2$ & 5.26 / 5.12 & 6.39 / 5.62 & 5.94 / 5.62 \\
$f_3$ & 6.83 / 7.37 & 6.17 / 6.87 & 1.04 / 1.25 \\
$f_4$ & 2.74 / 2.37 & 1.40 / 1.87 & 1.38 1.87 \\
$f_5$ & 1.09 / 1.25 & 1.15 / 1.25 & 1.14 / 1.25 \\
$f_6$ & 2.04 / 1.25 & 2.10 / 1.25 & 2.01 / 1.25 \\
$f_7$ & 10.22 / 10.25 & 10.49 / 11.25 & 14.4 / 16.87 \\
$f_8$ & 10.62 / 12.25 & 10.58 / 11.25 & 5.37 / 5.62 \\

\hline
\end{tabular}
}
\end{table}

\begin{table}
\center
\caption{Experimental versus theoretical average flow transmission rate (units in Mbps) for Section. \ref{ssec:routing} when using TCP Cubic.}
\label{tab:route_cubic}
\scalebox{0.8}{
\begin{tabular} {  c  c  c  c  c  c}
\hline
Flow & Shortest path & Longer path & Flow & Shortest path & Longer path 

\\ \hline
$f_1$ &  0.917 / 1.428  & 0.962 / 1.666  & $f_{14}$ &  1.841 / 1.666  & 1.864 / 1.666  \\
$f_2$ &  1.296 / 1.666  & 1.278 / 1.666   & $f_{15}$ &  1.284 / 1.666  & 1.230 / 1.666  \\
$f_3$ &  1.202 / 1.428  & 1.315 / 1.666   & $f_{16}$ &  1.294 / 1.666  & 1.292 / 1.666  \\
$f_4$ &  0.897 / 1.428  & 0.9685 / 1.666   & $f_{17}$ &  2.035 / 2.142  & 2.132 / 2.142  \\
$f_5$ &  1.186 / 1.428  & 1.336 / 1.666   & $f_{18}$ &  2.097 / 2.142  & 2.141 / 2.142  \\
$f_6$ &  2.227 / 3.000  & 2.126 / 2.500   & $f_{19}$ &  3.792 / 2.142  & 4.065 / 2.142  \\
$f_7$ &  1.716 / 1.428  & 1.966 / 1.666   & $f_{20}$ &  2.101 / 2.142  & 2.115 / 2.142  \\
$f_8$ &  1.211 / 1.428  & 1.333 / 1.666   & $f_{21}$ &  2.195 / 3.000  & 2.170 / 2.500  \\
$f_9$ &  1.000 / 2.142  & 1.022 / 2.142   & $f_{22}$ &  4.168 / 3.000  & 3.767 / 2.500  \\
$f_{10}$ &  1.291 / 1.666  & 1.296 / 1.666  & $f_{23}$ &  2.189 / 3.000  & 2.046 / 2.500  \\
$f_{11}$ &  1.411 / 2.142  & 1.397 / 2.142  & $f_{24}$ &  2.242/ 3.000  & 2.104 / 2.500  \\
$f_{12}$ &  0.984 / 2.142  & 0.999 / 2.142  & $f_{25}$ &  1.679 / 1.428  & 1.377 / 2.500  \\
$f_{13}$ &  1.276 / 1.666  & 1.252 / 1.666  \\ 

\hline
\end{tabular}
}
\end{table}

\begin{table}
\center
\caption{Flow completion time (seconds) for Section. \ref{ssec:clos} when using TCP Cubic.}
\label{tab:clos_cubic}
\scalebox{0.9}{
\begin{tabular} {  c  c  c  c  c  c  c  c}
\hline
Flow & $\tau = 1$ & $\tau = 4/3$ & $\tau = 2$ & Flow & $\tau = 1$ & $\tau = 4/3$ & $\tau = 2$ 

\\ \hline
$f_1$ &  108  & 120  & 139 & $f_7$ &  206  & 166  & 156 \\
$f_2$ &  208  & 149  & 135 & $f_8$ &  210  & 143  & 146 \\
$f_3$ &  187  & 166  & 164 & $f_9$ &  102  & 107  & 140 \\
$f_4$ &  80  & 127  & 139 & $f_{10}$ &  190  & 163  & 158 \\
$f_5$ &  176  & 170  & 159 & $f_{11}$ &  220  & 161  & 131 \\
$f_6$ &  206  & 162  & 145 & $f_{12}$ &  80  & 149  & 138 \\
\hline
max() & & & & & 220 & 170 & 164 \\
\hline
\end{tabular}
}
\end{table}

\section{Jain's Fairness Index Results} 
\label{app:jainfidx}
Jain's index \cite{jain-fairness-index-journals/corr/cs-NI-9809099} is a metric that rates the fairness of a set of values $x_1, x_2,...,x_n$ according to the following equation:
$$
\mathcal{J}(x_1,x_2, ...,x_n) = \frac{(\sum_{i=1}^{n}x_i)^2}{n\cdot\sum_{i=1}^{n}x_i^2} = 
\frac{\ \overline{\mathbf{x}}^2\ }{\ \overline{\mathbf{x}^2}\ }
$$
The index value ranges from $\frac{1}{n}$ (worst case) to $1$ (best case). As suggested in \cite{jain-fairness-index-journals/corr/cs-NI-9809099}, for multi-link networks the value $x_i$ must be normalized to an optimal fairness allocation. Throughout this paper, we normalize $x_i$ as the ratio $f_i/O_i$, where $f_i$ is the rate of flow $f_i$ achieved through the experiments and $O_i$ is its expected max-min fair throughput. This provides an index that qualitatively measures how closely the rates obtained from the experiments are to the theoretical rates predicted by the bottleneck structure of the network. The closer this index is to 1, the more accurate the mathematical model is to the experimental results. Table. \ref{tab:fairnessidx_timebound} shows the Jain's fairness index we obtained for all the experiments presented in this paper (Sections. \ref{ssec:timebound}, \ref{ssec:routing} and \ref{ssec:clos}). 

\begin{table}
\caption{Jain's Fairness index for all the experiments}
\label{tab:fairnessidx_timebound}
\scalebox{0.9}{
\begin{tabular}{  c  c  c  c  }
\hline
Algorithm & \ref{ssec:timebound}:Experiment 1 & \ref{ssec:timebound}:Experiment 2 & \ref{ssec:timebound}:Experiment 3 \\ \hline
BBR                               & 0.9926   &  0.9965 & 0.9985 \\
Cubic                             & 0.9353   &  0.9074 & 0.9218 \\
\hline
Algorithm & \ref{ssec:routing}:Experiment 1 & \ref{ssec:routing}:Experiment 2 & \\ \hline
BBR                              & 0.9954     &  0.9966 & \\
Cubic                            & 0.9077     &  0.8868 & \\
\hline
Algorithm & \ref{ssec:clos}:$\tau = 1$ &\ref{ssec:clos}:$\tau = 4/3$& \ref{ssec:clos}:$\tau = 2$ \\ \hline
BBR                               & 0.9987     &  0.9983 & 0.9939 \\
Cubic                             & 0.9903     &  0.9842 & 0.9957 \\
\hline
\end{tabular}
}
\end{table}

\section{Routing Configuration Used in Experiments}
\label{app:flowpaths} 
Table. \ref{tab:flowpaths_routing} presents the specific route configurations used for various experiments in Sections. \ref{ssec:routing} and \ref{ssec:clos}.

\begin{table}
\center
\caption{Path followed by each flow in the routing optimization experiments (Section \ref{ssec:routing})}
\label{tab:flowpaths_routing}
\scalebox{0.8}{
\begin{tabular} {  c  c  c  c  }
Experiment 1: \\
\hline
Flow & Links traversed & Flow & Links traversed
\\ \hline
$f_1$ &  $\{l_3,l_{15},l_{10},l_{18} \}$ & $f_{14}$ & $\{l_7,l_8 \}$ \\
$f_2$ & $\{l_5,l_7,l_8 \}$ & $f_{15}$ & $\{l_7,l_8,l_{19} \}$ \\
$f_3$ & $\{l_3,l_{15},l_{10} \}$ & $f_{16}$ & $\{l_7,l_8,l_{11} \}$ \\
$f_4$ & $\{l_3,l_{15},l_{10},l_{14} \}$ & $f_{17}$ & $\{l_{10},l_{18} \}$ \\
$f_5$ & $\{l_{15},l_{10},l_{18} \}$ & $f_{18}$ & $\{l_{10},l_{19} \}$ \\
$f_6$ & $\{l_{16},l_8 \}$ & $f_{19}$ & $\{l_{10} \}$ \\
$f_7$ & $\{l_{15},l_{10} \}$ & $f_{20}$ & $\{l_{10},l_{14} \}$ \\
$f_8$ & $\{l_{15},l_{10},l_{14} \}$ & $f_{21}$ & $\{l_8,l_9 \}$ \\
$f_9$ & $\{l_{13},l_6,l_{10},l_{18} \}$ & $f_{22}$ & $\{l_8 \}$ \\
$f_{10}$ & $\{l_{13},l_7,l_8 \}$ & $f_{23}$ & $\{l_8,l_{19} \}$ \\
$f_{11}$ & $\{l_{13},l_6,l_{10} \}$ & $f_{24}$ & $\{l_8,l_{11} \}$ \\
$f_{12}$ & $\{l_{13},l_6,l_{10},l_{14} \}$ & $f_{25}$ & $\{l_{15},l_{10} \}$ \\
$f_{13}$ & $\{l_7,l_8,l_9 \}$ & & \\
\hline
\end{tabular}
}
\scalebox{0.8}{
\begin{tabular} {  c  c  c  c  }
Experiment 2: \\
\hline
Flow & Links traversed & Flow & Links traversed
\\ \hline
$f_1$ &  $\{l_3,l_{15},l_{10},l_{18} \}$ & $f_{14}$ & $\{l_7,l_8 \}$ \\
$f_2$ & $\{l_5,l_7,l_8 \}$ & $f_{15}$ & $\{l_7,l_8,l_{19} \}$ \\
$f_3$ & $\{l_3,l_{15},l_{10} \}$ & $f_{16}$ & $\{l_7,l_8,l_{11} \}$ \\
$f_4$ & $\{l_3,l_{15},l_{10},l_{14} \}$ & $f_{17}$ & $\{l_{10},l_{18} \}$ \\
$f_5$ & $\{l_{15},l_{10},l_{18} \}$ & $f_{18}$ & $\{l_{10},l_{19} \}$ \\
$f_6$ & $\{l_{16},l_8 \}$ & $f_{19}$ & $\{l_{10} \}$ \\
$f_7$ & $\{l_{15},l_{10} \}$ & $f_{20}$ & $\{l_{10},l_{14} \}$ \\
$f_8$ & $\{l_{15},l_{10},l_{14} \}$ & $f_{21}$ & $\{l_8,l_9 \}$ \\
$f_9$ & $\{l_{13},l_6,l_{10},l_{18} \}$ & $f_{22}$ & $\{l_8 \}$ \\
$f_{10}$ & $\{l_{13},l_7,l_8 \}$ & $f_{23}$ & $\{l_8,l_{19} \}$ \\
$f_{11}$ & $\{l_{13},l_6,l_{10} \}$ & $f_{24}$ & $\{l_8,l_{11} \}$ \\
$f_{12}$ & $\{l_{13},l_6,l_{10},l_{14} \}$ & $f_{25}$ & $\{l_{16},l_{8},l_{19},l_{20} \}$ \\
$f_{13}$ & $\{l_7,l_8,l_9 \}$ & & \\
\hline
\end{tabular}
}
\end{table}

\begin{table}
\center
\caption{Path followed by each flow in the fat-tree networks experiments (Section \ref{ssec:clos})}
\label{tab:flowpaths_clos}
\scalebox{1}{
\begin{tabular} {  c  c  }
\hline
Flow & Experiment 1,2,3:Links traversed  
\\ \hline
$f_1$ & $\{l_1,l_2 \}$ \\
$f_2$ & $\{l_1,l_5,l_6,l_3 \}$ \\
$f_3$ & $\{l_1,l_5,l_6,l_4 \}$ \\
$f_4$ & $\{l_2,l_1 \}$ \\
$f_5$ & $\{l_2,l_5,l_6,l_3 \}$ \\
$f_6$ & $\{l_2,l_5,l_6,l_4 \}$ \\
$f_8$ & $\{l_3,l_6,l_5,l_2 \}$ \\
$f_9$ & $\{l_3,l_4 \}$ \\
$f_{10}$ & $\{l_4,l_6,l_5,l_1 \}$ \\
$f_{11}$ & $\{l_4,l_6,l_5,l_2 \}$ \\
$f_{12}$ & $\{l_4,l_3 \}$ \\
\hline
\end{tabular}
}
\end{table}

\section{Performance of Flow $f_{50}$ in the High-throughput Routing Problem} 
\label{app:routing}
\begin{figure}[t]
\centering
\includegraphics[width=0.6\columnwidth]{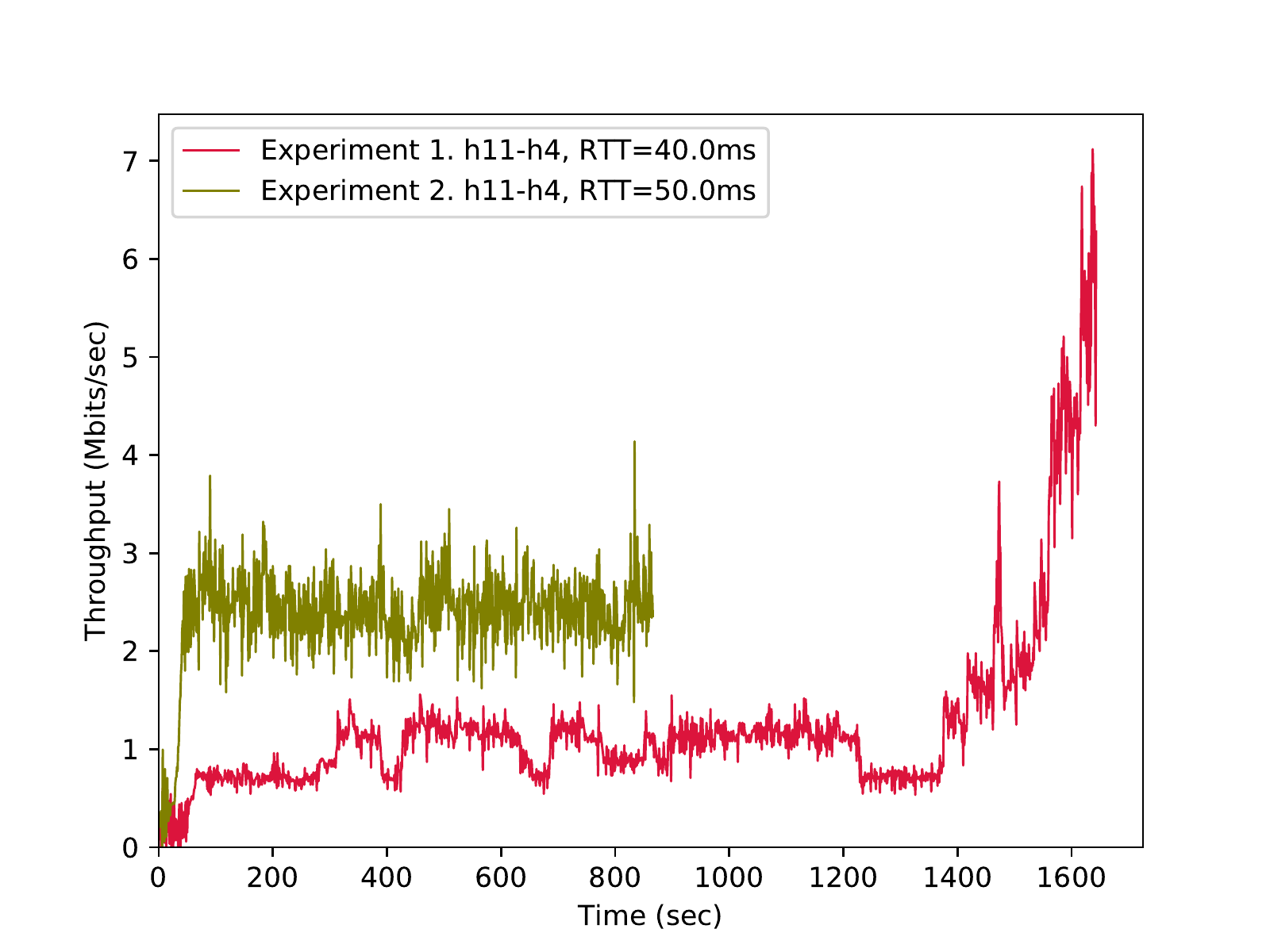}
\caption{Acceleration of flow $f_{50}$ by routing it through the high-bandwidth path.}
\label{fig:app_routing_flow50}
\end{figure}

Fig. \ref{fig:app_routing_flow50} shows the performance of flow $f_{50}$ in the experiment presented in Section \ref{ssec:routing}. As expected, its performance is very similar to flow $f_{25}$ shown in Fig. \ref{fig:ex_routing_flow25}.

\section{Performance of Flows $f_{26}$-$f_{50}$ in the High-throughput Routing Problem} 
\label{app:routing_flows_2650}
Table. \ref{tab:route_flows_2650} presents the performance of flows $f_{26}$-$f_{50}$ for the experiment described in Section. \ref{ssec:routing}.

 \begin{table}[ht]
 \center
 \caption{Experimental versus theoretical average flow transmission rate (units in Mbps).}
 \label{tab:route_flows_2650}
\scalebox{0.8}{
 \begin{tabular} {  c  c  c  c  c  c}
 \hline
 Flow & Shortest path & Longer path & Flow & Shortest path & Longer path 

 \\ \hline
 $f_{26}$ &  1.513 / 1.428  & 1.678 / 1.666  & $f_{39}$ &  1.452 / 1.666  & 1.443 / 1.666  \\
 $f_{27}$ &  1.580 / 1.666  & 1.572 / 1.666   & $f_{40}$ &  1.537 / 1.666  & 1.636 / 1.666  \\
 $f_{28}$ &  1.449 / 1.428  & 1.515 / 1.666   & $f_{41}$ &  1.576 / 1.666  & 1.564 / 1.666  \\
 $f_{29}$ &  1.523 / 1.428  & 1.595 / 1.666   & $f_{42}$ &  1.813 / 2.142  & 1.855 / 2.142  \\
 $f_{30}$ &  1.327 / 1.428  & 1.494 / 1.666   & $f_{43}$ &  1.813 / 2.142  & 1.903 / 2.142  \\
 $f_{31}$ &  2.605 / 3.000  & 2.230 / 2.500   & $f_{44}$ &  1.793 / 2.142  & 1.824 / 2.142  \\
 $f_{32}$ &  1.249 / 1.428  & 1.384 / 1.666   & $f_{45}$ &  1.831 / 2.142  & 1.852 / 2.142  \\
 $f_{33}$ &  1.340 / 1.428  & 1.449 / 1.666   & $f_{46}$ &  2.637 / 3.000  & 2.182 / 2.500  \\
 $f_{34}$ &  2.203 / 2.142  & 2.226 / 2.142   & $f_{47}$ &  2.465 / 3.000  & 2.108 / 2.500  \\
 $f_{35}$ &  1.536 / 1.666  & 1.522 / 1.666  & $f_{48}$ &  2.553 / 3.000  & 2.158 / 2.500  \\
 $f_{36}$ &  1.924 / 2.142  & 1.971 / 2.142  & $f_{49}$ &  2.630 / 3.000  & 2.235 / 2.500  \\
 $f_{37}$ &  2.065 / 2.142  & 2.253 / 2.142  & $f_{50}$ &  1.240 / 1.428  & 2.359 / 2.500  \\
 $f_{38}$ &  1.520 / 1.666  & 1.511 / 1.666  \\ 

\hline
\end{tabular}
}

\end{table}

\section{Using QTBS in Production Networks} 
\label{app:g2prodnws}

To construct the gradient graph of a network, only the information about a network $ \mathcal{N} = \langle \mathcal{L}, \mathcal{F}, $ $ \{c_l,\forall l \in \mathcal{L} \} \rangle$ is needed. The set of flows $\mathcal{F}$ can be obtained from traditional network monitoring tools such as NetFlow \cite{netflowCisco} or sFlow \cite{sflow01}. For each flow, the \textit{GradientGraph()} procedure (Algorithm \ref{al:GradientGraph}) needs to know the set of links it traverses. This information can also be obtained from NetFlow or sFlow provided that traffic sampling is performed at all the switches and routers of a network, as is often the case with production networks. If that is not the case, then the set of links traversed by each flow can also be derived by looking up the routing tables, for instance using a BGP collector \cite{BGP-RFC} or traceroute-like route discovery applications. The set of links $\mathcal{L}$ and their capacity $\{c_l,\forall l \in \mathcal{L} \} $ can be derived from protocols like SNMP \cite{snmpSpec} or simply from network topology information usually available to the network operator.

\end{document}